\documentclass[showpacs,amsmath,nofootinbib,amssymb,twocolumn,superscriptaddress]{revtex4-1}
\usepackage{array}
\usepackage{color}
\usepackage{multirow}
\usepackage{dcolumn}
\usepackage{bm}
\usepackage{epsf}
\usepackage{graphicx}
\usepackage{amsmath} 
\usepackage{dsfont} 
\usepackage{tikz}
\usepackage{hyperref}

\newcommand{\beq}{\begin{equation}}
\newcommand{\eeq}{\end{equation}}
\newcommand{\bea}{\begin{eqnarray}}
\newcommand{\eea}{\end{eqnarray}}

\begin{document}
\topmargin-0.5in
\textheight 8.7in 
\bibliographystyle{apsrev}
\title{Neutrino non-radiative decay and the diffuse supernova neutrino background}
\author{Pilar Iv\'a$\tilde{\rm n}$ez-Ballesteros}
\email{ivanez@apc.in2p3.fr}
\affiliation{Universit\'e Paris Cit\'e, Astroparticule et Cosmologie, F-75013 Paris, France}

\author{M. Cristina Volpe}
\email{volpe@apc.in2p3.fr (corresponding author)}
\affiliation{CNRS, Universit\'e Paris Cit\'e, Astroparticule et Cosmologie, F-75013 Paris, France}

\begin{abstract}
We revisit the possibility that neutrinos undergo non-radiative decay. We investigate the potential to extract information on the neutrino lifetime-to-mass ratio from 
the diffuse supernova neutrino background. To this aim, we explicitly consider the current uncertainties on the core-collapse supernova rate and the fraction of failed supernovae. 
We present predictions in a full $3 \nu$ framework in the absence and presence of neutrino non-radiative decay, for the Super-Kamiokande+Gd, the JUNO, the Hyper-Kamiokande, and the DUNE experiments, 
that should observe the diffuse supernova neutrino background in the near future. Our results show the importance of a {$3 \nu $ treatment of neutrino decay} and of identifying the neutrino mass ordering to break possible degeneracies between DSNB predictions in the presence of decay and standard physics. 
\end{abstract}
\date{\today}


\maketitle

\section{Introduction}
\noindent
The vacuum oscillation discovery \cite{Super-Kamiokande:1998kpq} and the solution of the solar neutrino problem \cite{Super-Kamiokande:1998kpq,SNO:2001kpb,KamLAND:2002uet}
represented a breakthrough in neutrino physics. Another milestone was the observation of neutrinos from the explosion of the blue supergiant Sanduleak, giving SN1987A, in the Large Magellanic Cloud \cite{Kamiokande-II:1987idp,Bionta:1987qt,Alekseev:1988gp}. 
This unique observation brought crucial progress on the longstanding open issue of the supernova explosion mechanism as well as 
on non-standard neutrino properties, particles, and interactions.

Past supernovae emitted huge amounts of neutrinos of all flavors which formed a diffuse supernova neutrino background (DSNB) 
(see \cite{Ando:2004hc,Beacom:2010kk,Lunardini:2010ab} for reviews).This background, integrated over cosmological time,  depends, on one hand, on the still uncertain core-collapse supernova rate and the debated fraction of failed supernovae and, on the other, on flavor mechanisms and unknown neutrino properties. Currently, for the DSNB, we only have upper limits.

The Super-Kamiokande (SK) experiment set the first limit on the $\bar{\nu}_e$ flux, i.e. $1.2$ $\bar{\nu}_e$ cm$^{-2}$s$^{-1}$ ($E_{\nu} > 19.3~$MeV, 90 $\%$ C.L.) \cite{Super-Kamiokande:2002hei}. This result was superseded by a subsequent analysis \cite{Super-Kamiokande:2013ufi} and by the combined analysis of SK-I to SK-IV data which gives the upper limit of the DSNB $\bar{\nu}_e$ flux around 2.7 cm$^{-2}$s$^{-1}$ ($E_{\nu} > 17.3~$MeV, 90 $\%$ C.L.) \cite{Super-Kamiokande:2021jaq}. 
The KamLAND experiment and the Borexino Collaboration also obtained limits in the window $[8.3, 31.8]$ MeV  \cite{KamLAND:2011bnd} and $[7.8, 16.8]$ MeV 
\cite{Borexino:2019wln}, respectively.
As for the relic $\nu_e$ flux, the ensemble of SNO results provides the upper limit of 19 $\nu_e$ cm$^{-2}$s$^{-1}$ in the window $[22.9, 36.9]$ MeV (90 $\%$ C.L.) \cite{SNO:2006dke}. Neutrino-nucleus coherent scattering in dark matter detectors could lower the current limits of $\phi_{\nu_x, \bar{\nu}_x} < (1.3$-$1.8) \times 10^3$ cm$^{-2}$s$^{-1}$ ($E_{\nu} > 19$ MeV) \cite{Lunardini:2008xd} to $\phi_{\nu_x, \bar{\nu}_x} < 10$ cm$^{-2}$s$^{-1}$ (for $x = \mu, \tau$ flavors)\cite{Suliga:2021hek}.
 
Numerous predictions
 \cite{Ando:2002ky,Galais:2009wi,Chakraborty:2010fft,Priya:2017bmm,Moller:2018kpn,Horiuchi:2017qja,Kresse:2020nto} of the DSNB 
rates are close to the current SK sensitivity limit \cite{Super-Kamiokande:2021jaq}, whereas the most conservative ones lie below by a factor of 2 \cite{Tabrizi:2020vmo,Lunardini:2009ya} or 3 to 5 \cite{Nakazato:2015rya,Horiuchi:2020jnc}.  
Beacom and Vagins \cite{Beacom:2003nk} suggested adding gadolinium (Gd) to SK (SK-Gd) to substantially improve the background suppression. Its inclusion introduces
better neutron tagging through the identification of the 8 MeV photons following neutron capture on Gd. 
The SK-Gd experiment is currently running.
With the development of new techniques for background suppression 
and the advent of the Jiangmen Underground Neutrino Observatory (JUNO) \cite{JUNO:2022lpc}, the Hyper-Kamiokande (HK) experiment \cite{Hyper-Kamiokande:2018ofw} and the Deep Underground Neutrino Experiment (DUNE) \cite{DUNE:2016hlj} the DSNB discovery should lie in the forthcoming future.

The DSNB detection constitutes a unique {\it harvest}. Complementary to the neutrino signals from a single supernova, it is sensitive to
the star-formation rate and the fraction of failed supernovae  \cite{Lunardini:2009ya}, evaluated in Ref.~\cite{Nakazato:2015rya} e.g. based on the metallicity evolution of galaxies. The DSNB receives a contribution from binaries \cite{Mathews:2014qba,Kresse:2020nto,Horiuchi:2020jnc,Schilbach:2018bsg} and has a sensitivity to the neutron star equation of state \cite{Moller:2018kpn} (see \cite{Mathews:2019klh} for a review). 

Moreover, the DSNB depends on neutrino flavor evolution in dense environments. This is a complex open problem that has triggered theoretical investigations for fifteen
years (see e.g. \cite{Duan:2010bg,Mirizzi:2015eza,Horiuchi:2018ofe,Volpe:2015rla} for reviews). 
The MSW effect  \cite{Wolfenstein:1977ue,Mikheev:1986wj} is routinely included in DSNB predictions. 
In contrast, shock waves, turbulence, and $\nu\nu$ neutral-current interactions, which impact the neutrino spectra,
have still received little attention in the context of the DSNB. For example, \cite{Galais:2009wi} implemented both shock waves  
and $\nu\nu$ interactions in the so-called {\it bulb} model and found that their effects
could modify the rates by 10-20 $\%$. Ref.\cite{Nakazato:2013maa} showed that the DSNB rates also depend on the shock wave revival time. 

The DSNB will be an interesting laboratory for the search for non-standard neutrino properties such as neutrino decay. This property has received attention in studies based on terrestrial experiments, astrophysical sources and on cosmological observables. From atmospheric and long-baseline experiments the lower bound $\tau_3/m_3 > 9.1 \times 10^{-11} $ s/eV (99 $\%$ C.L.) was deduced for example by Ref.\cite{Gonzalez-Garcia:2008mgl} in the framework of Majoron models. 
Ref.\cite{Berryman:2014qha} discussed model-independent bounds using solar neutrinos. SNO combined with other solar experiments
reported the limit $\tau_2/m_2 > 1.04 \times 10^{-3} $ s/eV (99 $\%$ C.L.) \cite{SNO:2018pvg}.
 If a supernova explodes at 10 kpc, the observation of the neutronisation burst can tell us if $\tau /m \leq 10^5$-$10^7$ s/eV with DUNE and HK \cite{deGouvea:2019goq}.
Limits on neutrino invisible two-body decay from SN1987A were also obtained \cite{Kachelriess:2000qc,Farzan:2002wx}. 
Several studies used CMB and BBN observations to infer lower bounds. Ref.\cite{Escudero:2019gfk} obtained the limit $\tau > 10^{-3}$ s (at 95.4 $\%$ C.L.)
to have a successful BBN. From Planck2018 data,
Ref.\cite{Barenboim:2020vrr} found the constraint $\tau \gtrsim (4\times10^5$-$4\times10^6)$ s $(m/0.05~ {\rm eV}^{-1})^5$ considering a massless daughter neutrino. For massive daughter neutrinos, weaker constraints are found \cite{Chen:2022idm}.

Usually, investigations of neutrino decay assume that the decaying and the mass eigenstates coincide. Instead,
Ref.\cite{Berryman:2014yoa} derived oscillation formulas with neutrino decay, including this mismatch. 
Ref.\cite{Chattopadhyay:2021eba} obtained compact expressions to implement it, using a resummation of the Zassenhaus expansion. The authors pointed out that the inclusion of this correction is relevant in precision experiments of neutrino vacuum oscillations. 

The DSNB  has a  unique sensitivity to neutrino non-radiative two-body decay for  $\tau /m \in [10^9, 10^{11}]$ s/eV \cite{Ando:2003ie}.
Ref.\cite{Fogli:2004gy} performed a detailed $3\nu$ flavor analysis of non-radiative decay and the DSNB, considering both normal and inverted mass ordering and different mass patterns for the neutrino decay.
Using one Fermi-Dirac distribution for the supernova neutrino spectra, the authors evaluated its impact on inverse beta-decay. 
With the same hypothesis on the supernova neutrino spectra, Ref.\cite{DeGouvea:2020ang} considered an effective case of $2\nu$, in which the heaviest neutrino decays into the lightest, and the intermediate remains stable. For normal ordering and strongly hierarchical mass pattern, the authors gave prospects for HK. 
Ref.\cite{Tabrizi:2020vmo} studied neutrino decay using a similar effective $2\nu$ framework but implementing some progenitor dependence. 
Combining DSNB rates from different detection channels in JUNO, DUNE, and HK, the authors showed the possibility to break some of the degeneracies between the no-decay and the decay cases.

The present manuscript presents a $3\nu$ flavor investigation of the DSNB including neutrino non-radiative two-body decay.
Our results go beyond previous works in several respects. First, we explicitly implement
the uncertainty coming from the evolving core-collapse supernova rate. Second, we include the progenitor dependence
of the supernova neutrino fluxes using inputs from one-dimensional supernova simulations (from the Garching group) 
and consider three different scenarios for the black-hole fraction. 
For flavor evolution, as in previous works, we consider the MSW effect only. Third, we show the influence of neutrino non-radiative decay on
the relic neutrino fluxes going from the $2 \nu$ to the $3 \nu$ framework, for the quasi-degenerate and
the strongly-hierarchical mass patterns in the normal or inverted neutrino mass ordering. 
While each of these aspects was considered individually in previous studies, we integrate all of them in the present work for the first time. 
We give our predictions of the DSNB (integrated) fluxes and the number of events for the running SK-Gd experiment and the upcoming 
HK, JUNO, and DUNE experiments. We discuss their potential to extract information on the neutrino lifetime-to-mass ratio. 

The manuscript is structured as follows. In Section II we introduce the theoretical framework for the DSNB with neutrino radiative two-body decay. 
We describe the different ingredients that influence the DSNB flux, in particular, the evolving core-collapse supernova rate and the black-hole contribution. Then we introduce
the formalism to include neutrino non-radiative decay.
Section III presents the numerical results on the DSNB fluxes with/without decay and the expected number of events in the four experiments. Section IV is the conclusion. 

\section{Theoretical framework}
\noindent
Let us introduce the astrophysical, cosmological, and particle physics aspects relevant to the DSNB.
We present here our choices for the evolving core-collapse supernova rate, the cosmological model, and the supernova neutrino fluxes that include a progenitor dependence.
Then we describe the $2\nu$ and $3\nu$ theoretical frameworks used to implement neutrino non-radiative two-body decay. 
 
\subsection{The DSNB flux and its ingredients}
\noindent
The DSNB flux is built up from the neutrino emission of past supernovae that left either a neutron star (NS) or a black hole (BH). 
In our calculations, we assume that neutrinos decay in vacuum, once they have been produced in the supernova core and have undergone spectral swapping, due to the Mikheev-Smirnov-Wolfenstein (MSW) effect, before reaching the star surface.

\subsubsection{Supernova neutrino fluxes without decay}
\noindent
At the neutrinosphere, the neutrino yields $Y_{\nu}$ are given by quasi-thermal neutrino spectra $ \phi^0_{\nu}(E_{\nu})$, normalized to unity ($\int dE_{\nu}  \phi^0_{\nu}(E_{\nu}) = 1$). These are characterized by
three inputs, i.e. the normalization, the neutrino average energy, and the pinching parameter $\alpha$. Explicitly,
one has 
\beq\label{eq:yield}
Y_{\nu} = { L_{\nu} \over{\langle E_{\nu} \rangle}}  \phi^0_{\nu}(E_{\nu})  \ ,
\eeq
with $L_{\nu}$ the total gravitational binding energy emitted by the supernova. 
The power-law distributions read \cite{Keil:2002in}
\beq\label{eq:PW}
 \phi^0_{\nu}(E_{\nu}) = {(\alpha + 1)^{\alpha + 1} \over{\langle E_{\nu} \rangle \Gamma(\alpha + 1)}}  \Big({ E_{\nu} \over {\langle E_{\nu} \rangle}} \Big)^{\alpha}{\rm  e}^{ -{ (1+ \alpha) E_{\nu} \over{\langle E_{\nu}} \rangle }} \ , 
\eeq
with $\alpha$ related to the first and second moments of the neutrino energy distribution through the relation
\beq\label{eq:alpha}
\alpha = {{\langle E_{\nu}^2 \rangle - 2 \langle E_{\nu} \rangle^2 }\over{\langle E_{\nu}\rangle^2  - \langle E_{\nu}^2 \rangle}} \ . 
\eeq

When neutrinos traverse the supernova they undergo flavor transformation due to neutrino interactions with the matter and experience
the MSW effect \cite{Wolfenstein:1977ue,Mikheev:1986wj,Dighe:1999bi}. More generally, 
the presence of shock waves, turbulence, and $\nu\nu$ interactions can trigger collective and non-collective flavor mechanisms 
investigated for many years (see \cite{Duan:2010bg,Duan:2009cd,Mirizzi:2015eza,Volpe:2015rla} for reviews). The complexity of this
problem is such that more work is needed to assess the final impact on the supernova neutrino fluxes. Therefore, we only include here the established 
MSW effect.
As a result, the neutrino yield at the star surface is\footnote{Since the supernova neutrino fluxes depend on the progenitor, Eqs.\eqref{eq:PW}-\eqref{eq:MSWIO} should have an explicit dependence on the progenitor mass ${\rm M}$. We have omitted it in this section, not to overburden the text.} 
\begin{align}\label{eq:MSWNO}
Y_{\nu_1} & = Y_{\nu_x} ~~Y_{\nu_2} = Y_{\nu_x} ~~Y_{\nu_3} = Y_{\nu_e} ~~~{\rm (NO)}  \ ,  \\
Y_{\nu_1} & = Y_{\nu_x}~~ Y_{\nu_2} = Y_{\nu_e} ~~Y_{\nu_3} = Y_{\nu_x} ~~~{\rm (IO)}  \nonumber \ , 
\end{align}
and the antineutrino yield is
\begin{align}\label{eq:MSWIO}
Y_{\bar{\nu}_1} & = Y_{\bar{\nu}_e} ~~Y_{\bar{\nu}_2} = Y_{\nu_x} ~~Y_{\bar{\nu}_3} = Y_{\nu_x} ~~~{\rm (NO)} \ , \\
Y_{\bar{\nu}_1} & = Y_{{\nu}_x~~} Y_{\bar{\nu}_2} = Y_{\nu_x} ~~Y_{\bar{\nu}_3} = Y_{\bar{\nu}_e} ~~~{\rm (IO)}  \nonumber \ , 
\end{align}
with NO standing for normal (i.e. $\Delta m^2_{31} > 0$) and IO for inverted (i.e. $\Delta m^2_{31} < 0$) $\nu$ mass ordering. 

\subsubsection{The DSNB flux without decay}
\noindent
The local relic supernova neutrino fluxes for the mass eigenstates $\nu_i$, including a progenitor dependence, read
\beq\label{eq:dsnbflux}
\phi_{\nu_i}(E_{\nu}) = {c} \int \int {\rm dM}~dz~ (1+z) \Big | {dt_{\rm c} \over {dz}} \Big |~R_{\rm SN}(z,{\rm ~M}) ~ {Y_{\nu_i}(E'_{\nu}, {\rm ~M})} \ ,
\eeq
with $E_{\nu} = E'_{\nu}(1 + z)^{-1}$ the redshifted neutrino energy, $c$ the speed of light and 
$z \in [0, z_{max}]$ the cosmological redshift. In our calculations, we take $z_{max} = 5$ and  ${\rm M} \in [8, 125] {\rm ~M}_{\odot}$ as mass range of the supernova progenitors. 

The first factor in Eq.\eqref{eq:dsnbflux} is the cosmic time that depends on the cosmological model. In this work, we assume the $\Lambda {\rm CDM}$ model\footnote{Note that Ref.\cite{Barranco:2017lug} investigated the influence of other cosmological models on the DSNB.} 
for which the expansion history of the Universe is given by 
\beq\label{eq:Hz}
\Big | {dz \over {dt_{\rm c} }} \Big | = H_0 ( 1 + z) \sqrt{ \Omega_{\Lambda} + (1+z)^3 \Omega_m} \ , 
\eeq
where $H_0$ is the Hubble constant, $\Omega_{\Lambda}$ and $\Omega_m$ the dark energy and the matter cosmic energy densities which we take equal to 0.7 and 0.3 respectively.
Concerning $H_0$, there is currently a tension between the Hubble constant value extracted with the "distance ladder method" and the Cosmological Microwave Background (CMB) \cite{DiValentino:2021izs}. The former gives $H_0 = 74.03 \pm 1.42 ~{\rm km~ s}^{-1} {\rm Mpc}^{-1}$, whereas the latter $H_0 = 67.4 \pm 0.5 ~{\rm km ~s}^{-1} {\rm Mpc}^{-1}$. For the present work we employ $H_0 = 67.4~{\rm km ~s}^{-1} {\rm Mpc}^{-1}$, while we have checked that 
the results are not sensitive to variations of $H_0$.

The second important input in Eq.\eqref{eq:dsnbflux} is the evolving core-collapse supernova rate (number per unit time per unit comoving volume) $R_{\rm SN}(z,{\rm ~M}) $ that is related to the star-formation rate history $\dot{\rho}_*(z) $ as
\beq\label{eq:CCSNrate}
R_{\rm SN}(z, {\rm M}) = \dot{\rho}_*(z) { \phi({\rm M}) d{\rm M} \over {\int^{125 {\rm ~M}_{\odot}}_{0.5 {\rm ~M}_{\odot}}  \phi({\rm M}){\rm M} d{\rm M}}} \ , 
\eeq
where $\phi(\rm M)$ is the initial mass function. The quantity $\phi(\rm M) d (\rm M)$ gives the number of stars in the mass interval\footnote{Note that changing the upper value of the integral from $100  {\rm ~M}_{\odot}$ to $125  {\rm ~M}_{\odot}$ does not introduce significant differences.} $[\rm M, \rm M + d \rm M]$.
We take the standard power-law introduced by Salpeter \cite{Salpeter:1955it} 
\beq\label{eq:IMF}
\phi(\rm M) \sim {\rm M}^{\chi} \ ,
\eeq
with $\chi =- 2.35$ for ${\rm M} \ge 0.5~ {\rm M}_{\odot}$ (for a discussion on the universality of $\phi({\rm M})$ at higher masses, see for example \cite{Ziegler:2022ivq}). 

\begin{figure}
\begin{center}
\includegraphics[scale=0.4]{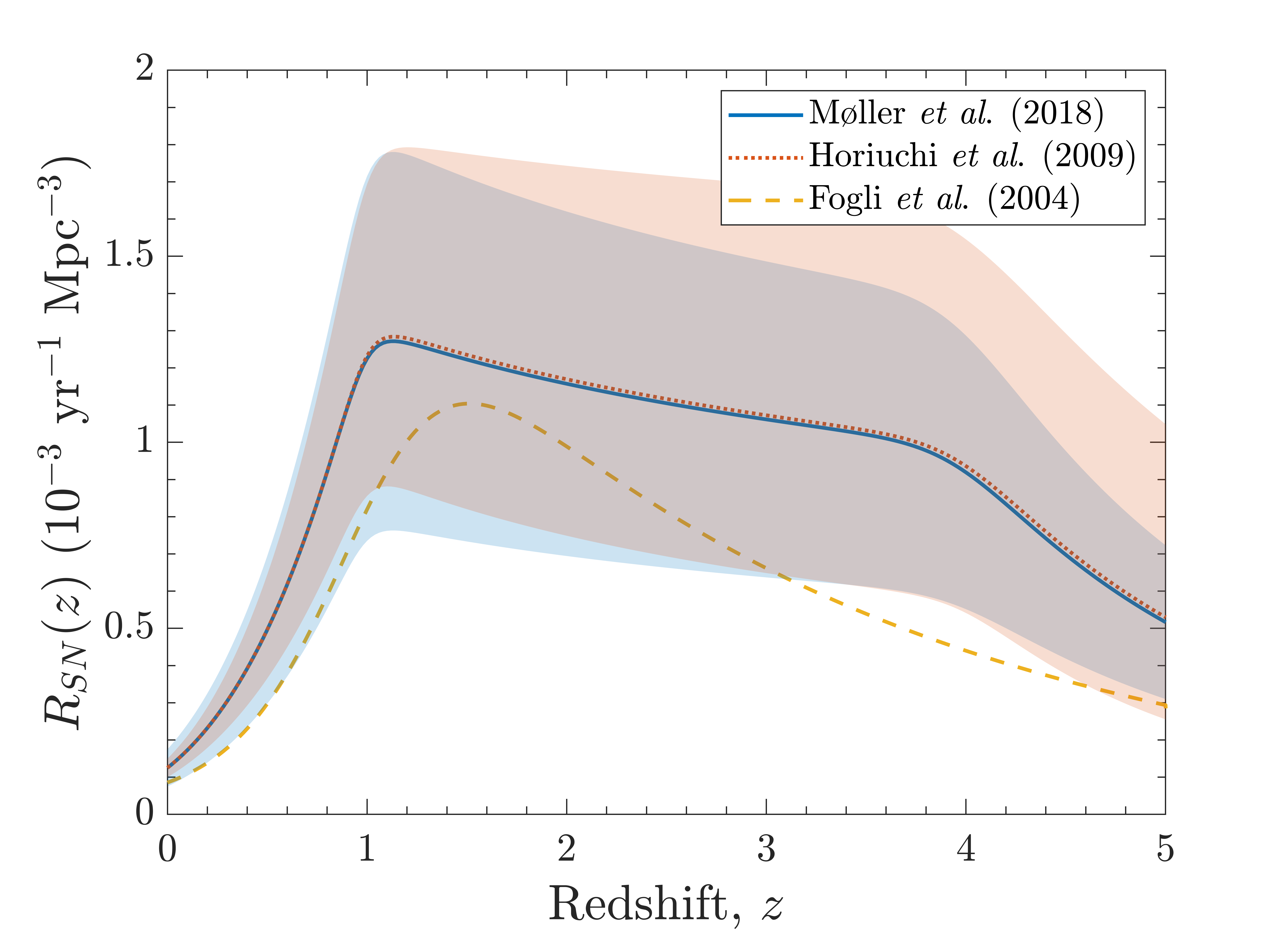}
\caption{Core-collapse supernova rate as a function of redshift. The figure shows the piecewise parametrization by \cite{Yuksel:2008cu} and \cite{Horiuchi:2011zz}, used in \cite{Moller:2018kpn,Priya:2017bmm} (blue), and the one from \cite{Horiuchi:2008jz} (pink) with the band showing the core-collapse supernova rate uncertainty. The dashed line shows the older evolving core-collapse supernova rate, employed in the DSNB study of \cite{Fogli:2004gy}, including neutrino non-radiative decay.}
\label{fig:CCSN}
\end{center}
\end{figure}

For the star-formation rate history, we employ the piecewise continuous form of a broken power law by \cite{Yuksel:2008cu} (see also \cite{Madau:2014bja})
\beq\label{eq:SFR}
\dot{\rho}_*(z) = \dot{\rho}_{0} \Big[(1+z)^{\alpha \eta} + \Big({1+z \over{B}}\Big)^{\beta \eta} + \Big({1+z \over{C}}\Big) ^{\gamma \eta}  \Big]^{-1/\eta}  \ ,
\eeq
with $\alpha = 3.4, \beta = -0.3, \gamma = -3.5 $ the logarithmic slopes at low, intermediate and high redshift, $\eta = -10$ the smoothing function
and $B = 5000$, $C = 9$ the constants defining the redshift breaks (Figure \ref{fig:CCSN})\footnote{Note that Ref.\cite{Baldry:2003xi} introduced a modified  broken power law for the IMF with $\chi =- 1.5$ at $ 0.1 ~{\rm M}_{\odot} \le {\rm M} \le 0.5~ {\rm M}_{\odot} $ and $\chi = - 2.12$ for ${\rm M} > 0.5~ {\rm M}_{\odot} $. It gives a similar $R_{\rm SN}(z, {\rm M})$ \cite{Horiuchi:2008jz}}.

Table~\ref{tab:Rsnpar} presents the values of $\dot{\rho}_{0}$ and of $R_{SN}(0) = \int^{125 {\rm ~M}_{\odot}}_{8 {\rm ~M}_{\odot}} R_{\rm SN}(0, {\rm M}) d{\rm M}$ (see Eq.\eqref{eq:CCSNrate}).
The evolving core-collapse supernova rate impacts the DSNB normalization and currently constitutes the largest source of uncertainties for the DSNB. 

It is to be noted that several parametrizations of the star-formation rate are available in the literature. The one given by \eqref{eq:SFR} that we adopt, was also used in \cite{Moller:2018kpn} (Figure \ref{fig:CCSN})). It is very close but does not present the kinks of the one employed by \cite{Priya:2017bmm}. The one used in \cite{Fogli:2004gy} has been superseded. Note also that Ref. \cite{Mathews:2014qba} obtained a modified parametrization\footnote{Note that their parametrization does not hold at $z > 4$ since the authors do not include GRB data, contrary to \cite{Yuksel:2008cu}.}, compared to the one of \cite{Yuksel:2008cu}.Their difference comes from the fact that the authors of Ref.\cite{Mathews:2014qba}  considered only the subset of the star-formation rate data corrected for extinction by dust\footnote{Moreover, they argued that the core-collapse supernova rate, deduced from the star-formation rate history, could agree with the one from direct core-collapse supernova observations (the two disagree by a factor 2 at $0 \le z \le 1$ \cite{Horiuchi:2011zz}), thus solving the "supernova rate problem", if one included a contribution from binaries, failed supernovae, and from (electron-capture) ONeMg supernovae.}
   
\begin{table}[tp]
\centering
\begin{tabular}{lclclcl}
\hline
 & $R_{SN}(0)$  &  ~~$\dot{\rho}_{0} $ \\
\hline
Low  & 0.75   & 0.0054 \\
Fiducial & 1.25 & 0.0089 \\
High & 1.75 &  0.0125  \\
\hline
\end{tabular}
\caption{Local core-collapse supernova 
rate (in units of $10^{-4}$ yr$^{-1}$ Mpc$^{-3}$) Eq.\eqref{eq:CCSNrate} and normalization of the star-formation rate history (in units of M$_{\odot}$ yr$^{-1}$ Mpc$^{-3}$) Eq.\eqref{eq:SFR}.}
\label{tab:Rsnpar}
\end{table}

The last important factor in Eq.\eqref{eq:dsnbflux} is the neutrino fluxes from a single supernova.
The neutrino flux emitted depends on the outcome of the collapse: either NS or BH. Considering explicitly the contribution from NS-forming and BH-forming collapses, we can rewrite Eq.\eqref{eq:dsnbflux} as
\begin{equation}
\begin{aligned}
    \phi_{\nu_i}(E_{\nu}) & = {c} \int  dz~ (1+z) \Big | {dt_{\rm c} \over {dz}} \Big | \times \\
    & \left[ 
\int_\Omega~{\rm dM}~R_{\rm SN}(z,{\rm ~M}) ~ {Y_{\nu_i}^{\rm NS}(E'_{\nu}, {\rm~M})} \right. \\
+  & \left. \int_\Sigma~{\rm dM}~R_{\rm SN}(z,{\rm ~M}) ~ {Y_{\nu_i}^{\rm BH}(E'_{\nu}, {\rm~M})}
\right] ,
\end{aligned}
\end{equation}
where $\Omega$ and $\Sigma$ indicate the range of masses for which a collapse forms a NS or a BH, respectively.
Thus, the fraction of BH-forming collapses can be defined as
\begin{equation}
    f_{BH} = \frac{\int_\Sigma d \rm M \phi(\rm M)}{\int_{8 \rm M_\odot}^{125 \rm M_\odot} d \rm M \phi(\rm M)}.
\end{equation}

Although dark collapses are subdominant, their contribution to the DSNB can be significant, as pointed out by Lunardini \cite{Lunardini:2005jf}.
In fact, the compression of baryonic matter, during black hole formation, generates large neutrino fluxes with higher average energies 
and larger differences among flavors (than optical supernovae) \cite{Sumiyoshi:2007pp}, depending on the (soft or stiff) equation of state.
Therefore, the black hole contribution impacts the tail of the DSNB flux (see Figure~\ref{fig:nodecay}).
Note that, in the present work, we neglect the dependence of the DSNB flux on the galaxy metallicity, considered for example in \cite{Nakazato:2015rya}.

\subsubsection{Scenarios for the fraction of failed supernovae}
\noindent
Let us now describe three scenarios for the fraction of failed supernovae and introduce what we refer to as {\it Fiducial}, {\it Low}, and {\it High}.
For the supernova neutrino spectra at the neutrinosphere Eq.~\eqref{eq:yield}-\eqref{eq:alpha}, we use fluences of one-dimensional simulations by the Garching group \cite{Hudepohl:2013zsj,Priya:2017bmm}, with the Lattimer-Swesty equation of state giving the matter compressibility parameter $K = 220$ (in agreement with nuclear measurements). The progenitors, with solar metallicity, are from Woosley and Weaver. The parameters defining the neutrino fluxes are given in Table~\ref{tab:fluxpar} (Appendix A).
On the way to the star's surface, the spectra 
are modified by the MSW effect, depending on the neutrino mass ordering Eqs.\eqref{eq:MSWNO}-\eqref{eq:MSWIO}. 

\begin{figure}
\begin{center}
\includegraphics[scale=0.27]{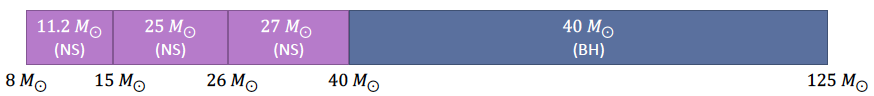}
\includegraphics[scale=0.27]{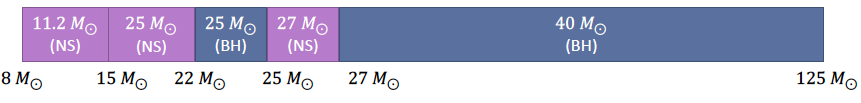}
\includegraphics[scale=0.27]{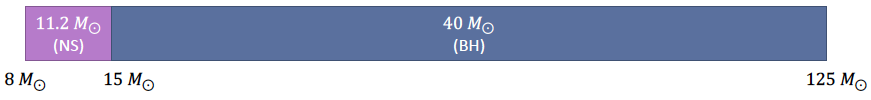}
\caption{Scenario I to III (top to bottom) for the BH fraction as well as the progenitor dependence of a supernova that left either a neutron star or a black hole. The parameters (neutrino luminosity, average energies and pinching) of the corresponding fluences are given in Table~\ref{tab:fluxpar} (Appendix A).}
\label{fig:fBH}
\end{center}
\end{figure}

In our scenarios for the black-hole fraction, we follow\footnote{Note that BH fractions used in \cite{Priya:2017bmm,Moller:2018kpn} differ from each other.} \cite{Priya:2017bmm} and \cite{Moller:2018kpn}, and combine the progenitors used in the two works. 
Here are the three scenarios:
\begin{itemize}\label{scen}
 \item[I-]   $f_{\rm BH} = 0.09$ is obtained when all stars that have $ {\rm M} \ge 40~ M_{\odot}$ become BH;  
 \item[II-]  $f_{\rm BH} = 0.21$ both stars with $ {\rm M} \in [22, 25] ~M_{\odot}$ and ${\rm M} \ge 27~ M_{\odot}$ collapse into a BH;
 \item[III-]  $f_{\rm BH} = 0.41$ is an extreme case where all stars with $ {\rm M} \ge 15 ~ M_{\odot}$ turn into a BH. 
\end{itemize}
For clarity, we show in Figure \ref{fig:fBH} the progenitors used and the corresponding mass intervals for which they were used as templates. A detailed description is given in Appendix A.

\noindent
As for our {\it Fiducial} DSNB model we employ
\beq\label{eq:FM}
f_{\rm BH} = 0.21 ~~~R_{SN}(0) = 1.25 \times 10^{-4} {\rm yr}^{-1} {\rm Mpc}^{-3} \ ;
\eeq
whereas the {\it Low} and {\it High} scenarios correspond to the variability of the local core-collapse supernova rate 
\begin{align}\label{eq:LHM}
R_{SN}(0) & = 0.75 \times 10^{-4} {\rm yr}^{-1} {\rm Mpc}^{-3}~~(Low) \\ \nonumber
R_{SN}(0) & = 1.75\times 10^{-4} {\rm yr}^{-1} {\rm Mpc}^{-3}~~(High). 
\end{align}

Obviously 
a more detailed dependence on the progenitor masses would be desirable. For example, \cite{Horiuchi:2017qja,Kresse:2020nto,Horiuchi:2020jnc} performed
extensive supernova simulations to make DSNB predictions. Since our focus here is to investigate  
non-standard neutrino properties we stick to a simpler, but still detailed progenitor dependence, which improves against \cite{Fogli:2004gy,deGouvea:2019goq} that used one power law spectrum and to \cite{Tabrizi:2020vmo} that included either one Fermi-Dirac spectrum, or only one value for the BH fraction.

\subsection{The DSNB flux in presence of neutrino non-radiative two-body decay}
\noindent
Having presented the main ingredients of the DSNB flux, we now describe how to extend the standard framework
to include neutrino non-radiative two-body decay.

\subsubsection{Neutrino non-radiative two-body decay}
We consider the processes where a heavy neutrino $\nu_i$ decays into a lighter one $\nu_j$
and a massless, or almost massless, scalar particle $\phi$, i.e.
\beq\label{eq:decay}
\nu_i \rightarrow \nu_j  + \phi ~~~~{\rm or} ~~~~ \nu_i \rightarrow \bar{\nu}_j + \phi \ . 
\eeq

Neutrino decay to Majorons has
been discussed in the context of various models (see for example \cite{Kim:1990km}). 
The new degrees of freedom are singlets under the Standard Model gauge group.
In the case of Dirac neutrinos, the decay requires
dimension five (lepton-number zero) or four and six (lepton-number two) operators. For Majorana neutrinos, the minimal interaction that leads
to the neutrino decay has dimension six \cite{deGouvea:2019goq}
\beq\label{eq:Dim5}
{\cal L_{\rm Maj}}  \supset   {\tilde{g}_{ij} \over{2 \Lambda^2}}(L_i H)(L_j H) \phi  + {\rm h.c.}  \supset g_{ij} (\nu_L)_i (\nu_L)_j  \phi + {\rm h.c.} \ , 
\eeq
where ${\tilde{g}_{ij}  = \tilde{g}_{ji} }$ and $ g_{ij}= \tilde{g}_{ij} v^2 /{\Lambda^2}$, $L, H$ are the Standard Model lepton doublets and Higgs field,
$v$ is the vacuum expectation value of the neutral component of the Higgs field, and ${\nu}$ is the neutrino field.
Since here we do not wish to focus on specific models, 
we will keep our considerations general.

The (rest-frame) neutrino lifetime and associated decay rate receive contributions from both processes \eqref{eq:decay}, that is
\beq\label{eq:gamma}
\tau^{-1}_{\nu_i} =  \tilde{\Gamma}_{\nu_i} = \sum_{m_j < m_i} \tilde{\Gamma}(\nu_i \rightarrow \nu_j ) +\tilde{\Gamma}(\nu_i \rightarrow \bar{\nu}_j) \ .
\eeq

The related decay rate in the laboratory frame reads
\beq\label{eq:tau}
\Gamma_{\nu_i} =  {m_i \over{E_{\nu}}}~\tilde{\Gamma}_{\nu_i} \ ,
\eeq
with $m$  the absolute neutrino mass.
Since the value of $m$ is not known yet, studies on $\nu$ non-radiative decay give limits for $\tau/m$,
the lifetime-over-mass ratio. In the following, we shall present our results as a function of this parameter, which also
facilitates the comparison with previous works. Finally 
the branching ratios are
\beq\label{eq:BR}
B_{\nu_i} = \Gamma(\nu_i \rightarrow \nu_j)/\Gamma_{\nu_i} \ ,
\eeq
and similarly for $\nu_i \rightarrow \bar{\nu}_j + \phi$.

\subsubsection{Neutrino kinetic equations in presence of non-radiative decay}

\noindent
We now consider the kinetic equations for ultra-relativistic neutrinos
implementing neutrino radiative two-body decay. Their generic form is \cite{Fogli:2004gy}
\begin{align}\label{eq:Lkin}
{\cal L} [n_{\nu_i}(E_{\nu}, t) ] & =  {\cal C} [n_{\nu_i}(E_{\nu}, t) ] \ , 
\end{align}
where $n_{\nu_i}(E_{\nu}, t)$ is the relic number density of the $\nu_{i}$ mass eigenstates  (per unit energy and comoving volume) at time $t$\footnote{
It is related to the phase space distribution function $f$ through $n_{\nu_i}(E_{\nu}, t) = 4 \pi p^2 f (R(t)/R_0)^3$. The function $R(t)$ is the universe scale factor of the Friedman-Robertson-Walker metric
and $E_{\nu} \approx p = \vert \vec{p} \vert $.}.  

The Liouville operator then reads
\begin{align}\label{eq:Lkin2}
{\cal L}  [n_{\nu_i}(E_{\nu}, t) ] & =  \big[ \partial_t - H(t) E_{\nu} \partial_E - H(t) \Big]  n_{\nu_i}(E_{\nu}, t) \ ,
\end{align}
with $H(t)$ the Hubble constant. 
The explicit expression of the collision term in Eq.\eqref{eq:Lkin} for decaying neutrinos\footnote{ In this section, the explicit dependence on the progenitor mass ${\rm M}$ is not included not to overburden the text.} is 
\begin{align}\label{eq:Lkin3}
{\cal C}  [ n_{\nu_i}(E_{\nu}, t) ] & =  R_{\rm SN} (t) Y_{\nu_i} (E_{\nu}) + \sum_{m_j > m_i} q_{ji} (E_{\nu},t) \nonumber \\
& - \Gamma_{\nu_i} n_{\nu_i}(E_{\nu}, t) \ , 
\end{align}
with 
\begin{align}\label{eq:Lkin4}
q_{ji} (E_{\nu},t)  = \int_{E_{\nu}}^{\infty} dE_{\nu}' n_{\nu_j}(E_{\nu}', t) \Gamma_{\nu_j \rightarrow \nu_i} \psi_{ji}(E_{\nu}', E_{\nu}) \ , 
\end{align}
where $\psi_{ji}(E_{\nu}', E_{\nu})$ are the neutrino decay energy spectra.
The first contribution in the collision term Eq.\eqref{eq:Lkin3} is the usual one from core-collapse supernovae (without decay). The second source term accounts
for the feeding of the lighter states $\nu_i$ from the decay of the heavier ones $\nu_j$. The last is a sink term that implements the $\nu_i$ decay loss
with total decay rate $\Gamma_{\nu_i}$ Eq.\eqref{eq:gamma} which is present for the heavier neutrinos only.

After performing a change of variables from $(t, E_{\nu})$ to $(z, E_{\nu}')$ the redshift and the redshifted neutrino energies, one can rewrite Eqs.\eqref{eq:Lkin}-\eqref{eq:Lkin4} and obtain the general solution for the relic number of neutrinos (per unit of comoving volume and
of energy, at redshift $z$) that is \cite{Fogli:2004gy} 
\begin{align}\label{eq:kinsol}
n_{\nu_i} (E_{\nu},z)  & = {1 \over {1+z}} \int_z^{\infty} {dz' \over{H(z')} } \Big[ R_{SN}(z') Y_{\nu_i} \Big(E_{\nu} {1 + z' \over {1 + z}} \Big) \nonumber \\
 & + \sum_{ m_j > m_i} q_{ji} \Big(E_{\nu} {1 + z' \over {1 + z}}, z'\Big)  \Big]  e^{- \Gamma_{\nu_i}[\chi(z') - \chi(z)](1 + z)} \ , 
\end{align}
where the auxiliary function $\chi(z)$ is 
\beq\label{eq:chi}
\chi(z) = \int_0^z dz' H^{-1}(z')(1+z')^{-2} \ . 
\eeq
This result reduces to the standard expression Eq.\eqref{eq:dsnbflux} when $\Gamma =  0$ (in the limit $\tau \rightarrow \infty$).
To determine the DSNB fluxes and the associated rates in the full $3 \nu$ framework, one exploits 
the general solution Eq.\eqref{eq:kinsol} for $z=0$, with Eqs.\eqref{eq:Lkin4} and \eqref{eq:chi} (see Appendix B).

\subsubsection{Neutrino decay patterns}
\noindent
It is our goal to perform a detailed investigation of the impact of neutrino non-radiative decay, considering not only the astrophysical uncertainties, but also neutrino properties that remain unknown. Since for the neutrino mass ordering, we only have indications that are statistically not significant enough, in our analysis we shall consider both normal (NO) and inverted mass ordering (IO).

Moreover, depending on the lightest absolute neutrino mass, the neutrino mass patterns can be either quasi-degenerate (QD) or strongly hierarchical (SH). Following the 3$\nu$ study of \cite{Fogli:2004gy},  
we consider these extreme possibilities:
\begin{itemize}
\item[{\it i)}] QD mass pattern if $m_h \simeq m_l \gg m_h - m_l$;
\item[{\it ii)}] SH mass pattern if $m_h - m_l \gg m_l \simeq 0$. 
\end{itemize}

\begin{figure}
\begin{center}
\includegraphics[scale=0.3]{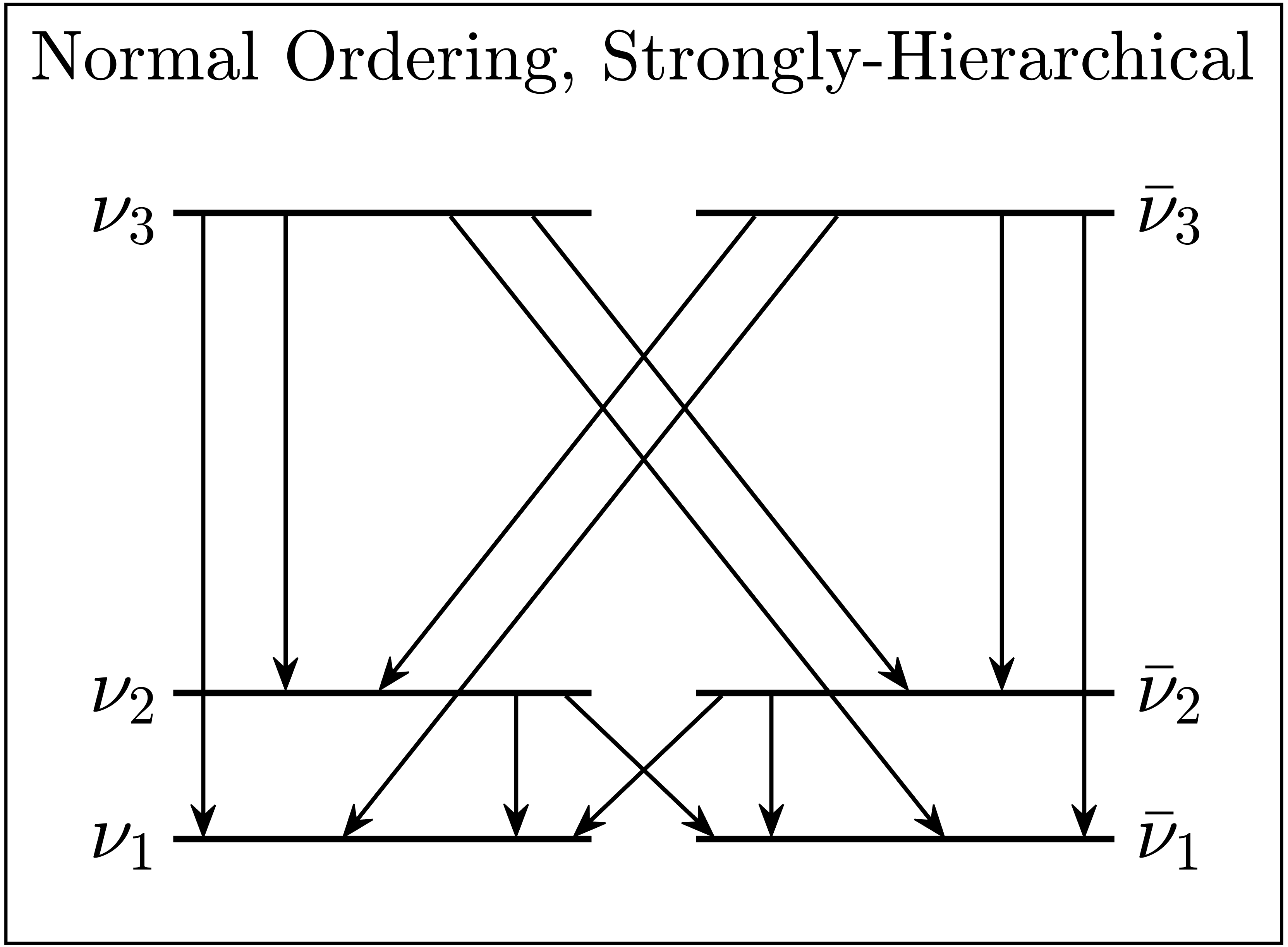}
\includegraphics[scale=0.3]{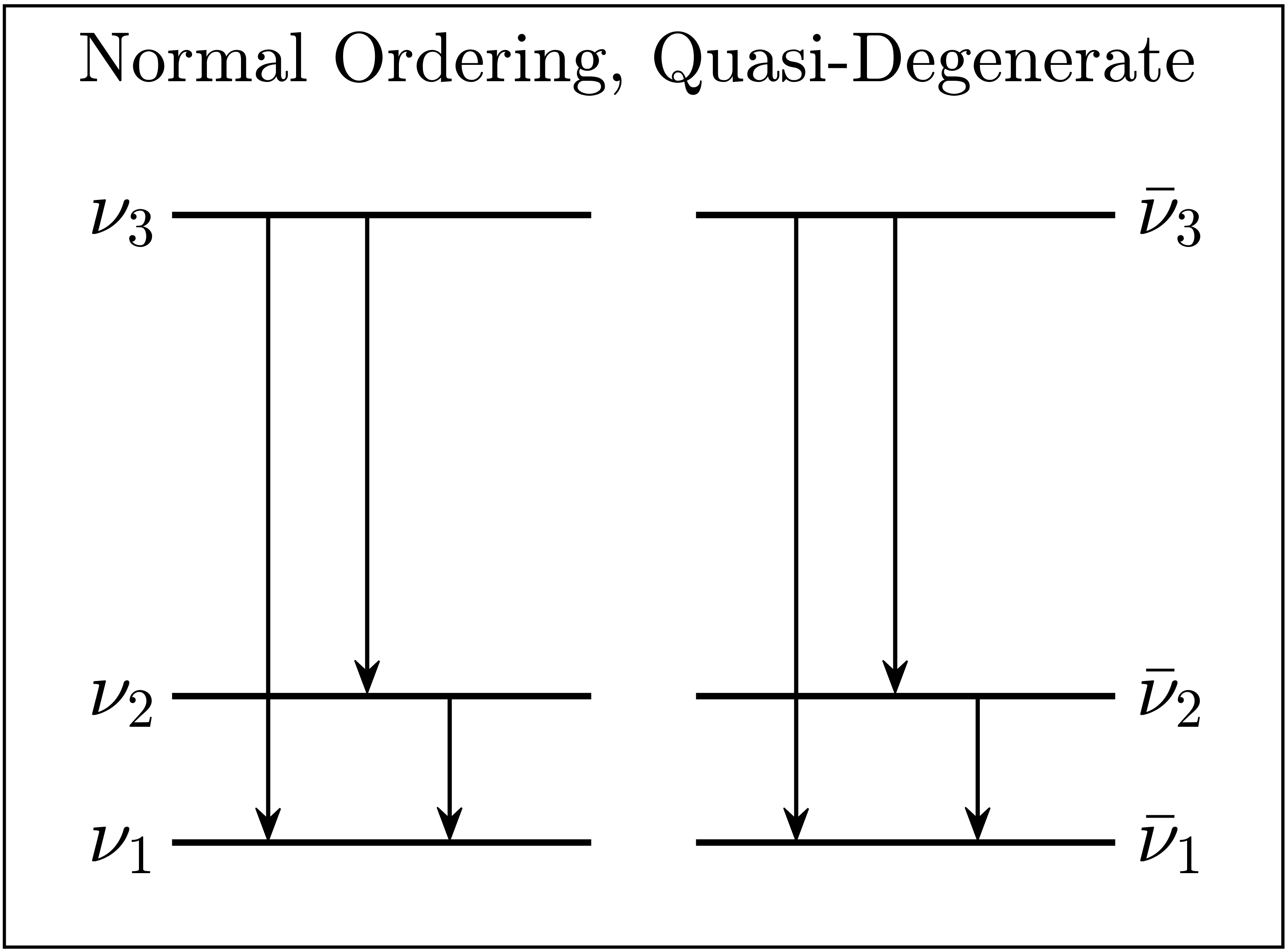}
\includegraphics[scale=0.3]{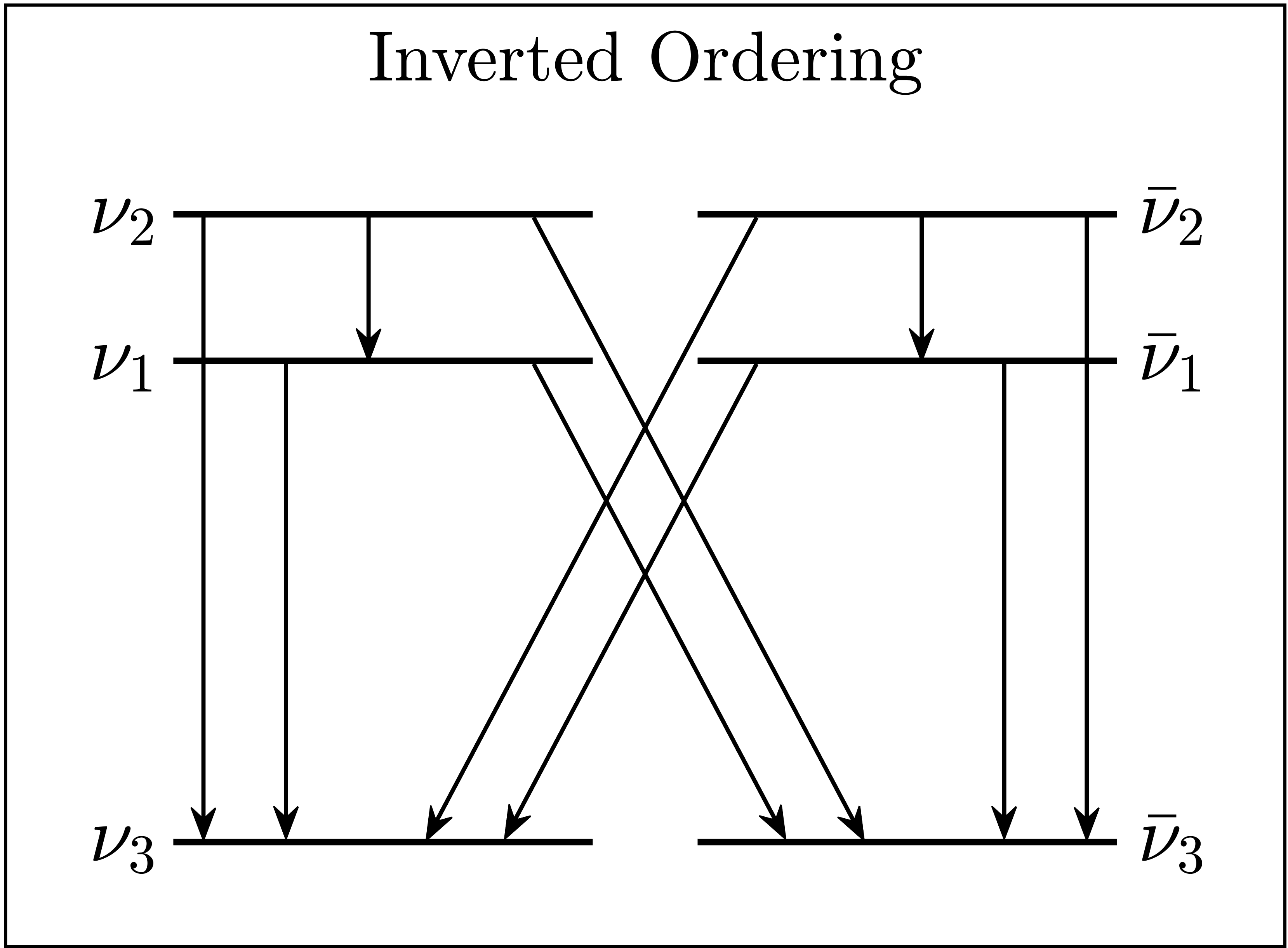}
\caption{Decay patterns for $3 \nu$ flavors. Upper left: Normal mass ordering in the strongly-hierarchical (SH) case. The branching ratios are equal to 1/4 for $\nu_3$ ($\bar{\nu}_3$) and 1/2 for $\nu_2$ ($\bar{\nu}_2$). Upper right:  Normal mass ordering in the quasi-degenerate case (QD). The branching ratios are equal to 1/2 for $\nu_3$ or $\bar{\nu}_3$.  Lower:  Inverted mass ordering case. The branching ratios are equal to 1/3 for $\nu_2$ ($\bar{\nu}_2$) and 1/2 for $\nu_1$ ($\bar{\nu}_1$). 
In all cases the lifetime-to-mass ratio of the decaying eigenstates is taken equal, i.e. $\tau_2/m_2 = \tau_3/m_3 $ (NO) or $\tau_2/m_2 = \tau_1/m_1 $ (IO).}
\label{fig:decaypatterns}
\end{center}
\end{figure}

Figure~\ref{fig:decaypatterns} presents the decay schemes and the associated branching ratios Eq.\eqref{eq:BR} for $3 \nu$ flavors. The figure shows the cases of IO and of NO, either with SH or with QD mass patterns. For IO the decay scheme comprises $m_1$ and $m_2$ as quasi-degenerate and strongly hierarchical with respect to $m_3$. For the computations  we use a {\it democratic} hypothesis for $B(\nu_i \rightarrow \nu_j)$ (see the caption of Figure~\ref{fig:decaypatterns}) and
assume equal lifetime-to-mass ratio for the decaying eigenstates. This choice does not employ specific {\it ansatz} and has the advantage of reducing 
the number of free parameters.

The last piece that needs to be specified in Eq.~\eqref{eq:Lkin4} are the neutrino decay energy spectra.
In the QD case, one has \cite{Fogli:2004gy,deGouvea:2019goq}
\beq\label{eq:psiQD}
\psi (E_{\nu_h}, E_{\nu_l}) = \delta(E_{\nu_h} -E_{\nu_l}) \ . 
\eeq
In the SH case, both helicity conserving (${\rm h.c.}$) and helicity flipping (${\rm h.f.}$) decays contribute to the neutrino decay rate with neutrino spectra given by
\beq\label{eq:psiSH}
\psi_{\rm h.c.} (E_{\nu_h}, E_{\nu_l}) = { 2 E_{\nu_l} \over{E_{\nu_h}^2}} ~~~~\psi_{\rm h.f.} (E_{\nu_h}, E_{\nu_l}) =  {2 \over {E_{\nu_h}}} \Big(1 - { E_{\nu_l} \over {E_{\nu_h}}}   \Big) \ .
\eeq
One can see that h.c. contributions produce neutrinos with harder spectra than h.f. contributions.

Finally, the total decay rate\footnote{Note that there is a factor of 2 missing in Eq. (2.6) of \cite{deGouvea:2019goq}. Note also that $g_{ij}$ should be $g_{ij}/2$ in Eq.(2.1) of \cite{Beacom:2002cb}.} for the process \eqref{eq:decay}, e.g. for a $\nu_3$ decaying to $\nu_1$, in the laboratory frame is  
\beq\label{eq:Trate}
\Gamma(E_{\nu_3}) = {g^2 \over{32 \pi }} {m_3^2 \over{E_{\nu_3}  }} = {1 \over {\tau_3}} {m_3 \over{E_{\nu_3} }} \ .
\eeq
 
We give in Appendix B the explicit solutions for the DSNB fluxes from the solution of the neutrino kinetic equations with decay Eq.\eqref{eq:kinsol}, for the three cases considered in Figure~\ref{fig:decaypatterns}.
 
The $2 \nu$ flavor solutions can obviously be obtained as special cases.
In particular, one has either a QD mass pattern, with $B(\nu_2 \rightarrow \nu_1) = 1 $ and $B(\nu_2 \rightarrow \bar{\nu}_1) =0$,
or a SH one, with $B(\nu_2 \rightarrow \nu_1) = B(\nu_2 \rightarrow \bar{\nu}_1) = 1/2$ (see for example \cite{Kim:1990km}).
We remind that it is the $2 \nu$ framework in NO and with a SH mass pattern which was used in Refs.\cite{deGouvea:2019goq,Tabrizi:2020vmo}.

The numerical results we will present are valid if neutrinos are Majorana or Dirac particles, with different assumptions on the new degrees of freedom\footnote{In this discussion, we assume that the heavy neutrinos are relativistic and the limit $m_1/E_{\nu_1} \rightarrow 0$.} (see also the discussion in Ref.\cite{deGouvea:2019goq}).

Let us discuss each case individually, having in mind that
in our calculations we have always assumed that the h.f. contributions are active. First, we remind that, for NO and the QD mass pattern, the results hold independently from the neutrino nature, since there is no $\nu \leftrightarrow \bar{\nu}$ decay in this case. 

If neutrinos are Majorana particles, it is not meaningful to assign a lepton number to $\phi$ and Eq.\eqref{eq:Dim5}
mediates both processes \eqref{eq:decay} with h.f. decays, namely $\nu_{L} \rightarrow \nu_R + \phi $  or $\nu_{R} \rightarrow \nu_L + \phi $ and h.c. decays, i.e. $\nu_{L} \rightarrow \nu_L + \phi $ or $\nu_{R} \rightarrow \nu_R + \phi $.
In this case the final states are active neutrinos that are visible in detectors ($\nu_R$ are antineutrinos and $\nu_L$ are neutrinos). 

If neutrinos are Dirac particles, the new degrees of freedom can be classified with respect to the conserved global (lepton-number) symmetry $U(1)_L$.
For Dirac neutrinos, if one has some combination of lepton-number zero ($\phi_0$) and two ($\phi_2$) new degrees of freedom, then for NO and SH one
has an h.c. contribution from $\nu_L \rightarrow \nu_L + \phi_0$ and an h.f. contribution from $\nu_L \rightarrow \bar{\nu}_R + \phi_2$.
Both final states are visible. 

Finally, if neutrinos are Dirac particles, if there is only one new degree of freedom (either $\phi_0$ or $\phi_2)$ and nature has opted either for IO or for NO and the SH mass pattern, then one should reconsider the impact of neutrino decay including  ``wrong helicity" -- sterile -- contributions due to h.f. (for initial decaying neutrinos) and h.c. (for initial decaying antineutrinos) contributions.

\section{Numerical results}
\noindent
We present now our results in the $2 \nu$ and $3 \nu$ frameworks for the DSNB fluxes and number of events, in the absence and presence of $\nu$ non-radiative two-body decay. 
We make predictions for the running SK-Gd and upcoming water Cherenkov detector HK, the JUNO scintillator, and the DUNE liquid argon detectors. 
We consider three values of the lifetime-over-mass ratio, namely 
\begin{itemize}
\item $(\tau/m) _{\it short} = 10^9$ s/eV; 
\item $(\tau/m) _{\it medium}= 10^{10}$ s/eV; 
\item $(\tau/m) _{\it long}= 10^{11}$ s/eV; 
\end{itemize}
The $(\tau/m) _{\it short}$ case corresponds to almost complete neutrino decay, whereas $(\tau/m) _{\it long}$ is close
to the upper bound that one gets with (the rule of thumb)  \cite{Fogli:2004gy} 
\beq
\tau/m \le H_0^{-1} \sim O(10^{11})~ {\rm s/eV}  
\eeq
for typical supernova neutrino energies.

\subsection{DSNB fluxes with and without decay} 
\noindent
Let us first look at the results on the relic supernova neutrino fluxes of flavor $\alpha$ that are connected to the ones in the mass eigenstate basis according to
\beq\label{eq:phiflavor}
\phi_{\nu_{\alpha}}(E_{\nu})  = \sum_i \vert U_{\alpha i} \vert^2 \phi_{\nu_i}(E_{\nu}) \ ,
\eeq
where $U$ is the Pontecorvo-Maki-Nakagawa-Sakata unitary matrix that relates the neutrino flavor
and mass basis, i.e. $\vert \nu_{\alpha} \rangle = \sum_{i} U_{\alpha i}^* \vert \nu_i \rangle$ ($\alpha = e, \mu, \tau, ~i = 1, 2, 3$). For $3 \nu$ flavors the matrix depends on three neutrino mixing angles, one Dirac and two Majorana CP violating phases. The latter are still unknown. For our calculations we employ $\theta_{23} \approx 45^{\circ} $, $\theta_{12} \approx 34^{\circ}$ and $\theta_{13} \approx 8.5^{\circ} $ as values of the neutrino mixing angles\footnote{Note that in the $3\nu$ flavor study of \cite{Fogli:2004gy} $\theta_{13} = 0^{\circ}$ and therefore $U_{e3} = 0$.} \cite{ParticleDataGroup:2020ssz}. Note that, there are hints at 2.5 $\sigma$ in favor of the normal mass ordering and for a Dirac phase such that $\sin \delta < 0$ (both at 90 $\%$ C.L.) \cite{Capozzi:2021fjo}.

\subsubsection{DSNB fluxes in absence of neutrino decay}
\begin{figure}[!thb] 
    \centering
    \begin{minipage}{.5\textwidth}
        \includegraphics[scale=0.376]{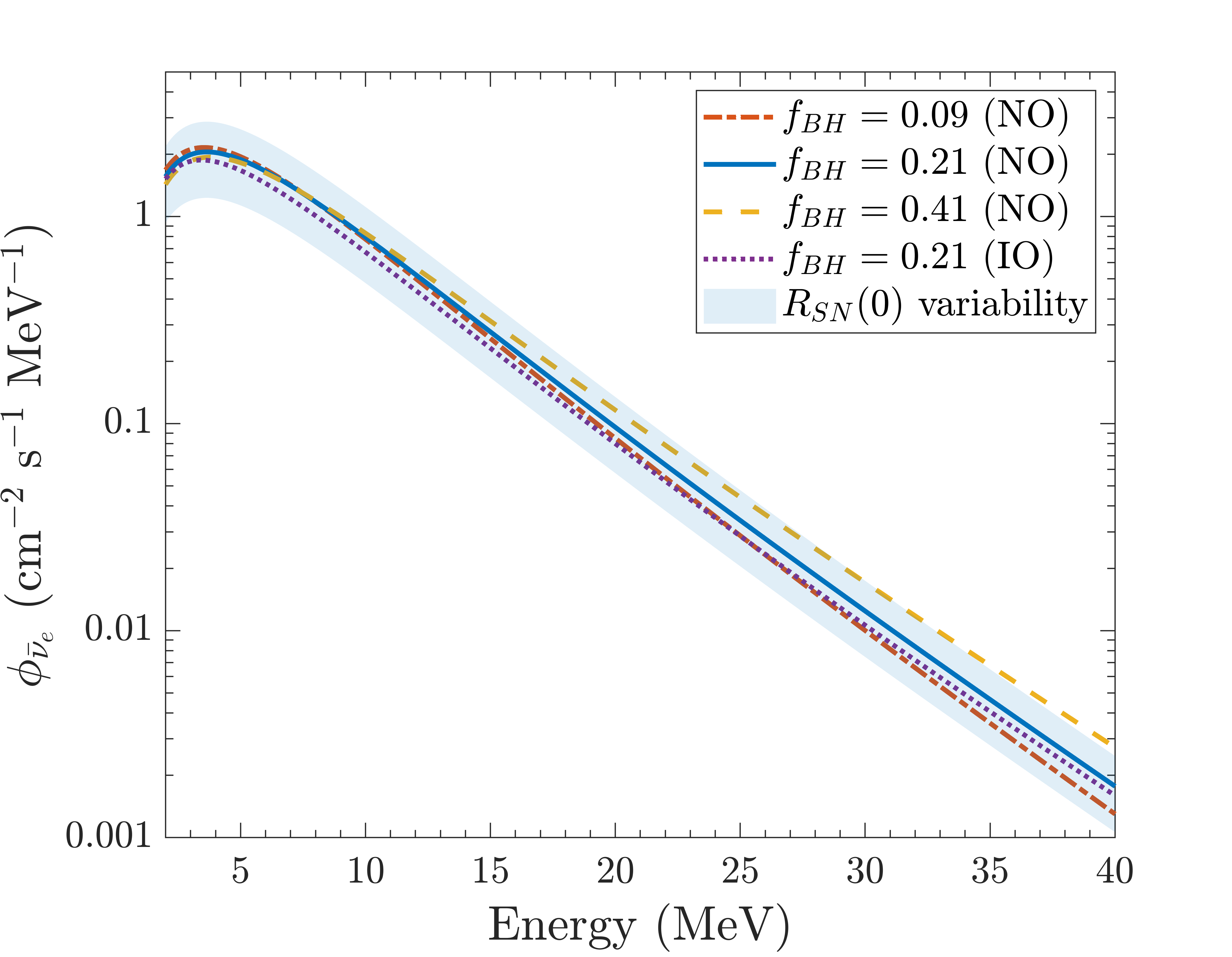}
        \includegraphics[scale=0.376]{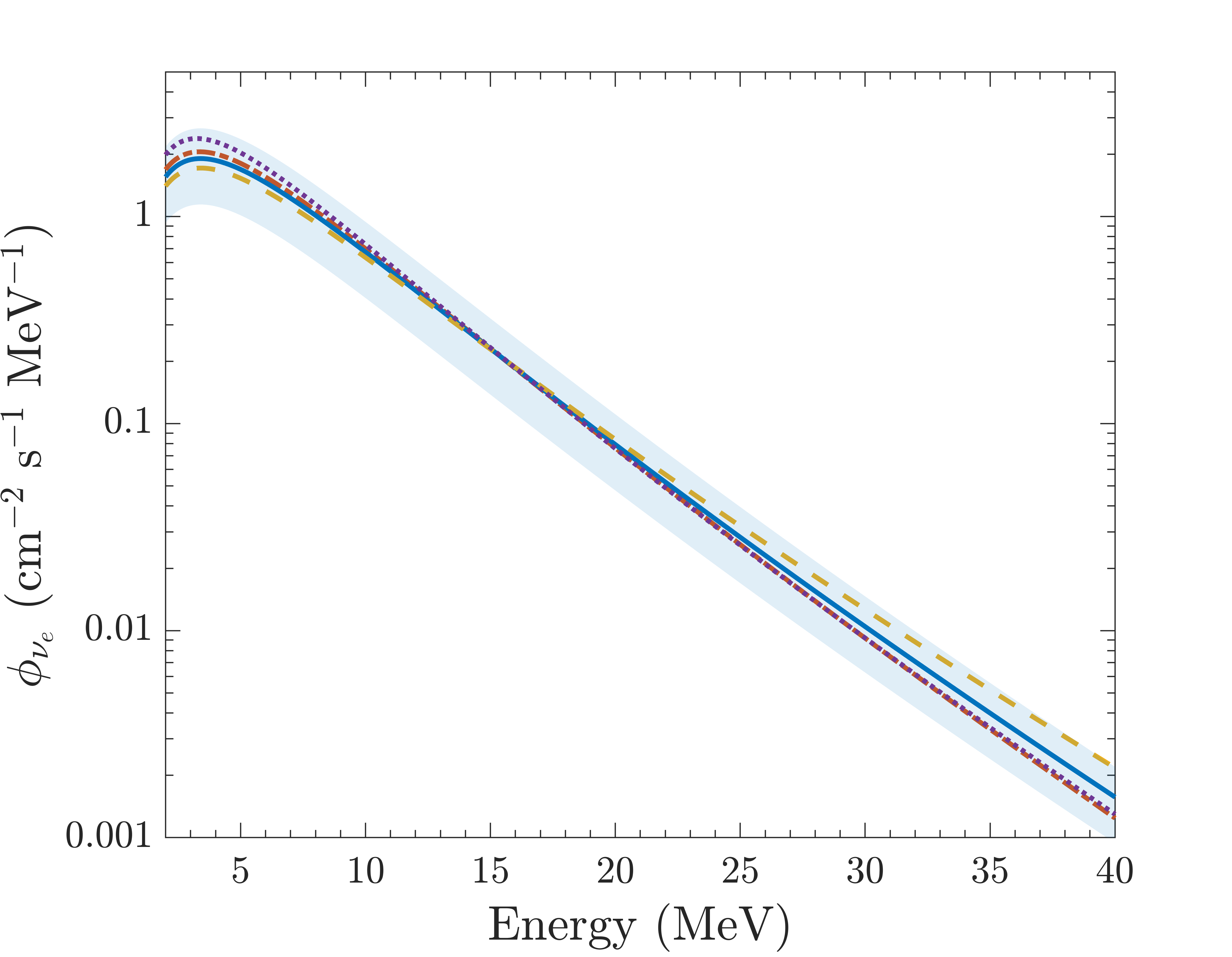}
               \includegraphics[scale=0.376]{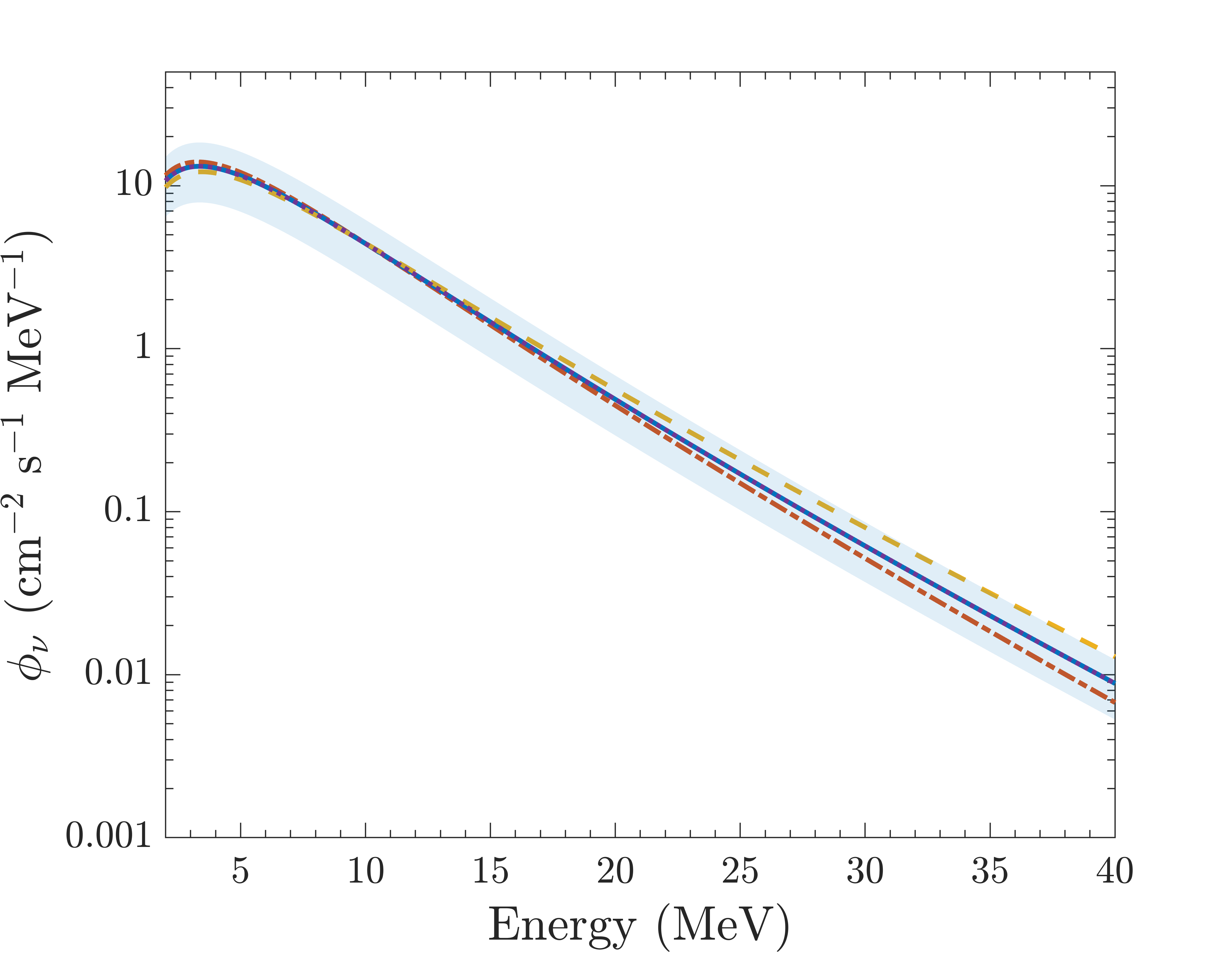}
        \caption{No $\nu$ decay: DSNB fluxes for electron neutrinos (top), electron anti-neutrinos (middle) and all neutrino flavors added (bottom) as a function of neutrino energy, with the $\Lambda$CDM model and the core-collapse supernova rate Eq.\eqref{eq:CCSNrate}. Three different scenarios are taken for the BH fraction. The NO results correspond to the {\it Fiducial} model ($f_{\rm BH} = 0.21 $), the band showing the uncertainty from $R_{SN}$ (see Table~\ref{tab:Rsnpar}). The IO results are for the {\it Fiducial} model without the $R_{SN}$ uncertainty.}
 \label{fig:nodecay}
    \end{minipage}
\end{figure}

\noindent
Let us first consider the DSNB fluxes for $\nu_e$, $\bar{\nu}_e$ and of all flavors added,  in absence of neutrino decay for NO and IO (Figure~\ref{fig:nodecay}).
These results are obtained with the core-collapse supernova rate Eq.\eqref{eq:CCSNrate}-\eqref{eq:SFR} and the three scenarios for the BH fractions  $f_{\rm BH} = 0.09, 0.21, 0.41$ described above. The parameters defining the neutrino fluences at the neutrinosphere are shown in Table~\ref{tab:fluxpar}. The band in Figure~\ref{fig:nodecay} corresponds to the uncertainty in the evolving core-collapse supernova rate.

 As one can see from the figure, our results for the DSNB fluxes agree well with those of \cite{Moller:2018kpn} ({\it cfr.} Figure~\ref{fig:decaypatterns}). We remind that here we included the 25 $M_{\odot}$ NS and BH cases as well, as in \cite{Priya:2017bmm}.    

\subsubsection{DSNB fluxes in presence of neutrino decay in the $3\nu$ and effective 2$\nu$ formalism}
\noindent
Before giving our results when neutrinos decay in the 3$\nu$ framework, let us look at the differences that arise when an effective 2$\nu$ flavor formalism is considered\footnote{We show the results for $\bar{\nu}_e$. The difference between $\nu_e$ and $\bar{\nu}_e$ relic fluxes is discussed in $3\nu$ framework.}, i.e. the decaying mass eigenstate $\nu_2$ for NO or $\nu_1$ for IO is considered as stable. 

\begin{figure}[!thb] 
    \centering
    \begin{minipage}{.5\textwidth}
          \includegraphics[scale=0.376]{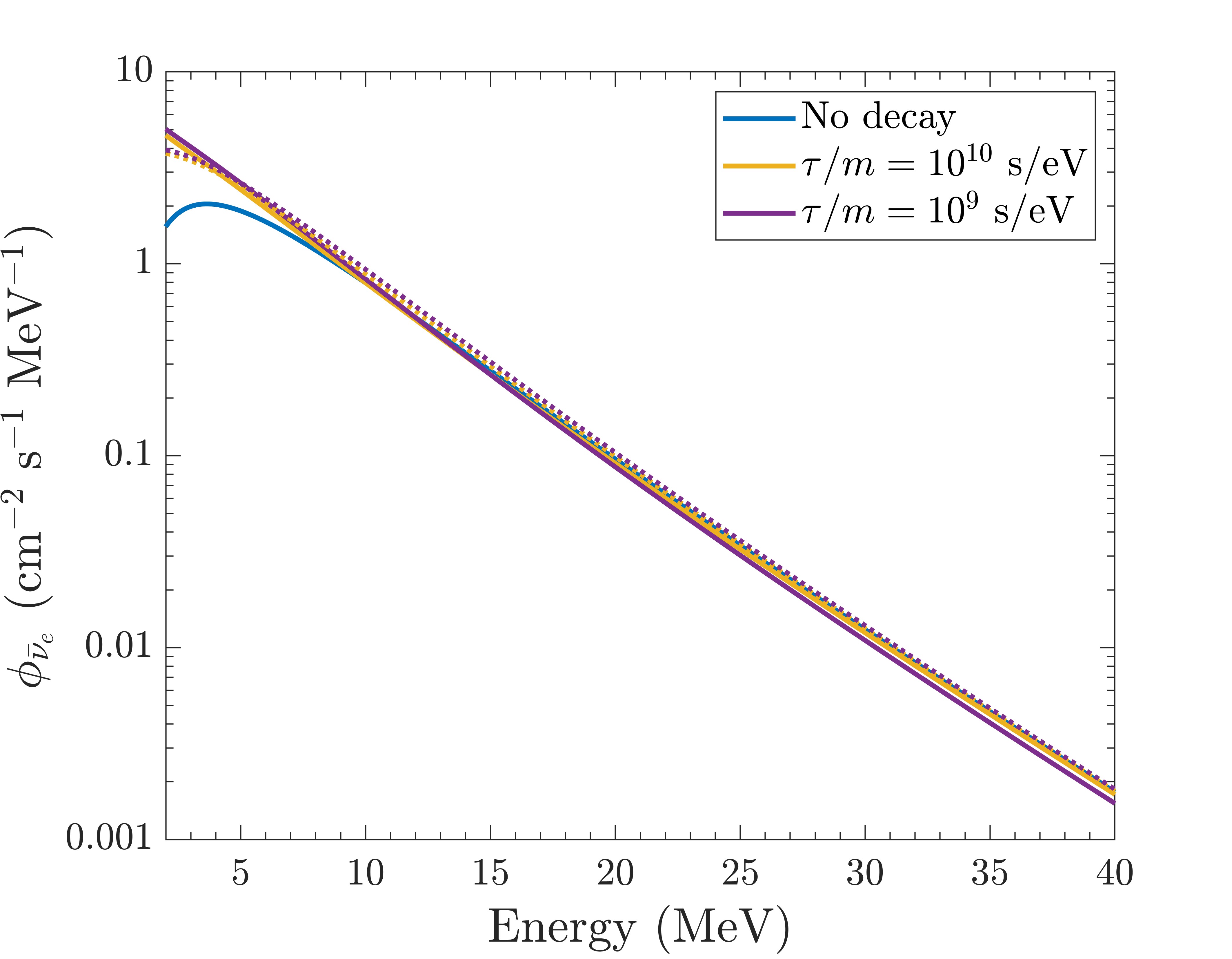}    
        \includegraphics[scale=0.376]{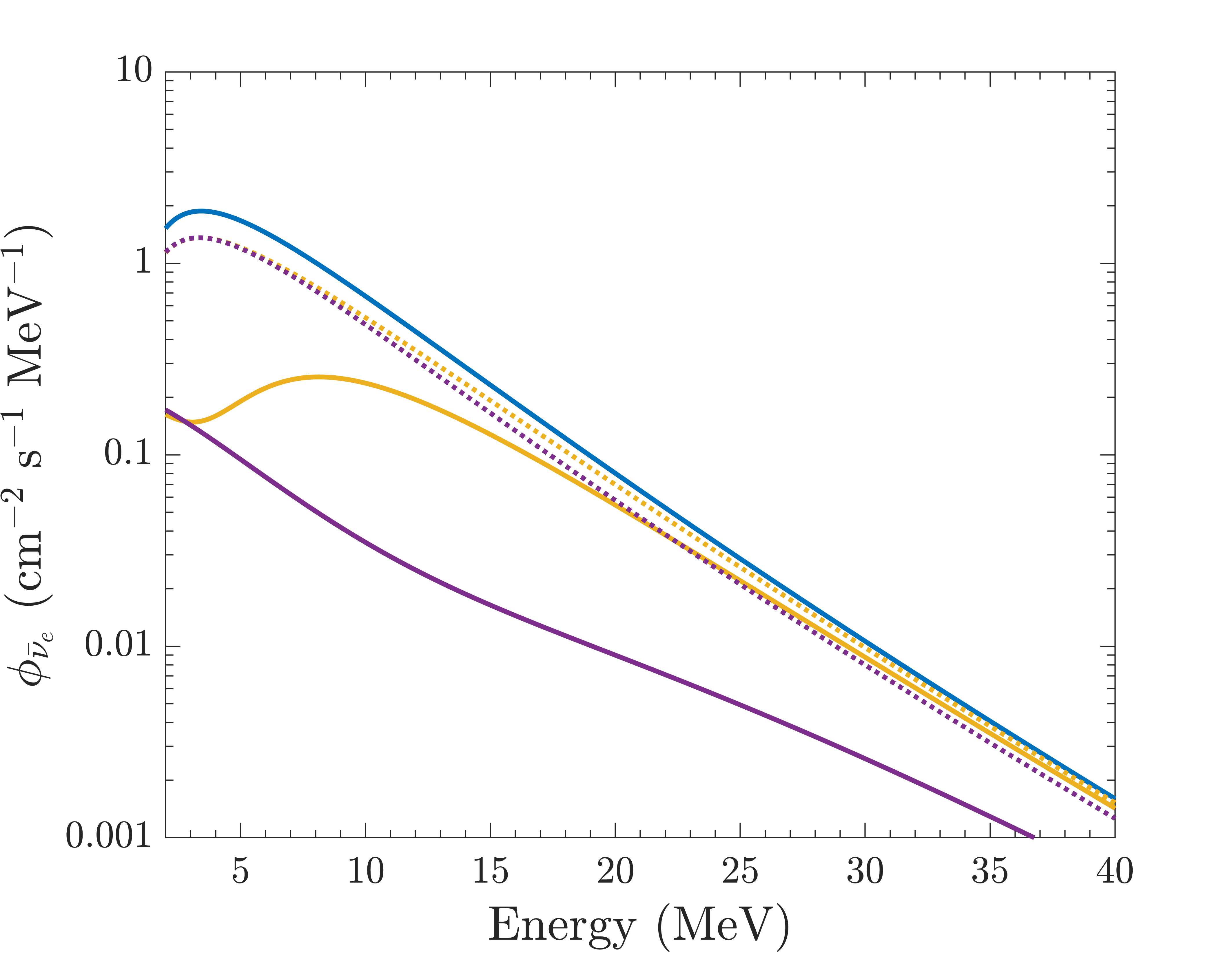}
        \caption{Comparison of the DSNB $\bar{\nu}_e$ fluxes in presence of $\nu$ decay, with $2\nu$ (dotted) or 3$\nu$ (full lines), for NO,  SH (top) and IO (bottom). Only $(\tau/m)_{short}$ and $(\tau/m)_{medium}$ are presented for clarity.  The DSNB fluxes in absence of decay are also shown. The results correspond to the {\it Fiducial} model.}
 \label{fig:flux2fvs3f}
    \end{minipage}
\end{figure}

Figure~\ref{fig:flux2fvs3f} compares the DSNB $\bar{\nu}_e$ fluxes without decay with those corresponding to the shortest and medium lifetime-over-mass ratios, for NO (top) or IO (bottom). As one can see, for the NO and SH scenario, predictions for the DSNB $\bar{\nu}_e$ fluxes from the effective $2 \nu$ or the $3 \nu$ frameworks are indistinguishable. This result supports the findings of, for example, \cite{deGouvea:2019goq,Tabrizi:2020vmo} where $2 \nu$ flavors are considered with NO and SH pattern only.
We also find that, for NO and a QD mass pattern, the DSNB flux predictions are practically the same with $2 \nu$ or $3 \nu$. 

As for IO, the results based on $2 \nu$ or $3 \nu$ flavors are very close when considering $(\tau/m)_{\it medium}$ (above 15 MeV) and $(\tau/m)_{\it long} $ (not shown). On the contrary, for $(\tau/m)_{\it medium}$ (below 20 MeV) and $(\tau/m)_{\it short}$, there are significant differences between the two frameworks, as one can see from Figure~\ref{fig:flux2fvs3f}.

\begin{figure}[!thb] 
    \centering
    \begin{minipage}{.5\textwidth}
        \includegraphics[scale=0.376]{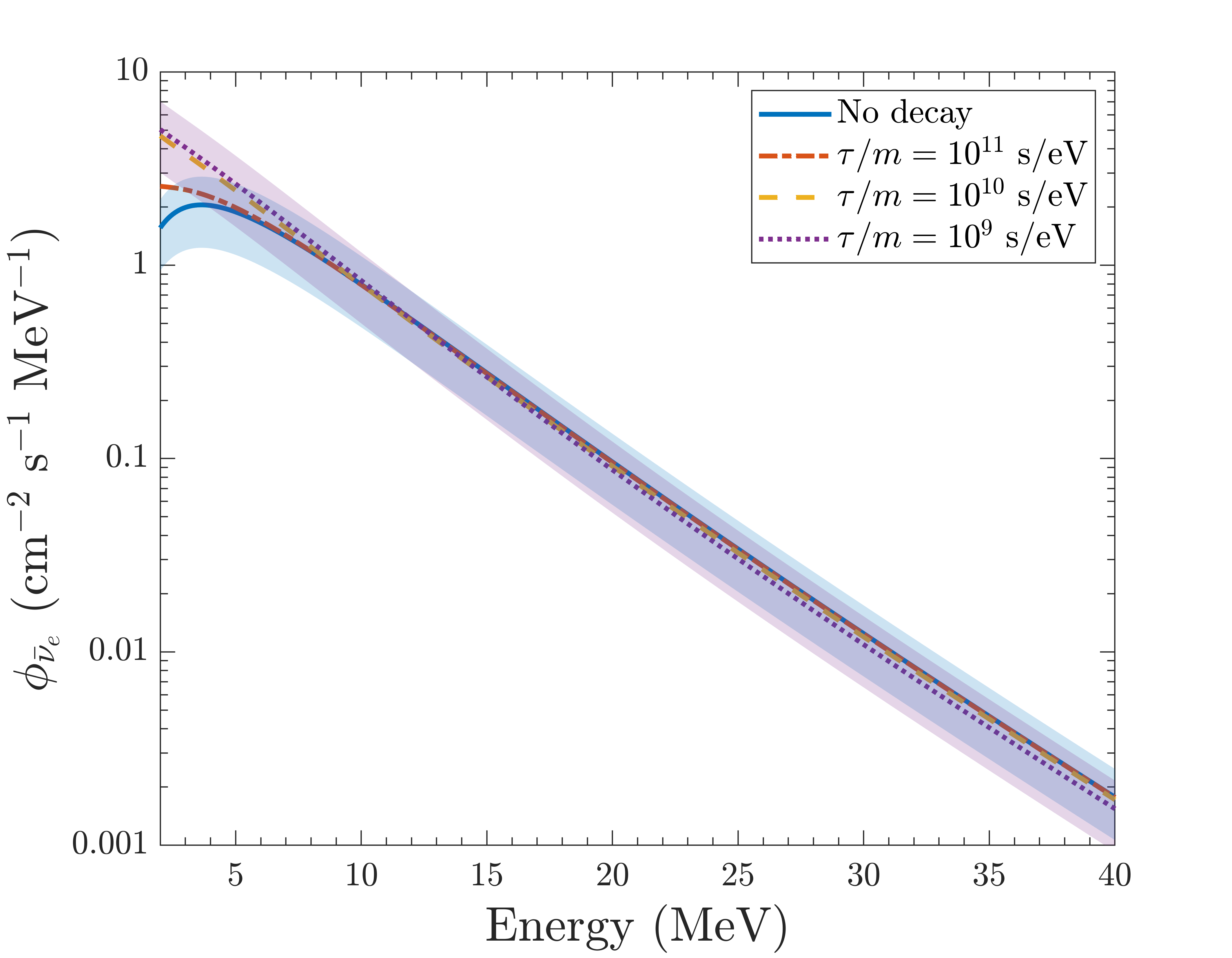}
        \caption{Neutrino decay in $3\nu$ framework: DSNB $\bar{\nu}_e$ fluxes for NO with the SH decay pattern. Three values of the lifetime-over-mass ratio are considered. 
        The lines show the results for the {\it Fiducial} model, whereas the bands come from the uncertainty on core-collapse supernova rate.  The DSNB fluxes in absence of neutrino decay are given for comparison.}
 \label{fig:3fNOSH_ea}
    \end{minipage}
\end{figure}

\begin{figure}[!thb] 
    \centering
    \begin{minipage}{.5\textwidth}
        \includegraphics[scale=0.376]{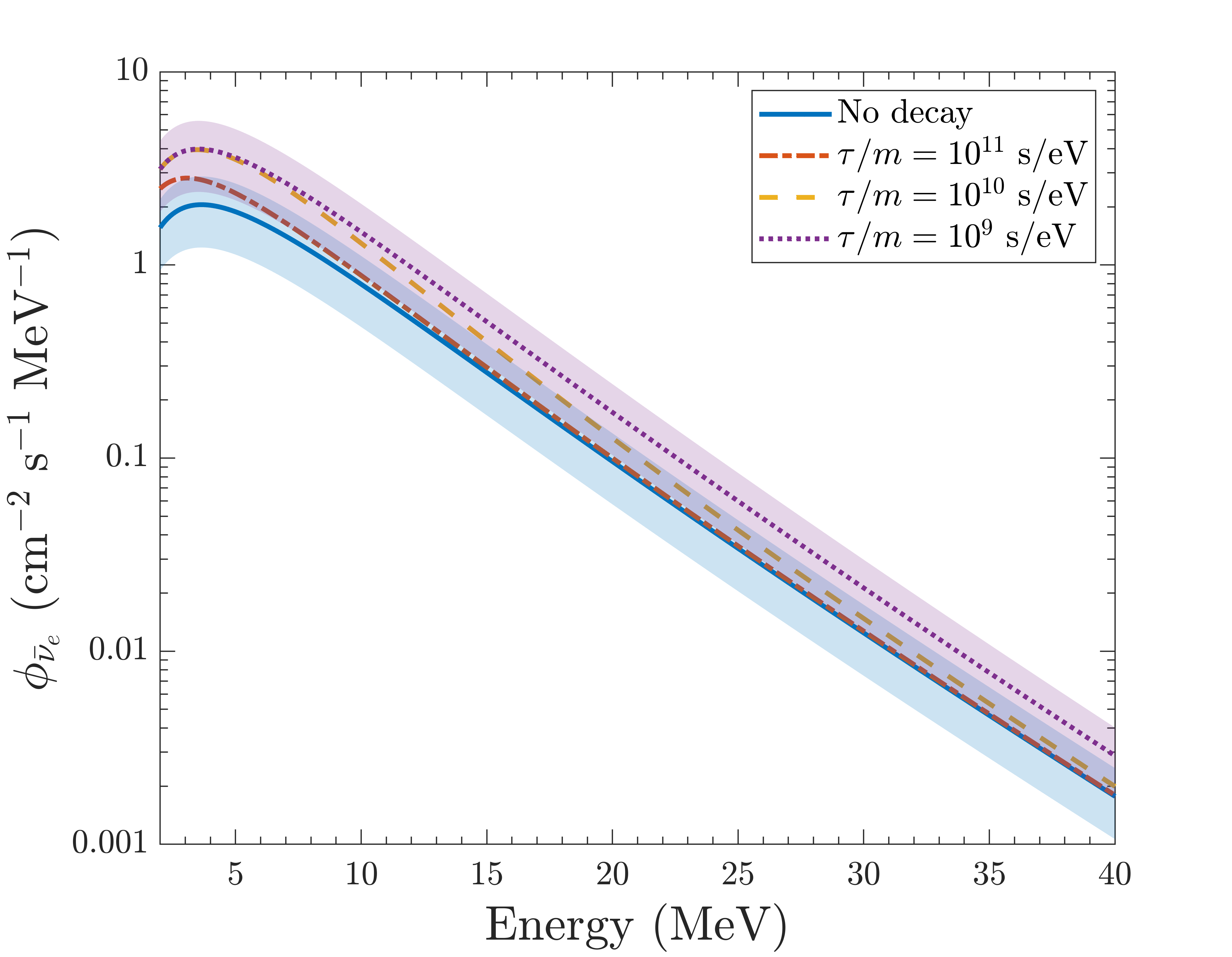}
            \includegraphics[scale=0.376]{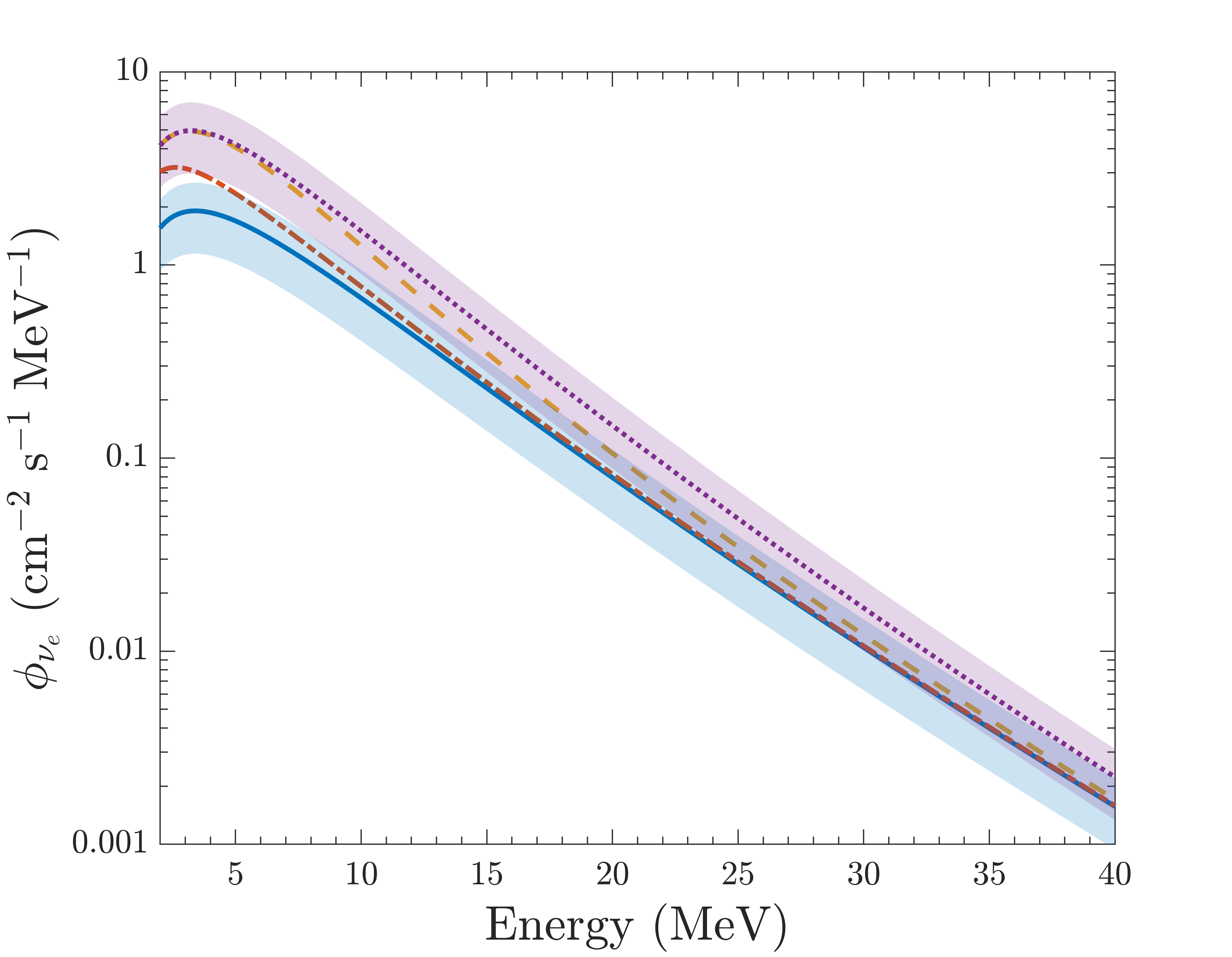}
        \caption{Neutrino decay in $3\nu$ framework: DSNB $\bar{\nu}_e$ and $\nu_e$ fluxes in NO, with the QD decay pattern. Three values of the lifetime-over-mass ratio are considered. 
        The lines show the results for the {\it Fiducial} model, whereas the bands come from the uncertainty on core-collapse supernova rate. The DSNB fluxes in absence of neutrino decay are given for comparison.}
 \label{fig:3fNOQD}
    \end{minipage}
\end{figure}

Let us now consider the DSNB fluxes in presence of neutrino decay in the 3$\nu$ framework.
Figures~\ref{fig:3fNOSH_ea}, \ref{fig:3fNOQD} and \ref{fig:3fIO_ea} present the results obtained by solving Eq.\eqref{eq:kinsol} (Appendix B). The DSNB flux behaviors we find are in concordance with those of \cite{Fogli:2004gy} although the authors employed an older core-collapse supernova rate (see Figure~\ref{fig:CCSN}),
one effective Fermi-Dirac distribution for the supernovae neutrino spectra and no progenitor dependence\footnote{This {\it ansatz } was common in the predictions at that time.}.

Figure~\ref{fig:3fNOSH_ea} shows the DSNB $\bar{\nu}_e$ fluxes for NO in the SH case. The $\nu_e$ ones are close and follow a similar behavior. As in \cite{Fogli:2004gy}, we find an enhancement of the DSNB fluxes for $(\tau/m)_{\it short}$ at low energies.  
In fact, at low energy, the $\nu_e$ and $\bar{\nu}_e$ fluxes receive a significant contribution from $\nu_2$ and $\nu_3$ decays that dominate over the contribution of the first term of Eq.\eqref{eq:kinsol}. Even if this enhancement, present for $(\tau/m)_{\it short} $ and for $(\tau/m)_{\it medium} $, is interesting, it appears well below the current energy thresholds. Unfortunately, it will be hard to see such enhancement, even considering lower energy thresholds due to improvements in background suppression (from e.g. the reduction of atmospheric spallation products like $^9$Li in SK-Gd and HK).  For $(\tau/m)_{\it short} $ the flux is slightly suppressed at higher energies.

The NO and the QD case show different flux behaviors, compared to the SH one (Figure~\ref{fig:3fNOQD}) since the first term dominates over the second in Eq.\eqref{eq:kinsol}. 
The DSNB $\nu_e$ flux differs from the $\bar{\nu}_e$ one, only below 10 MeV. This difference comes from the different spectra at the neutrinosphere. 
One can also see that, when the uncertainty in the evolving core-collapse supernova rate is included, the results for the {\it Fiducial} model 
with no decay significantly overlap, in the DSNB detection window, with those for $(\tau/m)_{\it short}$. Clearly, with the present knowledge, flux modifications due to neutrino non-radiative decay would be hidden by such uncertainty in NO with SH and QD mass patterns.  

The situation is different in IO. The corresponding DSNB fluxes with neutrino decay present interesting features, as can be seen from Figure~\ref{fig:3fIO_ea}. 
We give the results for $\bar{\nu}_e$, since the DSNB $\nu_e$ fluxes ({\it Fiducial} model) are the same. First of all, one can see a significant suppression of the fluxes for $(\tau/m)_{\it short}$, whereas the ones for $(\tau/m)_{\it long}$ 
are equivalent (above 10 MeV) to the no decay case. The results with $(\tau/m)_{\it medium}$ are close to the no decay results above 15 MeV, whereas below they show a suppression up to a factor of 6, compared to the fiducial no decay case. 

Interestingly, for IO, the $(\tau/m)_{\it short}$ and no decay cases differ significantly in the full DSNB detection window, even considering the current core-collapse supernova normalization uncertainty. This suppression is due to the fact that, for IO, the DSNB $\bar{\nu}_e$ flux receives a small contribution ($\vert U_{e3} \vert^2= 2 \times 10^{-2}$) from the stable $\bar{\nu}_3$ and large from $\nu_1$ ($\bar{\nu}_1$)  and $\nu_2$ ($\bar{\nu}_2$). We shall discuss its implication for the DSNB events in the following section.    

\subsubsection{Integrated DSNB fluxes and current bounds}
\noindent
Let us now discuss the DSNB integrated $\bar{\nu}_e$  and ${\nu}_e $ fluxes for NO (IO)
without decay, in comparison with current bounds (Table~\ref{tab:Fluxlim}). We consider both the {\it Fiducial} model and an optimistic prediction with  $f_{\rm BH} = 0.41$. 
The theoretical errors correspond to the core-collapse supernova rate uncertainty.

Our values are below the $\bar{\nu}_e$ upper limit, obtained from the combined analysis of SK-I to SK-IV data \cite{Super-Kamiokande:2021jaq}, by a factor 2 to 4. 
Note that the KamLAND experiment obtained the upper value of  139 $\bar{\nu}_e$ cm$^{-2}$s$^{-1}$ (90 $\%$ C.L.) in the window $[8.3, 31.8]$ MeV \cite{KamLAND:2011bnd}; slightly improved in the interval $[7.8, 16.8]$ MeV by the model-dependent limit of 112.3 $\bar{\nu}_e$ cm$^{-2}$s$^{-1}$ (90 $\%$ C.L.) of the Borexino Collaboration \cite{Borexino:2019wln}.
For the integrated DSNB ${\nu}_e $ flux, the predictions are lower by about two orders of magnitudes than the current bound from the ensemble of SNO data \cite{SNO:2006dke}.

\begin{table}[tp]
\centering
\begin{tabular}{cccc}
\hline \hline
 &  NO  & IO & Upper limits \\
\hline
$~\bar{\nu}_e~$  & 0.77 $\pm$ 0.30  & 0.63 $\pm$ 0.25  & 2.7 (SK) \\
 &  [$1.02 \pm 0.41$]  & [0.75 $\pm$ 0.3] & \\
$~\nu_e~$ & 0.20 $\pm$ 0.08  & 0.18 $\pm$ 0.08   &  19  (SNO) \\
&   [0.24 $\pm$ 0.9] & [0.23 $\pm$ 0.9] & \\
\hline \hline
\end{tabular}
\caption{Integrated DSNB fluxes (cm$^{-2}$ s$^{-1}$) for the {\it Fiducial} model and the optimistic case with $f_{\rm BH} = 0.41$ (in brackets), in comparison with the current upper limits.
Our results are for the case of no-decay and for the two mass orderings. The theoretical errors come from the core-collapse supernova rate uncertainty. The experimental upper limits  (90 $\%$ C.L.) are from SK-I to SK-IV \cite{Super-Kamiokande:2021jaq} and SNO \cite{SNO:2006dke} with the DSNB windows of $E_{\nu} > 17.3 $ MeV (positron energy) and [22.9, 36.9] MeV (neutrino energy) respectively.}
\label{tab:Fluxlim}
\end{table}

\begin{figure}[!thb] 
    \centering
    \begin{minipage}{.5\textwidth}
        \includegraphics[scale=0.376]{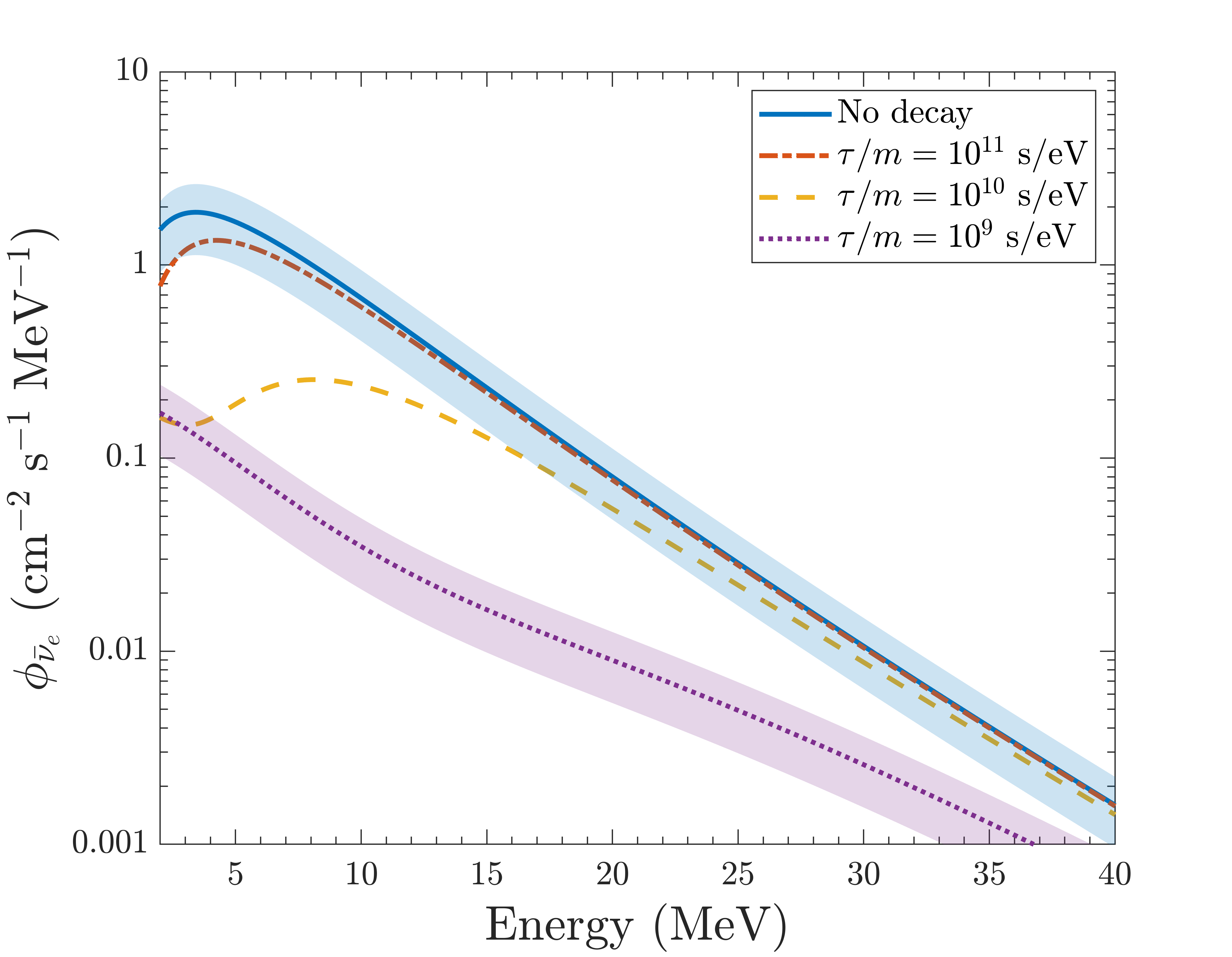}
        \caption{Neutrino decay in $3\nu$ framework: DSNB $\bar{\nu}_e$ fluxes for IO. Three values of the lifetime-over-mass ratio are considered. 
        The lines show the results for the {\it Fiducial} model, whereas the bands come from the uncertainty on core-collapse supernova rate.  The DSNB fluxes in absence of neutrino decay are given for comparison.}
 \label{fig:3fIO_ea}
    \end{minipage}
\end{figure}

Table~\ref{tab:integrated_flux} shows the integrated supernova relic fluxes for $\bar{\nu}_e $ and $\nu_e$ without/with decay for the SK-Gd, HK, JUNO and DUNE experiments and the related  DSNB detection windows.  Different values of $\tau/m$ are considered as well as the three decay patterns of Figure~\ref{fig:decaypatterns}. 

As expected  from the flux results shown in Figures ~\ref{fig:3fNOSH_ea}-~\ref{fig:3fIO_ea}, the integrated DSNB fluxes have little sensitivity to $\tau/m$ for NO and the SH pattern. On the contrary, for NO and the QD pattern, they increase by a factor of 1.8 from no decay to the $(\tau/m)_{\it short}$ case.  
For IO, a significant decrease appears when neutrino decay is considered, from a factor of about 6.7 in DUNE to about 14 in SK-Gd, HK and JUNO for $(\tau/m)_{\it short}$.  For $(\tau/m)_{\it medium}$, the suppression grows
from 40 $\%$ (DUNE) to a factor of 2 (HK).
    
  \begin{table}[]
\begin{tabular}{|l|cccc|}
\hline
\multicolumn{1}{|c|}{\multirow{2}{*}{}} &
  \multicolumn{4}{c|}{DSNB Flux (cm$^{-2}$ s$^{-1}$)} \\ 
  \cline{2-5} 
\multicolumn{1}{|c|}{}   &
  \multicolumn{1}{l|}{No decay} &
  \multicolumn{1}{l|}{$(\tau/m)_{\it long}$} &
  \multicolumn{1}{l|}{$(\tau/m)_{\it medium}$} &
  \multicolumn{1}{l|}{$(\tau/m)_{\it short}$} \\ 
  \hline
 \multicolumn{1}{|c|}{\begin{tabular}[c]{@{}c@{}}SK-Gd ($\bar{\nu}_e$) \\ (12.8, 30.8) \end{tabular}}  &
  \multicolumn{1}{c|}{\begin{tabular}[c]{@{}c@{}}2.05\\ (1.71)\end{tabular}} &
  \multicolumn{1}{c|}{\begin{tabular}[c]{@{}c@{}}2.16\\ {[}2.03{]}\\ (1.62)\end{tabular}} &
  \multicolumn{1}{c|}{\begin{tabular}[c]{@{}c@{}}2.87\\ {[}1.96{]}\\ (1.02)\end{tabular}} &
  \begin{tabular}[c]{@{}c@{}}3.72\\ {[}1.92{]}\\ (0.12)\end{tabular} \\ 
  \hline
  \multicolumn{1}{|c|}{\begin{tabular}[c]{@{}c@{}}HK ($\bar{\nu}_e$) \\ (17.3, 31.3)\end{tabular}} &
  \multicolumn{1}{c|}{\begin{tabular}[c]{@{}c@{}} 0.77 \\ (0.64)\end{tabular}} &
  \multicolumn{1}{c|}{\begin{tabular}[c]{@{}c@{}} 0.80 \\ {[}0.76{]}\\ (0.62)\end{tabular}} &
  \multicolumn{1}{c|}{\begin{tabular}[c]{@{}c@{}} 1.00\\ {[}0.73{]}\\ (0.45)\end{tabular}} &
  \begin{tabular}[c]{@{}c@{}} 1.37\\ {[}0.70{]}\\ (0.08)\end{tabular} \\ 
  \hline
  \multicolumn{1}{|c|}{\begin{tabular}[c]{@{}c@{}}JUNO ($\bar{\nu}_e$) \\ (11.3, 33.3)\end{tabular}} &
  \multicolumn{1}{c|}{\begin{tabular}[c]{@{}c@{}}2.85\\ (2.38)\end{tabular}} &
  \multicolumn{1}{c|}{\begin{tabular}[c]{@{}c@{}}3.03\\ {[}2.83{]}\\ (2.24)\end{tabular}} &
  \multicolumn{1}{c|}{\begin{tabular}[c]{@{}c@{}}4.1\\ {[}2.74{]}\\ (1.33)\end{tabular}} &
  \begin{tabular}[c]{@{}c@{}}5.2\\ {[}2.72{]}\\ (0.20)\end{tabular} \\ \hline
\multicolumn{1}{|c|}{\begin{tabular}[c]{@{}c@{}}DUNE (${\nu}_e$) \\ (19, 31)\end{tabular}} &
  \multicolumn{1}{c|}{\begin{tabular}[c]{@{}c@{}}0.43\\ (0.40)\end{tabular}} &
  \multicolumn{1}{c|}{\begin{tabular}[c]{@{}c@{}}0.45\\ {[}0.43{]}\\ (0.39)\end{tabular}} &
  \multicolumn{1}{c|}{\begin{tabular}[c]{@{}c@{}}0.55\\ {[}0.41{]}\\ (0.29)\end{tabular}} &
  \begin{tabular}[c]{@{}c@{}}0.77\\ {[}0.38{]}\\ (0.06)\end{tabular} \\ \hline
\end{tabular}
\caption{Integrated DSNB fluxes for the {\it Fiducial  } model with $f_{\rm BH} =0.21$, in the absence and presence of neutrino non-radiative decay. The results are obtained in the $3\nu$ framework. The first column gives the experiment and the expected DSNB detection window (in MeV) used in our calculations. For the results with neutrino decay, three values of the neutrino lifetime-over-mass ratio are shown: $\tau/m = 10^{9}$ s/eV ({\it short)}, $10^{10}$ s/eV ({\it medium}) and $10^{11}$ s/eV ({\it long}). For each experiment, the upper values correspond to NO, QD; the middle ones to NO, SH (in brackets), and the lower ones to IO (in parenthesis). For comparison, the results for stable neutrinos are also shown for NO and IO (in parenthesis).}
\label{tab:integrated_flux}
\end{table}

\subsection{Predictions of the DSNB events}
\noindent
The DSNB total rates in a detector on Earth are 
\beq\label{eq:events}
N_{\alpha}=  \epsilon N_{t} \int d E_{\nu} \phi_{\nu_{\alpha}}(E_{\nu}) \sigma(E_{\nu}) \ ,
\eeq
where $\epsilon$ is the detector efficiency (in a given detection channel), $N_{t}$ is the number of targets (active volume) and $ \sigma(E_{\nu}) $ is the reaction cross section of the associated neutrino detection channel. 

Let us remind that SK is a 50 kton water (22.5 kton fiducial volume) Cherenkov detector located in the Kamioka mine in Japan. 
SK-Gd is running since 2020 and has three phases, with a Gd concentration that increases from 0.01 $\%$ (phase I), 0.03 $\%$ (phase II) to the ultimate 0.1 $\%$ (phase III)
reaching 90 $\%$ efficiency in neutron tagging. It will be running until the start of HK (approximately ten years).

Located at Tochibora site, HK will be the largest water Cherenkov detector with 258 ktons and a fiducial volume 8.4 times the one of SK. Construction started in early 2020 and the detector is expected to start operating in 2027. Numerous aspects relevant to the DSNB are currently under study, such as the PMT coverage of the detector, or algorithms to reduce contributions from spallation due to atmospheric backgrounds.
The possibility to add Gadolinium is also under study \cite{BQ}.

JUNO, with 20 ktons, will be the largest underground scintillator detector. It will be located in Jiangmen, South China, and will be online in 2023 \cite{JUNO:2015zny}. Techniques are being developed for the DSNB flux detection, in particular, concerning background reduction with the pulse shape analysis \cite{JUNO:2022lpc}.

Finally, the DUNE experiment will comprise 40 ktons liquid argon (fiducial volume) with 4 far TPC modules at the Sanford Underground Research Facility in South Dakota \cite{DUNE:2016hlj}. The backgrounds and the efficiency of the detector relevant for low energy DSNB events are under investigation.

In order to study the role of neutrino decay on the DSNB rates, we consider the main detection channels for the Cherenkov and scintillator detectors, that is inverse beta-decay (IBD)
\beq
\bar{\nu}_e  + p \rightarrow n + e^+ \ ,
\eeq
with $E_{e^+} = E_{\bar{\nu}_e} - \Delta_{np} $, $\Delta_{np}= 1.293$ MeV and a low energy threshold $E_{\bar{\nu}_e} > 1.806$ MeV.
For the DUNE experiment, the main detection channel is the charged-current neutrino-argon interaction
\beq
{\nu}_e  + ^{40}{\rm Ar} \rightarrow e^- +  ^{40}{\rm K}^* .
\eeq
We employ the IBD cross section from \cite{Strumia:2003zx} for the former and the $\nu_e$-Ar cross section from \cite{GMP} for the latter.

Table~\ref{tab:exppari} gives the parameters (number of targets in the fiducial volume, efficiency, expected DSNB detection window) as well as the running time of the four experiments, which we use to predict the rates. The table also presents the expected DSNB number of events if neutrinos are stable.   
Note that for SK-Gd we consider two running periods due to the improved efficiency from increased Gd concentration. We keep the same DSNB detection
window for the two periods even though the threshold energy might be lowered thanks to Gd addition. For HK we consider conservative detection efficiencies and windows without and with Gd \cite{BQ}. For JUNO and DUNE, we follow \cite{JUNO:2015zny} and \cite{Tabrizi:2020vmo} for the efficiencies and the detection windows. 

\begin{table}[tp]
\centering
\begin{tabular}{cccccccccccccccccc}
\hline
& & & & \\
 &$N_t$ & $\epsilon$ & Time &  DSNB window & DSNB events\\
& ($10^{33}$) & & (years) & (MeV)  &  \\
& & & & \\
\hline
 & & & & \\
SK-Gd & $1.5 $ & $57.5 \%$  & 2 & (12.8, 30.8) & 2 (2)  $\bar{\nu}_e$ \\
" & " & $73.75\%$ & 8 &  " & 12 (10) $\bar{\nu}_e$ \\
HK & $12.5 $ & $25 \%$ & 20 & (17.3, 31.3) & 48 (40) $\bar{\nu}_e$ \\
HK-Gd & 12.5 & 40 $\%$ & 20 & (17.3, 31.3)  &  76 (64) $\bar{\nu}_e$  \\
JUNO & $1.21$ & $50 \%$ & 20 & (11.3, 33.3) & 20 (17) $\bar{\nu}_e$\\
DUNE & $0.602 $&  $86 \%$ & 20 &  $(19, 31) $ & 12 (11)  ${\nu}_e$\\ 
& & & & \\
\hline
\end{tabular}
\caption{The table shows the four experiments considered: the running SK-Gd, the upcoming HK Cherenkov detectors \cite{Super-Kamiokande:2021jaq}, the scintillator detector JUNO \cite{JUNO:2015zny} and the liquid argon TPC (LArTPC) far detectors in DUNE \cite{DUNE:2016hlj}. The second to fifth columns give the
number of targets (fiducial volume), the efficiency, the running time and the energy window in which the DSNB detection is expected. For SK-Gd the two efficiencies correspond to 0.01 $\%$ and 0.03 $\%$ Gd concentration (phases I and II). For HK we also consider a Gd-doped case, currently under study \cite{BQ}. The last column presents
the DSNB expected events for the case of no decay and NO;  the results for no decay and IO are  in parenthesis.}
\label{tab:exppari}
\end{table}

\subsubsection{DSNB events in the $3\nu$ and effective $2 \nu$ formalism for the decay} 
\begin{figure}[!thb] 
    \centering
                 \includegraphics[scale=0.35]{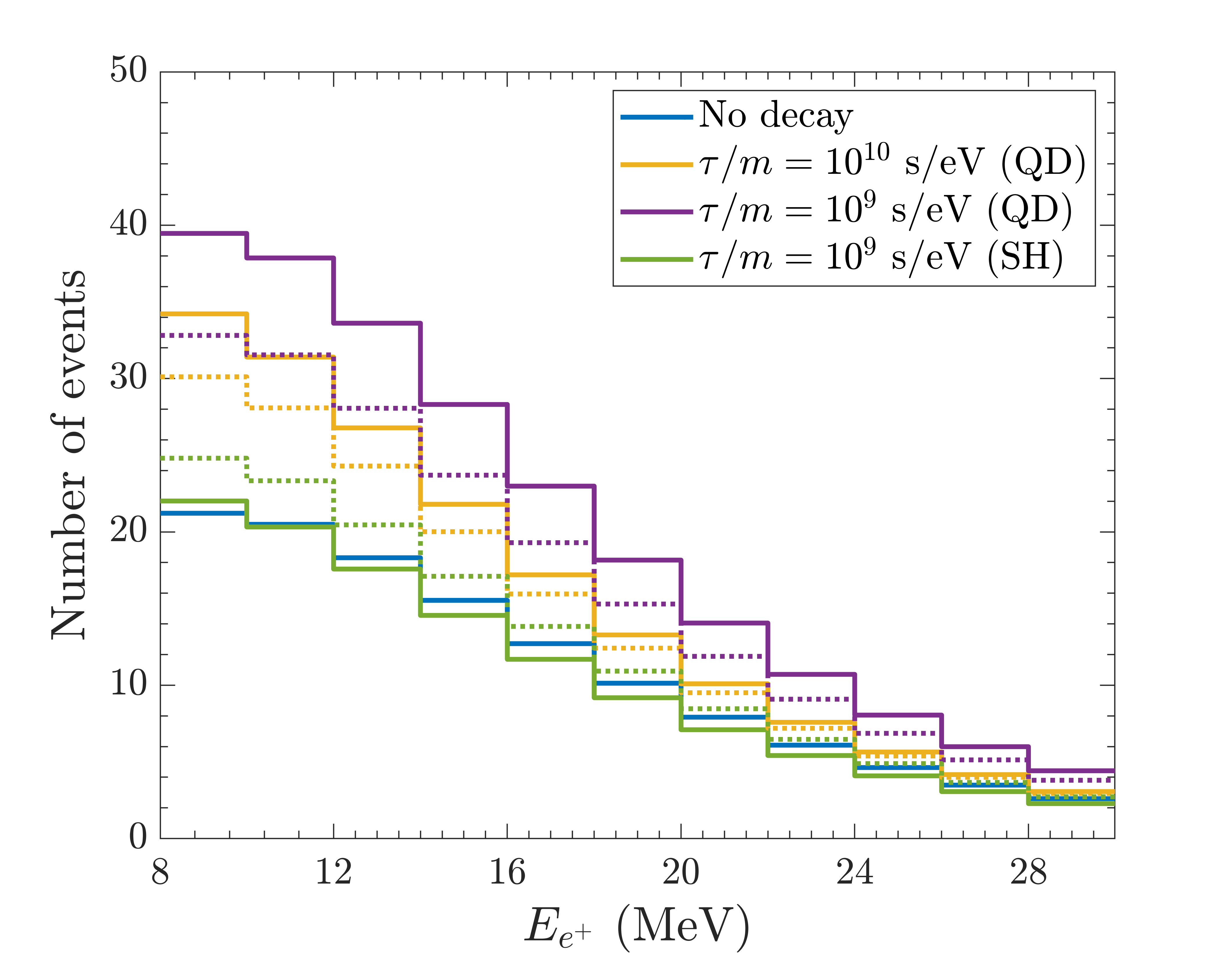}    
         \includegraphics[scale=0.35]{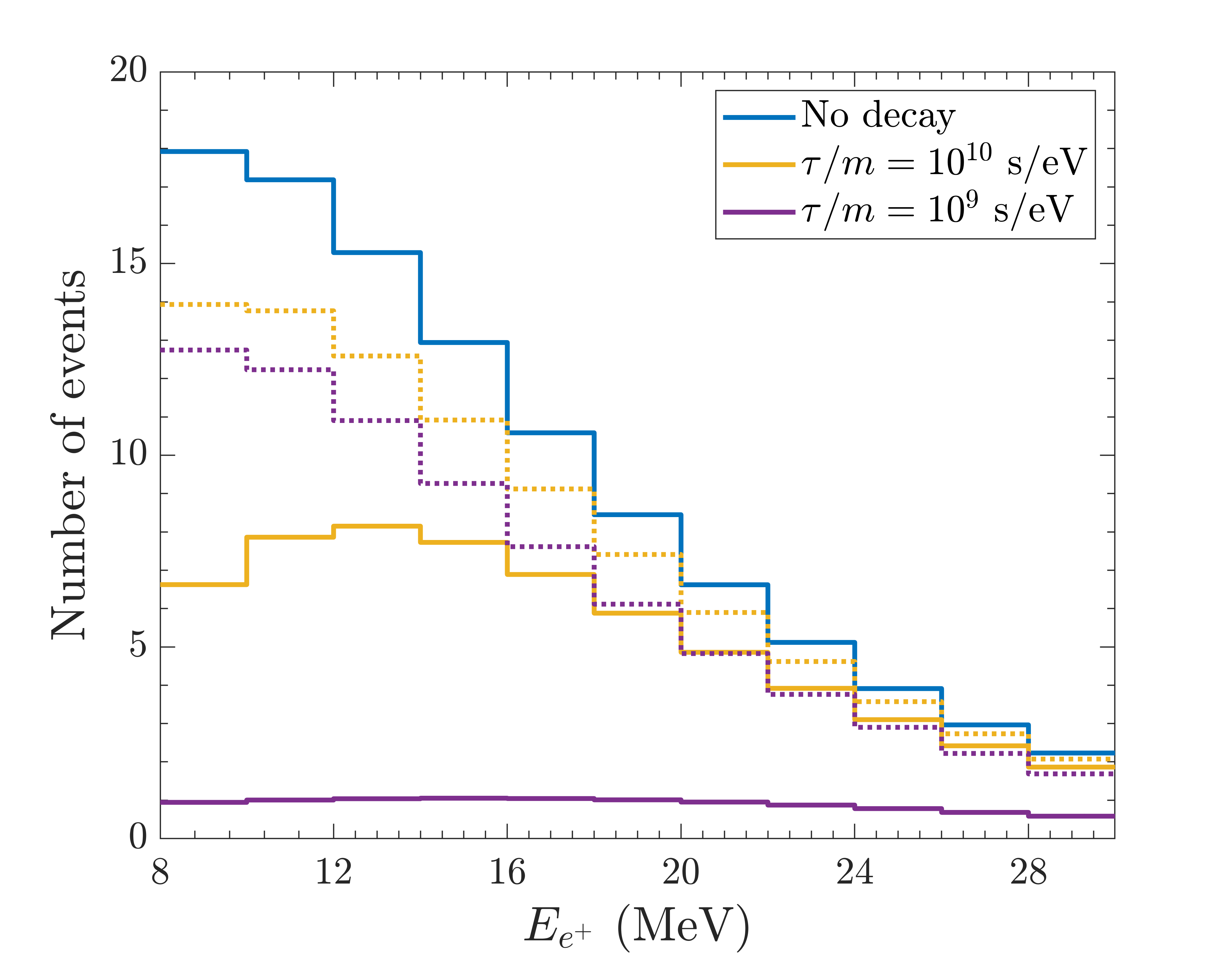}   
               \caption{Comparison of the expected DSNB events, as a function of positron energy, in the effective $2\nu$ (dotted) and the $3\nu$ frameworks (full lines). The results are for the HK detector and a running time of 20 years for NO (upper) and IO (lower figure).  The events for no decay are also shown for comparison.}
 \label{fig:ev2v3}
\end{figure}

\noindent
Let us first look at the differences  in the predictions of the DSNB rates from a $2 \nu$ instead of a $3 \nu$ framework. Figure~\ref{fig:ev2v3} shows a comparison of the expected number of events
with $2 \nu$ (dotted) and $3 \nu$ (full lines), as a function of positron energy, for the three lifetime-over-mass ratios and for the case of no decay. The results shown are for HK  as an example. (Similar trends are found for the other experiments.) 

For NO, one can see that while the events are underestimated in the QD case, they are overestimated in the SH one when considering a $2 \nu$ framework. The latter is on par with the different trends in the fluxes already observed in Figure~\ref{fig:flux2fvs3f}. On the contrary, the event predictions with 2$\nu$ and 3$\nu$ are strikingly different for IO, in particular for the shortest $\tau/m$. For $({\tau/m})_{medium}$, in the $2\nu$ case, the events below 16 MeV (positron energy) overestimate the ones with $3\nu$ decay by more than a factor of 2. 

More quantitatively, if one uses the effective $2 \nu$, instead of $3 \nu$ decay, for $({\tau/m})_{long}$ the total number of events differs by a few up to about $10 \%$ (for both mass orderings) in the four experiments. For $({\tau/m})_{medium}$, variations range from a few percent (SK-Gd, HK, JUNO) to 13 $\%$ for NO (SH or QD), or almost 30 $\%$ (SK-Gd, DUNE) and 50 $\%$ (HK) for IO. 

Moreover, as expected from the DSNB flux results shown in Figure~\ref{fig:flux2fvs3f}, the largest differences appear for $(\tau/m)_{short}$. In particular, these vary from 15-20 $\%$ in NO (SH or QD) to a factor of 4.5 (DUNE), 6 (JUNO), 7 (SK-Gd) and 8 (HK). Clearly, if nature has opted for the inverted mass ordering, one should employ a $3\nu$ treatment to learn about neutrino decay.   

\begin{table}[]
\begin{tabular}{|l|cccc|}
\hline
\multicolumn{1}{|c|}{\multirow{2}{*}{Experiment}} &
  \multicolumn{4}{c|}{Number of events} \\ \cline{2-5} 
\multicolumn{1}{|c|}{}   &
  \multicolumn{1}{l|}{No decay} &
  \multicolumn{1}{l|}{$(\tau/m)_{\it long}$} &
  \multicolumn{1}{l|}{$(\tau/m)_{\it medium}$} &
  \multicolumn{1}{l|}{$(\tau/m)_{\it short}$} \\ 
  \hline
\multicolumn{1}{|c|}{\begin{tabular}[c]{@{}c@{}}SK-Gd\\ (12.8 - 30.8) \\ 2 + 8 years \end{tabular}}  &
  \multicolumn{1}{c|}{\begin{tabular}[c]{@{}c@{}}2 \\ (2)\end{tabular}} &
  \multicolumn{1}{c|}{\begin{tabular}[c]{@{}c@{}}2\\ {[}2{]}\\ (2)\end{tabular}} &
  \multicolumn{1}{c|}{\begin{tabular}[c]{@{}c@{}}3\\ {[}2{]}\\ (1)\end{tabular}} &
  \begin{tabular}[c]{@{}c@{}}4\\ {[}2{]}\\ (0)\end{tabular} \\ \cline{3-5} 
   &
  \multicolumn{1}{c|}{\begin{tabular}[c]{@{}c@{}}12 \\ (10)\end{tabular}} &
  \multicolumn{1}{c|}{\begin{tabular}[c]{@{}c@{}}13\\ {[}12{]}\\ (10)\end{tabular}} &
  \multicolumn{1}{c|}{\begin{tabular}[c]{@{}c@{}}17\\ {[}12{]}\\ (7)\end{tabular}} &
  \begin{tabular}[c]{@{}c@{}}22\\ {[}11{]}\\ (1)\end{tabular} \\ \hline
\multicolumn{1}{|c|}{\begin{tabular}[c]{@{}c@{}}HK\\ (17.3 - 31.3) \\ 20 years\end{tabular}} &
  \multicolumn{1}{c|}{\begin{tabular}[c]{@{}c@{}} 48 \\ (40)\end{tabular}} &
  \multicolumn{1}{c|}{\begin{tabular}[c]{@{}c@{}} 49 \\ {[}47{]}\\ (39)\end{tabular}} &
  \multicolumn{1}{c|}{\begin{tabular}[c]{@{}c@{}} 61\\ {[}45{]}\\ (29)\end{tabular}} &
    \begin{tabular}[c]{@{}c@{}}84\\ {[}43{]}\\ (6)\end{tabular} \\ \hline
\multicolumn{1}{|c|}{\begin{tabular}[c]{@{}c@{}}HK-Gd\\ (17.3 - 31.3) \\ 20 years\end{tabular}} &
  \multicolumn{1}{c|}{\begin{tabular}[c]{@{}c@{}} 76 \\ (64)\end{tabular}} &
  \multicolumn{1}{c|}{\begin{tabular}[c]{@{}c@{}} 79\\ {[}73{]}\\ (62)\end{tabular}} &
  \multicolumn{1}{c|}{\begin{tabular}[c]{@{}c@{}} 98 \\ {[}73{]}\\ (46)\end{tabular}} &
  \begin{tabular}[c]{@{}c@{}} 135 \\ {[}69{]}\\ (9)\end{tabular} \\ \hline
\multicolumn{1}{|c|}{\begin{tabular}[c]{@{}c@{}}JUNO\\ (11.3 - 33.3) \\ 20 years\end{tabular}} &
  \multicolumn{1}{c|}{\begin{tabular}[c]{@{}c@{}}20 \\ (17)\end{tabular}} &
  \multicolumn{1}{c|}{\begin{tabular}[c]{@{}c@{}}21\\ {[}20{]}\\ (16)\end{tabular}} &
  \multicolumn{1}{c|}{\begin{tabular}[c]{@{}c@{}}28\\ {[}19{]}\\ (10)\end{tabular}} &
  \begin{tabular}[c]{@{}c@{}}37\\ {[}19{]}\\ (2)\end{tabular} \\ \hline
\multicolumn{1}{|c|}{\begin{tabular}[c]{@{}c@{}}DUNE\\ (19 - 31) \\ 20 years\end{tabular}} &
  \multicolumn{1}{c|}{\begin{tabular}[c]{@{}c@{}}12 \\ (11)\end{tabular}} &
  \multicolumn{1}{c|}{\begin{tabular}[c]{@{}c@{}}12\\ {[}11{]}\\ (10)\end{tabular}} &
  \multicolumn{1}{c|}{\begin{tabular}[c]{@{}c@{}}15\\ {[}11{]}\\ (8)\end{tabular}} &
  \begin{tabular}[c]{@{}c@{}}20\\ {[}10{]}\\ (2)\end{tabular} \\ \hline
\end{tabular}
\caption{Number of events associated with inverse-beta decay in SK-Gd, HK and JUNO as well as with $\nu_e$-$^{40}$Ar scattering in DUNE. The DSNB detection windows (in MeV) and running time are also shown under each experiment label. The events with decay  are given in the third to fifth columns.
For each experiment and $\tau/m$ value, the upper values are for NO, QD; the middle ones for NO, SH (in brackets) and the lower ones for IO (in parenthesis). The results for no decay (second column) for NO (IO in parenthesis) are given again for comparison. The predicted events correspond to the {\it Fiducial} model.}
\label{tab:events_decay}
\end{table}

\subsubsection{Expected DSNB rates in the $3 \nu$ formalism with decay} 
\begin{figure}[!thb] 
    \centering
                 \includegraphics[scale=0.3]{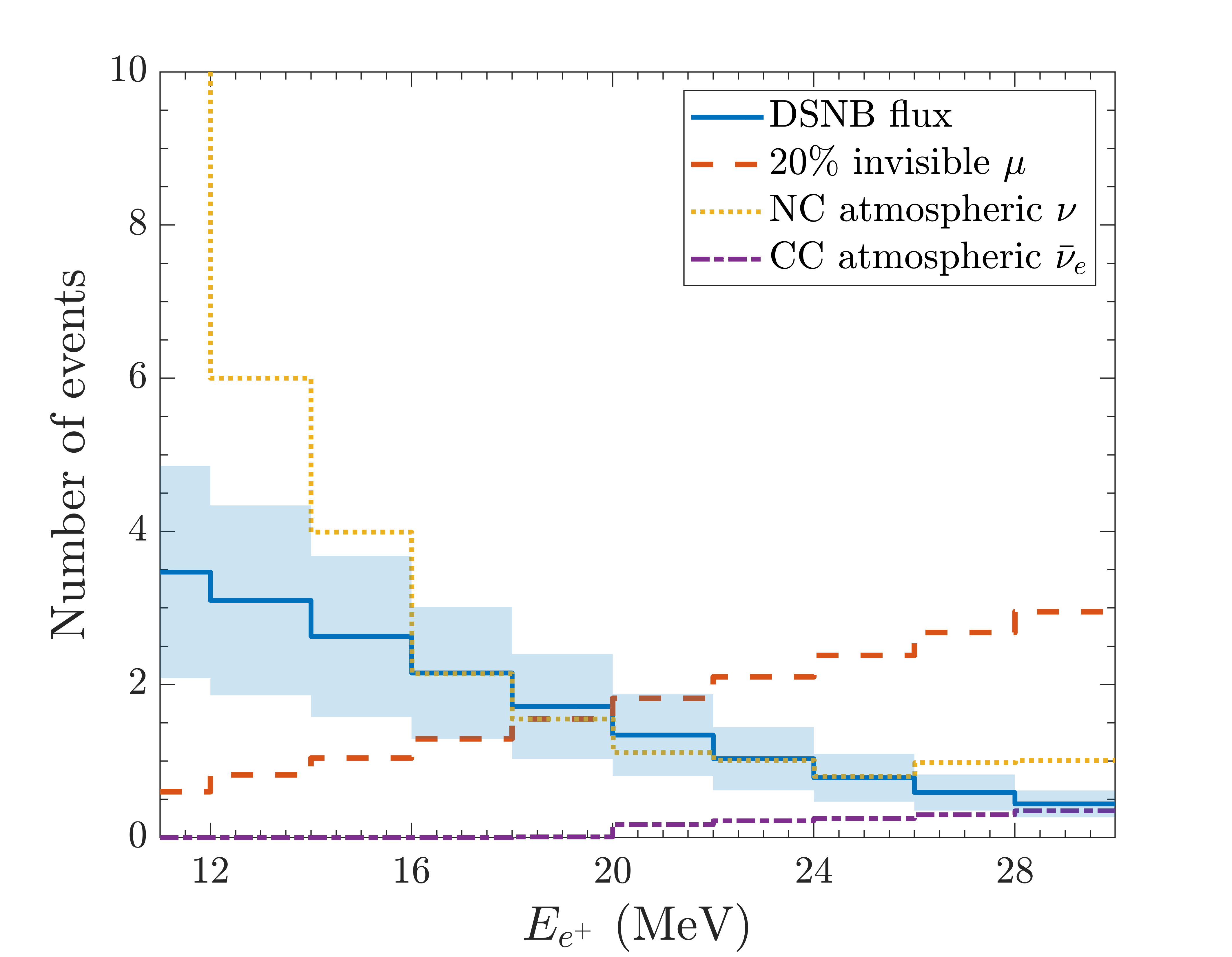}    
               \caption{No decay case: Expected DSNB events, as a function of positron energy, for the {\it Fiducial} model in the SK-Gd experiment and a running time of 10 years. 
               The band corresponds to the current uncertainty on $R_{SN}$. Backgrounds from invisible muons, NC and CC atmospheric neutrinos are shown
               (from \cite{KH}). Spallation due to cosmogenic backgrounds (producing for example $^{9}$Li) and accidentals  \cite{Super-Kamiokande:2021jaq} are not shown.}  
 \label{fig:NoDecay_SK-Gd}
\end{figure}

\begin{figure}[!thb] 
    \centering
                 \includegraphics[scale=0.3]{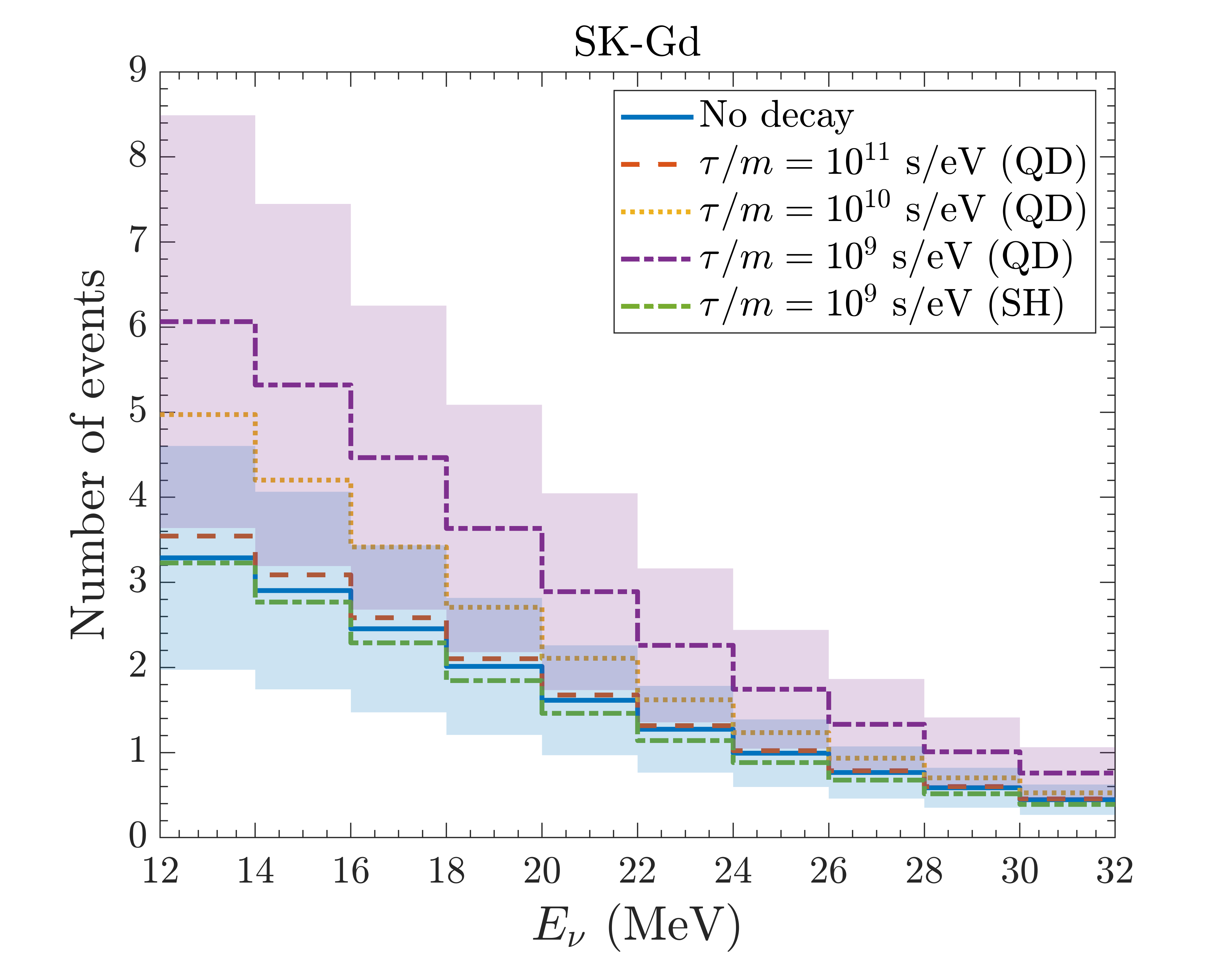}    
         \includegraphics[scale=0.3]{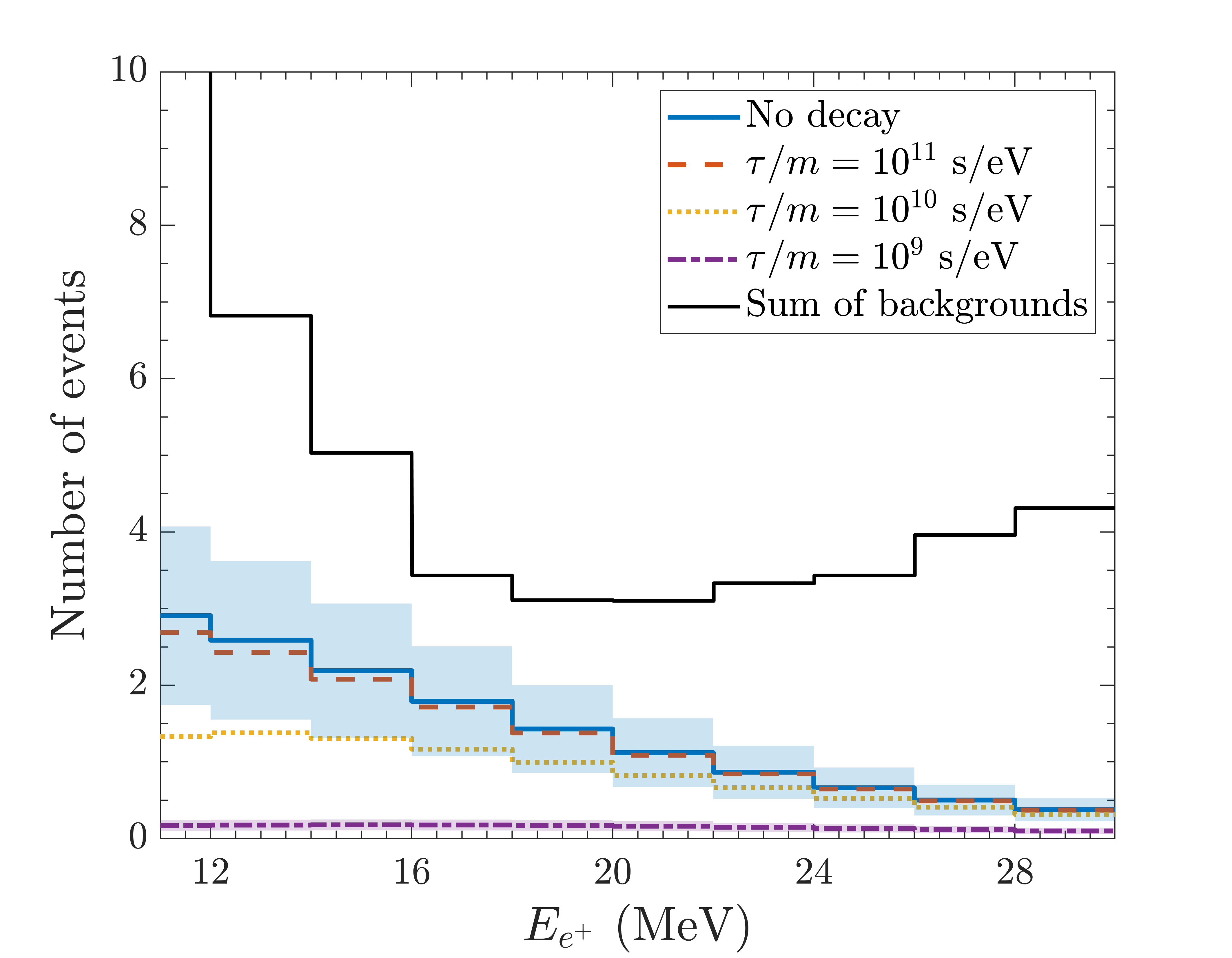}   
               \caption{Expected DSNB $\bar{\nu}_e$ events associated with inverse beta-decay, as a function of positron energy in SK-Gd for a running time of 10 years. The cases are NO (upper figure) and IO (lower figure). The results correspond to the {\it Fiducial} model with the shortest $\tau/m$ (dot-dashed line), the intermediate $\tau/m$ (dotted) and the long $\tau/m$ (dashed) (decay patterns in Figure~\ref{fig:decaypatterns}). For no decay and for the case of decay with $\tau/m = 10^9$ s/eV, the bands come from the uncertainty in the core-collapse supernova rate. The black line corresponds to the summed backgrounds shown in Figure~\ref{fig:NoDecay_SK-Gd}.} 
 \label{fig:Decay_SK-Gd}
\end{figure}

\begin{figure}[!thb] 
    \centering
                  \includegraphics[scale=0.3]{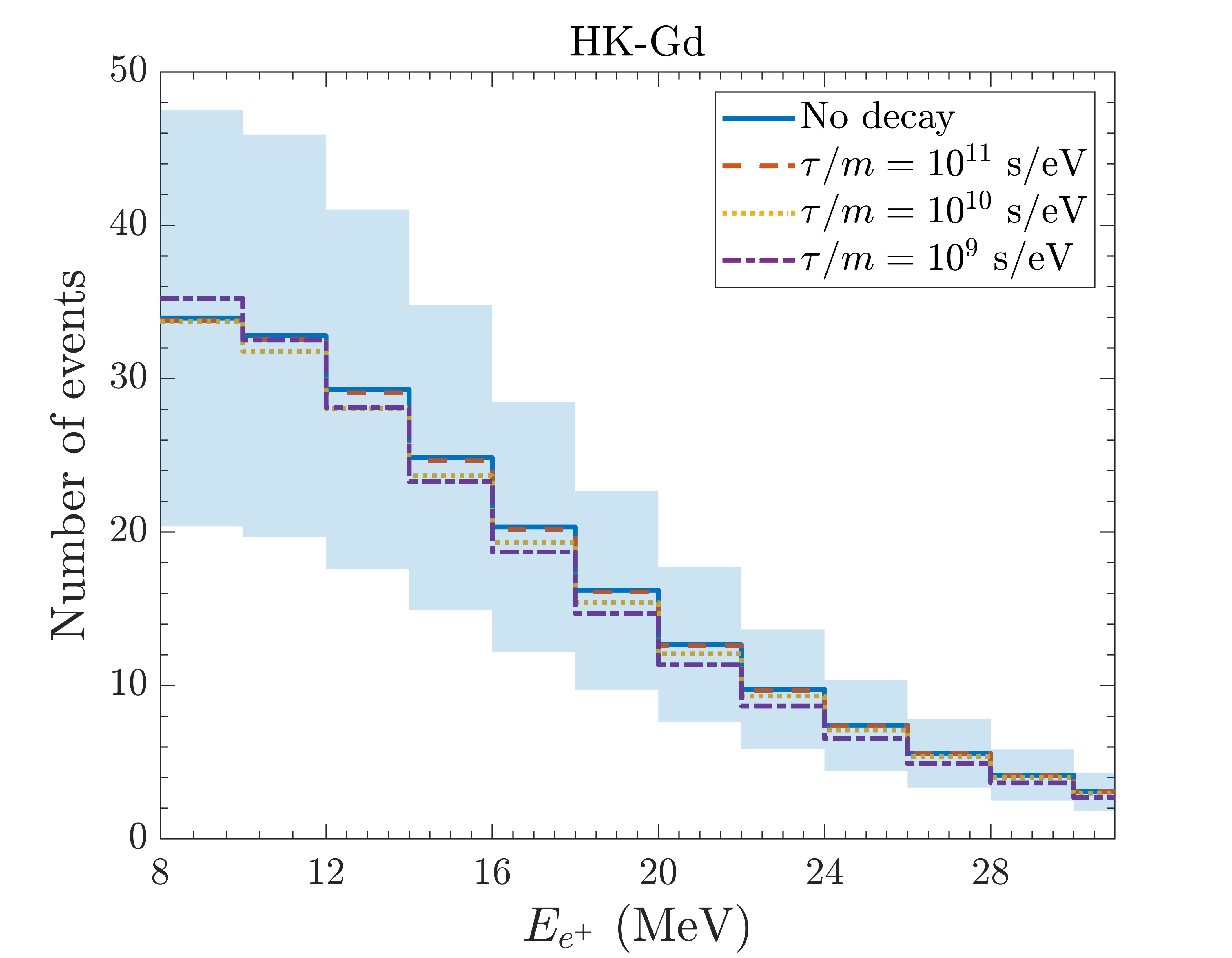}
         \includegraphics[scale=0.3]{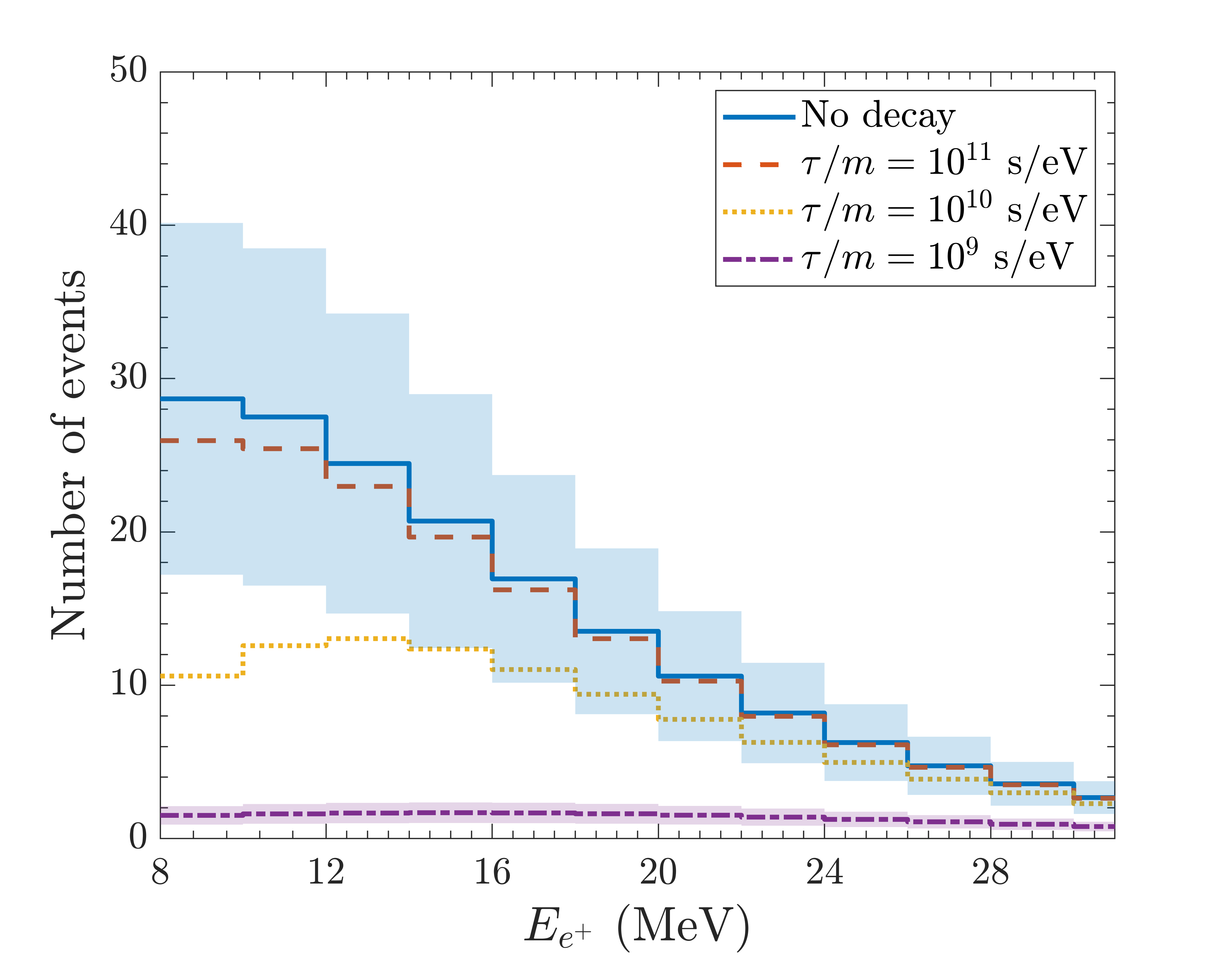}    
               \caption{Expected DSNB $\bar{\nu}_e$ events associated with inverse beta-decay, as a function of positron energy in HK-Gd for a running time of 20 years (see Table~\ref{tab:exppari}). The cases are NO and SH (upper figure) an IO (lower figure). The results correspond to the {\it Fiducial} model with the shortest $\tau/m$ (dot-dashed line), the intermediate $\tau/m$ (dotted) and the long $\tau/m$ (dashed). For the cases of no decay case and of $\tau/m = 10^9$ s/eV, the bands come from the uncertainty in the core-collapse supernova rate (see text). Backgrounds are not shown here.}
 \label{fig:Decay_HK-Gd}
\end{figure}

\begin{figure}[!thb] 
    \centering
                 \includegraphics[scale=0.3]{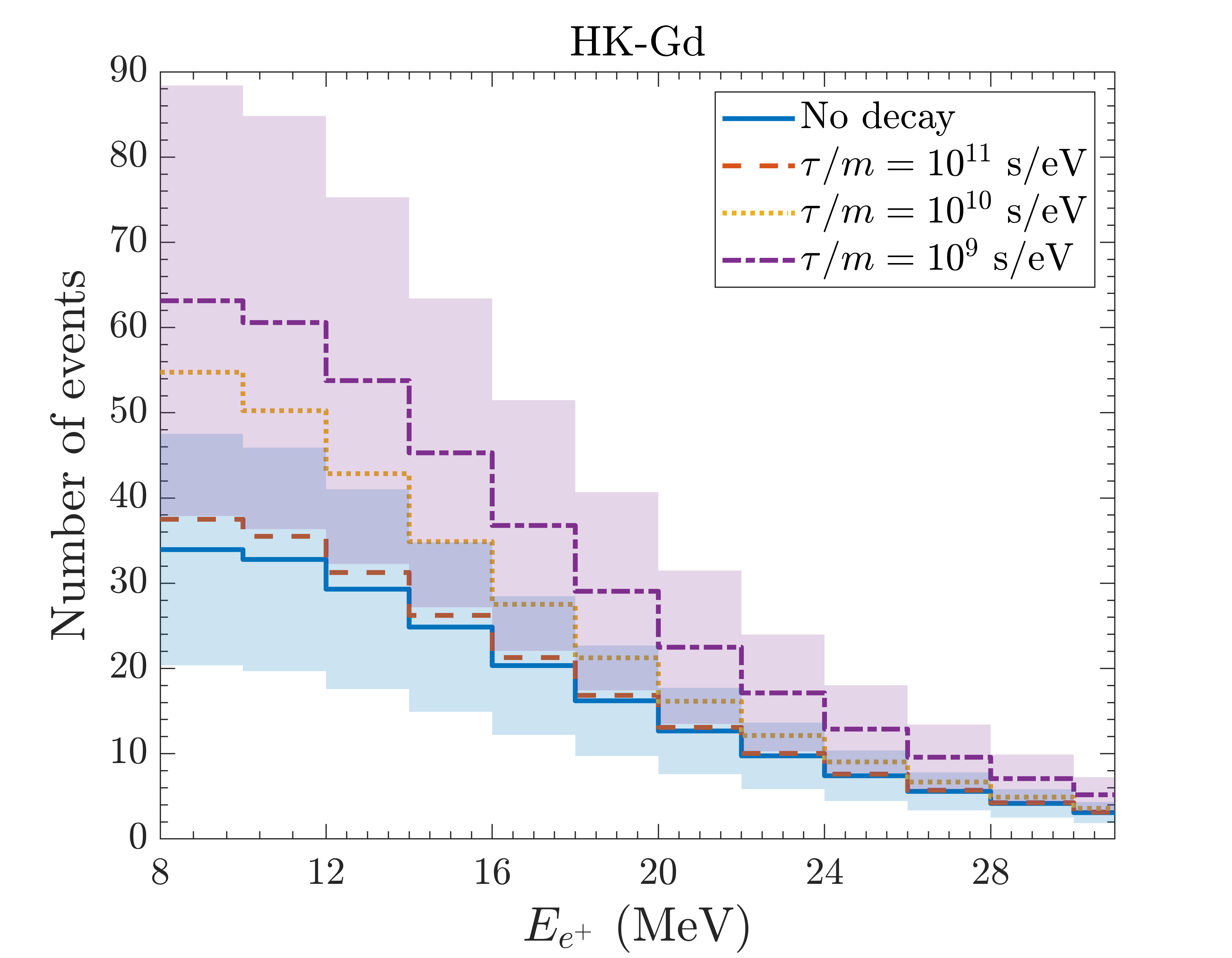}    
               \caption{Expected DSNB $\bar{\nu}_e$ events associated with inverse beta-decay, as a function of positron energy in HK-Gd for a running time of 20 years (see Table~\ref{tab:exppari}). The case is NO and QD decay pattern. The results correspond to the {\it Fiducial} model with the shortest $\tau/m$ (dot-dashed line), the intermediate $\tau/m$ (dotted) and the long $\tau/m$ (dashed). For the cases of no decay case and of $\tau/m = 10^9$ s/eV, the bands come from the uncertainty in the core-collapse supernova rate (see text). Backgrounds are not shown here.}  
 \label{fig:Decay_HK-Gd_QD}
\end{figure}

\begin{figure}[!thb] 
    \centering
                 \includegraphics[scale=0.3]{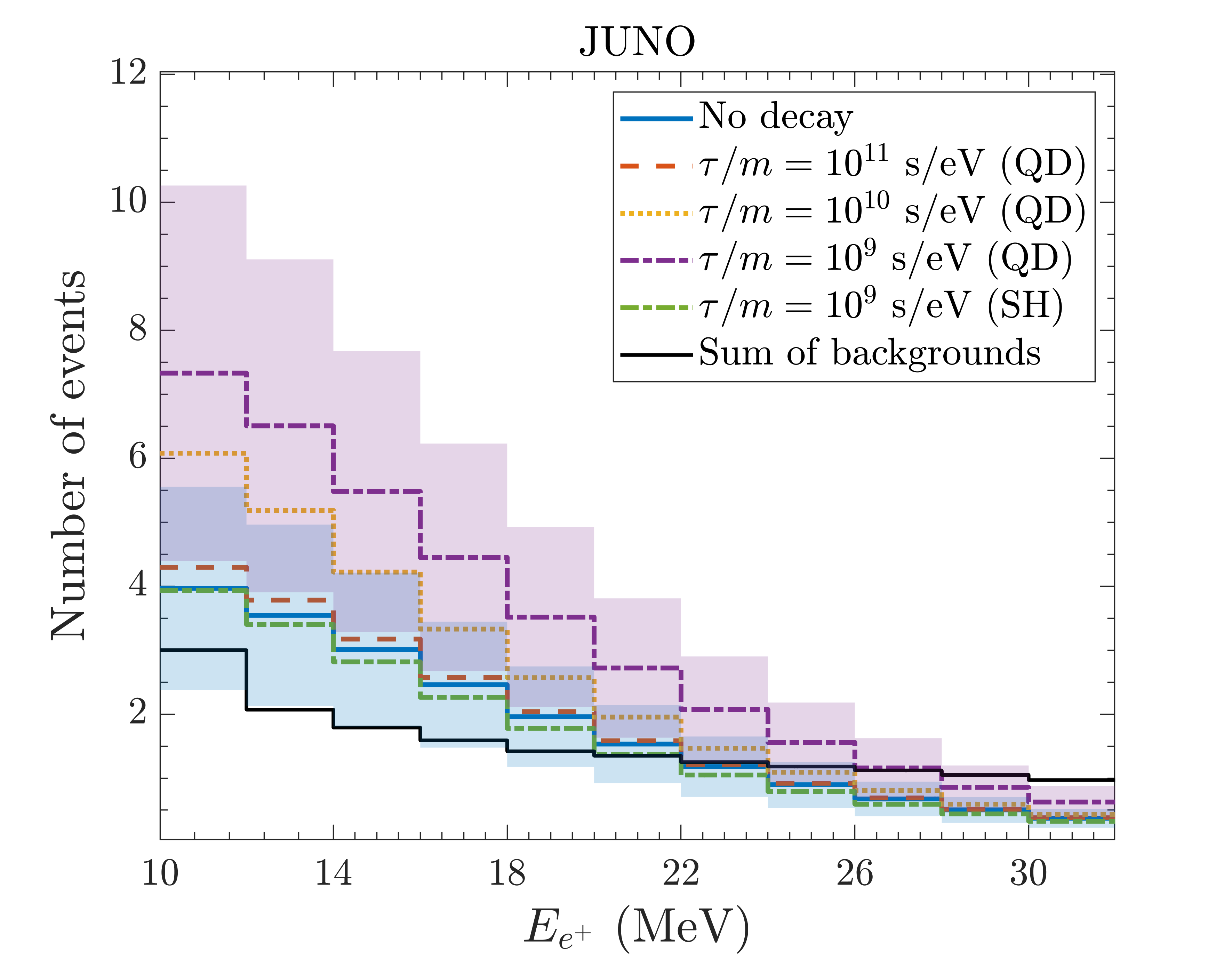}
         \includegraphics[scale=0.3]{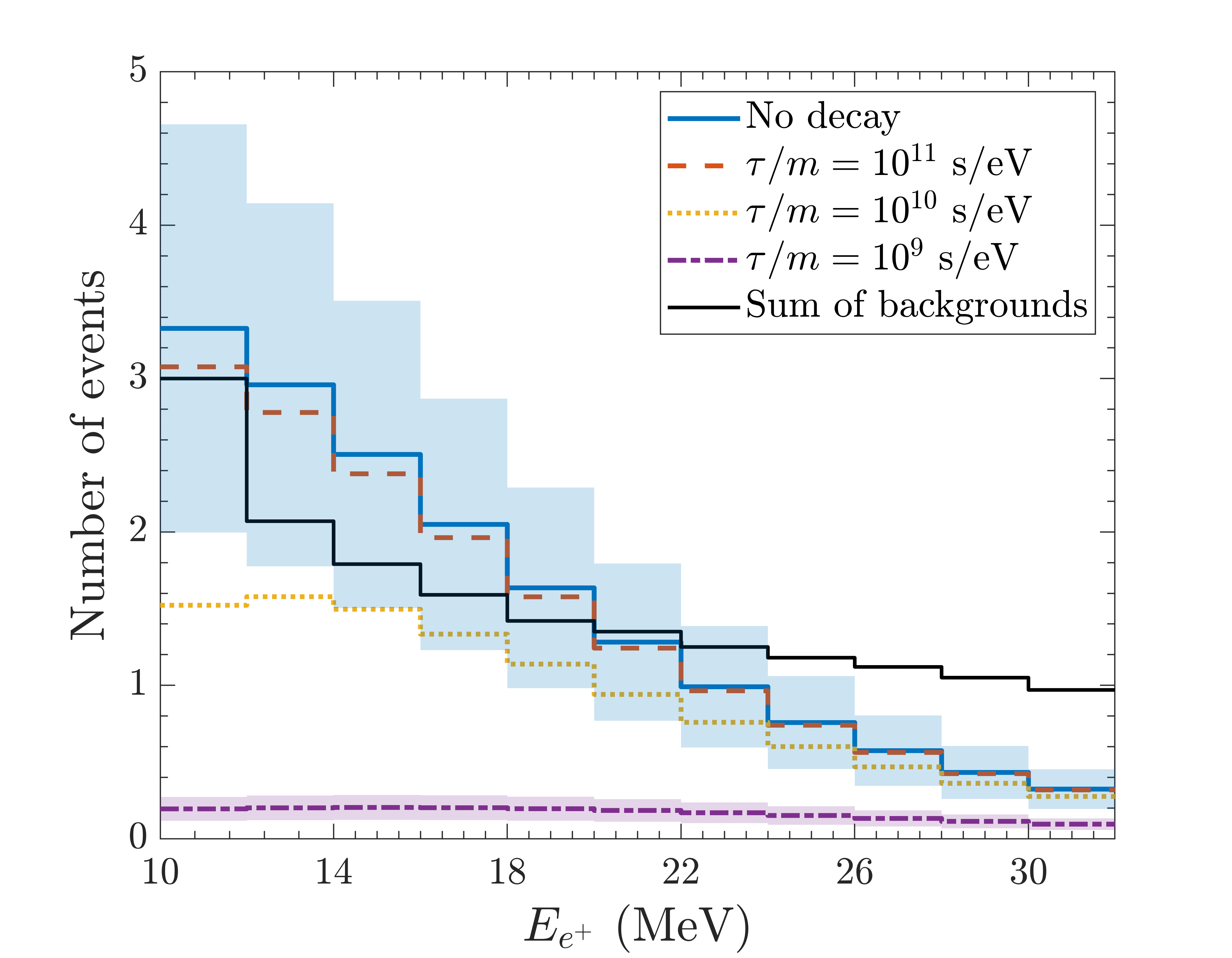}    
               \caption{Expected DSNB $\bar{\nu}_e$ events associated with inverse beta-decay, as a function of positron energy in JUNO for a running time of 20 years. The cases are NO (upper figure) and IO (lower figure). The results correspond to the {\it Fiducial} model with the shortest $\tau/m$ (dot-dashed line), the intermediate $\tau/m$ (dotted) and the long $\tau/m$ (dashed) (decay patterns in Figure~\ref{fig:decaypatterns}). For no decay and for the case of decay with $\tau/m = 10^9$ s/eV, the bands come from the uncertainty in the core-collapse supernova rate.}
 \label{fig:Decay_JUNO}
\end{figure}

\begin{figure}[!thb] 
    \centering
         \includegraphics[scale=0.3]{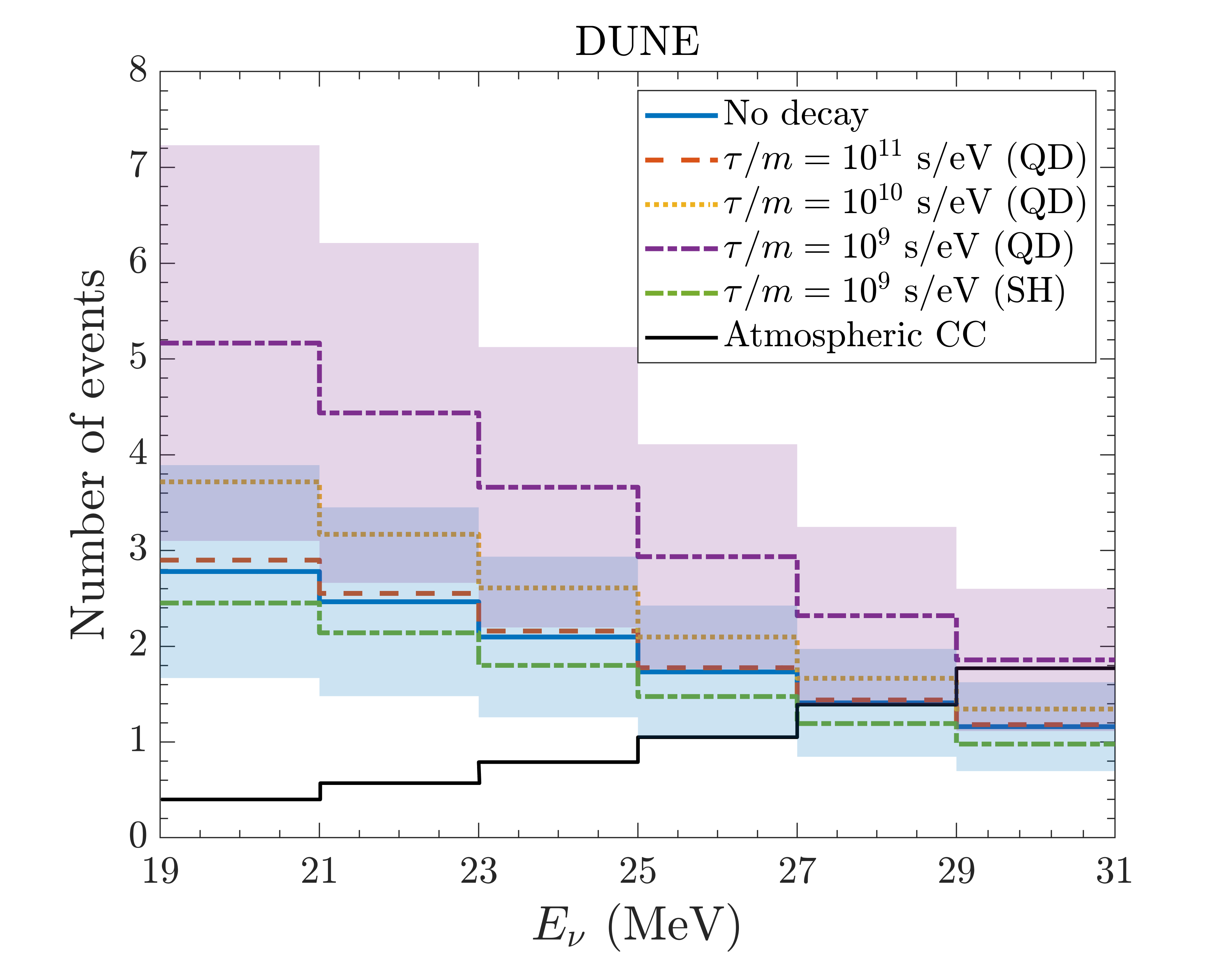}
         \includegraphics[scale=0.3]{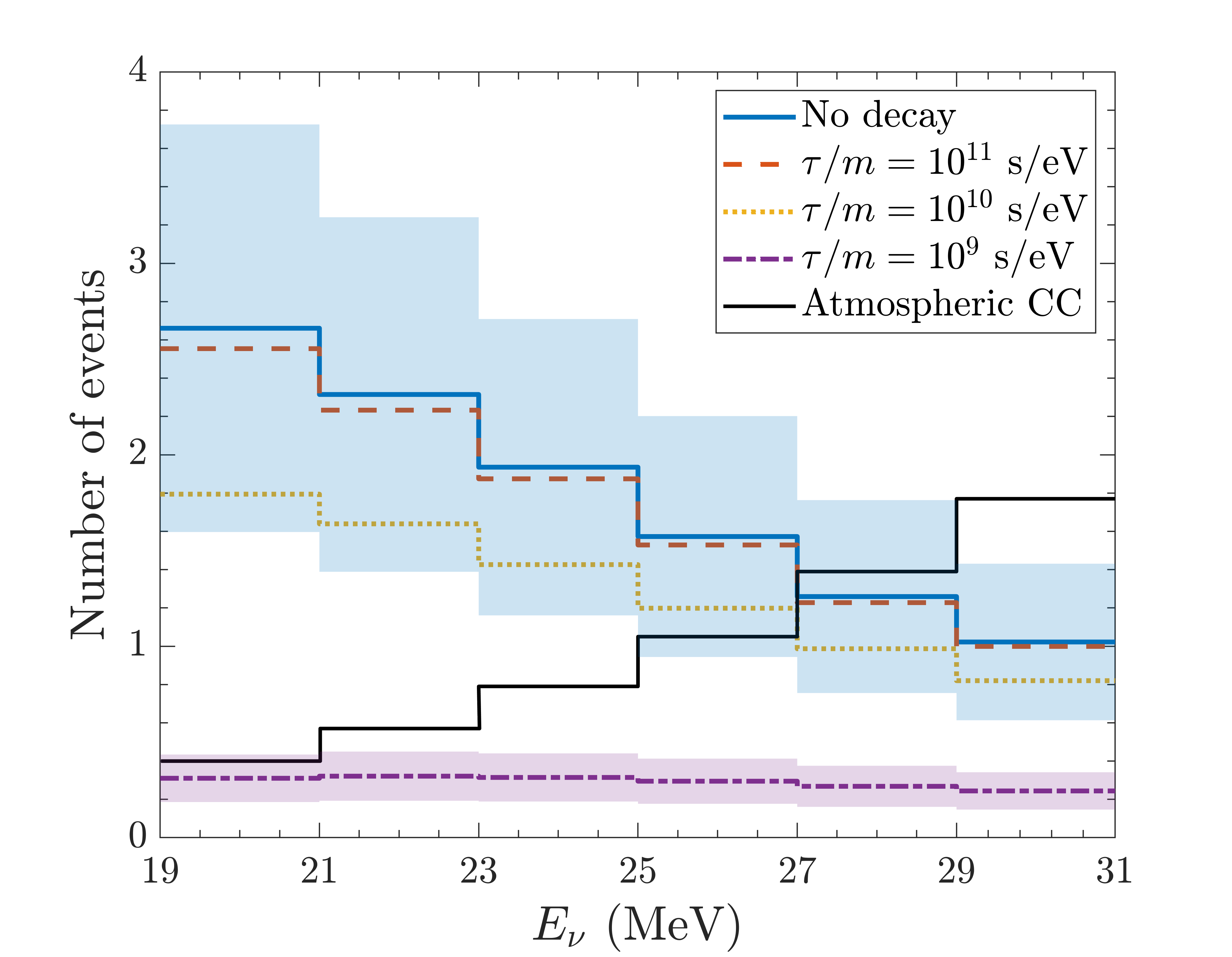}      
               \caption{Number of events associated with $\nu_e$ scattering on $^{40}$Ar, as a function of neutrino energy, for the DUNE detector. The running time is 20 years. The cases are NO (upper figure) and IO (lower figure).}
 \label{fig:Decay_DUNE}
\end{figure}

\noindent
Let us now look at our predictions on the DSNB differential number of $\bar{\nu}_e$  IBD events, as a function of energy. The results for SK-Gd are given in Figures~\ref{fig:NoDecay_SK-Gd} and~\ref{fig:Decay_SK-Gd}, for HK in Figures~\ref{fig:Decay_HK-Gd} and \ref{fig:Decay_HK-Gd_QD}) and for JUNO in Figure~\ref{fig:Decay_JUNO}. The expected events for $\nu_e$+$^{40}$Ar events in DUNE are shown in Figure~\ref{fig:Decay_DUNE}. 
Note that we do not show the very low energy range where reactor $\bar{\nu}_e$ and solar $\nu_e$ backgrounds dominate over the DSNB signal.
Obviously, the comparisons of the events for the different cases considered (mass ordering, mass patterns and values of $\tau/m$) show very similar trends in the four detectors. 

Figure~\ref{fig:NoDecay_SK-Gd} shows the predictions for SK-Gd for one-decade running time, for NO, with backgrounds from invisible muons and charged- and neutral-current
interactions induced by atmospheric neutrinos (taken from \cite{KH}). Note that the discovery of the DSNB by the SK-Gd experiment is challenging as pointed out by Ref.~\cite{Priya:2017bmm} due to neutral-current interactions which could hide the DSNB detection window\footnote{Note that, for the event calculations, we take the detection windows quoted by the Collaborations whenever possible. These can be at variance with the DSNB detection windows visible in our figures.}. 

The upper panels of Figures~\ref{fig:Decay_SK-Gd}, \ref{fig:Decay_HK-Gd}, \ref{fig:Decay_JUNO} and \ref{fig:Decay_DUNE}, and Figure~\ref{fig:Decay_HK-Gd_QD} present the expected events for the case of NO.
Following the flux behaviors for NO and the QD mass pattern,  visible in Figure~\ref{fig:3fNOQD}, the number of events is larger for shorter lifetimes. The fastest decay, $(\tau/m)_{short}$, gives the largest number of events. The events decrease for $(\tau/m)_{long}$ and for the case of no decay.
The events for NO and SH with $(\tau/m)_{short}$ are almost the same as in the case of no decay.
When including the uncertainty on the core-collapse supernova rate (bands), the two most different cases, $(\tau/m)_{short}$ and no decay, cannot be distinguished anymore.

We take HK as a typical example (upper Figure~\ref{fig:Decay_HK-Gd})
to show the comparison between the predicted events with no decay and with decay for SH, NO.
For the three $\tau/m$ values, the differential number of events is practically degenerate with the results in absence of decay. This is in concordance with the findings of \cite{Tabrizi:2020vmo}, but with the quantitative differences mentioned above;
that is, we find that the use of $2 \nu$ instead of the $3 \nu$ scenario gives a higher (and not lower) number of events by about a few tens of percent.

The IO case is presented in the lower panels of Figures~\ref{fig:Decay_SK-Gd}, \ref{fig:Decay_HK-Gd}, \ref{fig:Decay_JUNO} and \ref{fig:Decay_DUNE}). 
Note that the results are below the current backgrounds for three experiments. With the background shown, DUNE could have a sensitivity at the lower end of the DSNB detection window. For IO, the event trend is opposite to the one found for NO when going from $(\tau/m)_{long}$ (close to no decay) to
$(\tau/m)_{short}$. With $(\tau/m)_{medium}$ the results overlap significantly with no decay, if one includes the current knowledge on 
$R_{SN}$. On the contrary, the events could be clearly distinguishable if neutrinos decay with $\tau/m = 10^9$ s/eV. 

Table~\ref{tab:events_decay} presents the total number of events for the four experiments, with/without neutrino decay. One can see that
for NO when $\tau/m$ is short (long) and the mass pattern is SH (QD), the results are practically degenerate with the no decay case. For
$(\tau/m)_{short}$ and a QD decay pattern the number of events is always larger than for the no decay case. 

The largest differences in the number of events appear for IO for which
the values for the shortest $(\tau/m)$ are a factor of 6 (DUNE) to 10 (JUNO, SK-Gd, HK) smaller than in absence of decay. 
For $(\tau/m)_{medium}$, whatever is the mass ordering, the results are in between 
the ones for $(\tau/m)_{short}$ and $(\tau/m)_{long}$.


\section{Conclusions}
\noindent
In this work, we have investigated the impact of neutrino non-radiative decay on the DSNB. This is the first investigation where a $3\nu$ flavor framework is used with the main astrophysical uncertainties explicitly implemented. These comprise the evolving core-collapse supernova rate and the fraction of dark collapses. We have also implemented the progenitor dependence of the supernova neutrino spectra, using inputs from one-dimensional simulations by the Garching group. We have considered three scenarios for the black-hole fraction as well as the different possibilities due to the unknown neutrino mass ordering and mass patterns.

The results rely on the solution of the neutrino kinetic equations in presence of $\nu$ decay,
assuming the decaying eigenstates (considered equal to the mass eigenstates) have {\it democratic} branching ratios and the same
$\tau/m$. We have presented predictions for the DSNB (integrated) fluxes for $\nu_e$ and $\bar{\nu}_e$ in the presence/absence of $\nu$ decay
as well as the DSNB differential and the total number of events for the running SK-Gd and the upcoming JUNO, DUNE, and HK experiments.
Note that, for both the fluxes and the events, our results compare well with existing ones with no decay, with $2 \nu$ decay and $3 \nu$ decay within
the same approximations (e.g. using one Fermi-Dirac distribution for the neutrino spectra of a single supernova and/or no progenitor dependence, older core-collapse supernova rates). 

We have presented a detailed comparison of the results based on the $2\nu$ and $3\nu$ decay framework. If the neutrino mass ordering is normal and the $\nu$ mass pattern is strongly hierarchical or quasi-degenerate, the use of both frameworks gives similar predictions for the extreme $10^{11}$ s/eV. On the contrary, if $\tau/m = 10^{9}$ s/eV ($\tau/m = 10^{10}$) the expected number of events with $2 \nu$ decay is underestimated (overestimated) up to about 20$\%$ (30$\%$), depending on the experiments. The situation is strikingly different if the neutrino mass ordering is inverted, in which case the $3 \nu$ framework clearly gives lower predictions by large factors.

More generally, our $3 \nu$ results on the DSNB (integrated) fluxes and expected number of events for the four experiments show that,
for normal mass ordering and strongly hierarchical mass pattern, if $\tau/m = 10^{9}$ s/eV, the events will be essentially
degenerate with no decay. This is also the case for the quasi-degenerate mass pattern if $\tau/m = 10^{11}$ s/eV. In contrast, 
for normal mass ordering and quasi-degenerate mass pattern, if $\tau/m = 10^{9}$ s/eV, considering neutrinos as stable underestimates the events by almost a factor of 2, whereas $\tau/m = 10^{10}$ s/eV gives results intermediate between the two. 

Interestingly, if the neutrino mass ordering is inverted, the results on the events for the intermediate lifetime-to-mass ratio
are smaller by a factor of about 2 than in the case neutrinos are stable. For the short lifetime-to-mass ratio, the event predictions are much smaller and clearly distinguishable from no decay,
even considering astrophysical uncertainties.  

It is to be noted that current DSNB predictions can vary by similar factors due to standard physics, with rates
up to a factor of 5 smaller than the present SK-I to SK-IV sensitivity \cite{Super-Kamiokande:2021jaq}. 
In the unlucky case of non-observation we would not know if this is due to more conservative inputs based on standard physics, 
or to the fact that neutrinos undergo non-radiative two-body decay with $\tau/m = 10^{9}$ s/eV. 

One can envisage several improvements to the present study, such as a more detailed progenitor dependence of the supernova neutrino fluxes, or a specific decaying hypothesis from models. The present study provides another quantitative example of how much reducing uncertainties, such as the one from the evolving core-collapse supernova rate, is
crucial to extract the most from the DSNB observation.

The upcoming identification of the neutrino mass ordering constitutes a key step in restricting the possible scenarios for the impact of neutrino non-radiative two-body decay on the DSNB. If the mass ordering is normal, the possibility of a low DSNB rate due to neutrino invisible decay with a short  lifetime-over-mass ratio will be excluded, thus avoiding a potential degeneracy with standard physical inputs. However, we shall still need to disentangle the no decay from the decay case.

Finally our results show the necessity of using a $3 \nu$ framework for DSNB predictions with neutrino non-radiative two-body decay instead of an effective 2$\nu$ framework. Indeed in most of the scenarios considered for the mass ordering and mass patterns, we find significant differences between the two. For the others, where the variations are small, the trends obtained with the two frameworks, with respect to no decay, are opposite, making again the difference between the two sizable.

In conclusion, the discovery of the DSNB will bring crucial information for astrophysics and particle physics and will have a unique sensitivity to new physics, such as neutrino decay.

\begin{acknowledgments}
\noindent
The authors wish to thank  C. Lunardini, T. M\"uller, B. Quilain, K. Scholberg, A. Sieverding and Y. Wong for providing useful information.  
\end{acknowledgments}

\appendix{}
\section{Supernova neutrino flux parameters and black-hole fractions }
\noindent
We give here detailed information on the supernova neutrino fluxes used in our calculations.
Table~\ref{tab:fluxpar} gives their mass and type as well as the neutrino luminosities, average energies, and pinching. 

\begin{table}[tp]
\centering
\begin{tabular}{lclccclccclccc}
\hline
  &  & & & & & & & & & \\
Run Type & $\langle E_{\nu_e}\rangle $ &  $\langle E _{\bar{\nu}_e} \rangle $ &   $\langle E_{\nu_x} \rangle$ & $\alpha_{\nu_e}$ & 
$\alpha_{\bar{\nu}_e}$ & $ \alpha_{\nu_x}$ &  $L_{\nu_e}$ & $ L_{\bar{\nu}_e}$  & $ L_{\nu_x}$ & \\
  &  & & & & & & & & & \\
\hline
  &  & & & & & & & & & \\
s11.2c NS  & 10.43 & 12.89 & 12.93 & 2.99  & 2.61 & 2.30 & 3.56 & 3.09 & 3.02 \\
25.0c  NS & 12.67 & 15.5 & 15.41 & 2.61 & 2.61 & 2.30 & 7.18 & 6.78 & 6.02 \\
25.0c    BH & 15.32 & 18.2  &  17.62 & 3.21 & 3.21 & 2.16 & 7.08 & 6.51 & 3.7  \\
27 ~~~   NS   & 11.3 & 13.89 & 13.85  & 2.79 & 2.45 & 2.16 & 5.87 &5.43  &  5.1\\
40.0c    BH  & 15.72 & 18.72 & 17.63 & 2.79 & 2.79 & 1.92 & 9.38 & 8.6 & 4.8\\
  & & & & & & & & & & \\
\hline
\end{tabular}
\caption{The first two columns give the run and type of the 11.2 M$_{\odot}$, 25 M$_{\odot}$, 27 M$_{\odot}$ and 40 M$_{\odot}$ progenitors used in the DSNB predictions. The other columns provide the corresponding average energies (MeV), pinching parameters and total gravitational energy emitted by the supernova ($10^{52}$~erg). These parameters define the supernova neutrino fluences, Eq.\eqref{eq:PW}, of the different neutrino flavors. The values are obtained from one-dimensional supernova simulations of the Garching group \cite{Priya:2017bmm,Hudepohl:2013zsj}. }
\label{tab:fluxpar}
\end{table}

Moreover, for each scenario, we explain
the progenitor used as template and the corresponding progenitor mass intervals, which corresponds to the information
shown in Figure \ref{fig:fBH}.
Here are the three scenarios for the black-hole fraction considered in our work:

{\it Scenario} I:
This is the most conservative case that we take for comparison with the previous literature (see for example \cite{Priya:2017bmm,Mathews:2014qba}. 
In this case, we use 4 templates for the supernova progenitors, as \cite{Priya:2017bmm} does. These comprise a 11.2 ${\rm M}_{\odot}$ NS progenitor in the [8, 15] ${\rm M}_{\odot}$ interval, a 25 ${\rm M}_{\odot}$ and 27 ${\rm M}_{\odot}$ NS progenitors for the mass ranges [15,26) ${\rm M}_{\odot}$ and [26,40] ${\rm M}_{\odot}$ respectively, and a 40 ${\rm M}_{\odot}$ BH progenitor for ${\rm M}_{\odot} \ge 40$ ${\rm M}_{\odot}$. 

{\it Scenario} II: 
Detailed supernova simulations, such as the ones of \cite{Kresse:2020nto}, give the black hole fraction of 0.17-0.18, as conservative. In our calculations we employ $f_{BH}=0.21$ as typical value for this case. For the progenitors, we take\footnote{Note that the shortage of optical supernovae in the [17,25] ${\rm M}_{\odot}$ window could be related to the {\it red supergiant problem} (see \cite{Horiuchi:2008jz}).} the 11.2 ${\rm M}_{\odot}$ NS progenitor in the interval [8, 15] ${\rm M}_{\odot}$, the 25 ${\rm M}_{\odot}$ NS for [15, 22) ${\rm M}_{\odot}$, the 25 ${\rm M}_{\odot}$ BH progenitor in [22, 25] ${\rm M}_{\odot}$, the 27 ${\rm M}_{\odot}$ NS in the (25,27) ${\rm M}_{\odot}$ interval, and the 40 ${\rm M}_{\odot}$ BH progenitor above 27 ${\rm M}_{\odot}$.

{\it Scenario} III:
This is the most optimistic case, in agreement with simulations  \cite{Kresse:2020nto}. We implement the 11.2 ${\rm M}_{\odot}$ NS progenitor for the [8,15] ${\rm M}_{\odot}$ interval and the 40 ${\rm M}_{\odot}$ BH progenitor for ${\rm M}_{\odot} \ge 15$ ${\rm M}_{\odot}$.

\section{DSNB fluxes in the presence of neutrino non-radiative two-body decay}
\noindent
We present here the explicit equations used in the 3$\nu$ flavor calculations, for the different cases.

{\it NO and QD: }
considering Eq.~\ref{eq:kinsol} and Eq.~\ref{eq:psiQD}, we can explicitly write the DSNB flux on Earth (z~=~0)
\begin{equation}\label{eq:decayQD_n3}
n_{\nu_3} (E_{\nu}) = \int_0^{\infty} {dz \over{H(z)} } R_{SN}(z) Y_{\nu_3} \bigl(E_{\nu} (1 + z) \bigl) e^{- \Gamma_{\nu_3}\chi(z)} \ , 
\end{equation}
\begin{equation}\label{eq:decayQD_n2}
\begin{aligned}
n_{\nu_2}\left(E_\nu\right) &=\int_0^{\infty} \frac{d z}{H(z)}\Big[R_{SN}(z) Y_{\nu_2}\bigl(E_\nu(1+z)\bigl) \\
& + n_{\nu_3}(E_\nu (1+z), z) \Gamma_{\nu_3 \rightarrow \nu_2}\Big]\times e^{-\Gamma_{\nu_2} \chi(z)},
\end{aligned}
\end{equation}
\begin{equation}\label{eq:decayQD_n1}
\begin{aligned}
n_{\nu_1}\left(E_\nu\right) &=\int_0^{\infty} \frac{d z}{H(z)}\Big[R_{SN}(z) Y_{\nu_1}\bigl(E_\nu(1+z)\bigl) \\ 
& + n_{\nu_3}(E_\nu (1+z), z) \Gamma_{\nu_3 \rightarrow \nu_1}  \\
& + n_{\nu_2}(E_\nu (1+z), z) \Gamma_{\nu_2 \rightarrow \nu_1}\Big].
\end{aligned}
\end{equation}
Analogous expressions can be found for the flux of antineutrinos ($\nu_i \leftrightarrow \bar\nu_i$).

{\it NO and SH:}
using Eq.~\ref{eq:psiSH}, one obtains the following equations for the DSNB flux on Earth (z = 0)
\begin{equation}\label{eq:decaySH_n3}
n_{\nu_3} (E_{\nu}) = \int_0^{\infty} {dz \over{H(z)} } R_{SN}(z) Y_{\nu_3} \bigl(E_{\nu} (1 + z) \bigl) e^{- \Gamma_{\nu_3}\chi(z)} \ , 
\end{equation}
\begin{equation}\label{eq:decaySH_n2}
\begin{aligned}
n_{\nu_2}\left(E_\nu\right) &=\int_0^{\infty} \frac{d z}{H(z)}\Bigl\{R_{SN}(z) Y_{\nu_2}\bigl(E_\nu(1+z)\bigl) \\
&+\int_{E_\nu (1+z)}^{\infty} dE_\nu^{\prime} \Big[ n_{\nu_3}(E_\nu^{\prime}, z) \Gamma_{\nu_3 \rightarrow \nu_2} \psi_{\text{h.c.}}(E_\nu^{\prime}, E_\nu (1+z))\\
& +~n_{\bar{\nu}_3}(E_\nu^{\prime}, z) \Gamma_{\bar{\nu}_3 \rightarrow \nu_2} \psi_{\text{h.f.}}(E_\nu^{\prime},E_\nu (1+z)) \Big]\Bigr\} \\
& \times e^{-\Gamma_{\nu_2} \chi(z)},
\end{aligned}
\end{equation}
\begin{equation}\label{eq:decaySH_n1}
\begin{aligned}
n_{\nu_1}\left(E_\nu\right) &=\int_0^{\infty} \frac{d z}{H(z)} \Bigl\{R_{SN}(z) Y_{\nu_1}\bigl(E_\nu(1+z)\bigl) \\
&+\int_{E_\nu (1+z)}^{\infty} dE_\nu^{\prime} \Big[ n_{\nu_3}(E_\nu^{\prime}, z) \Gamma_{\nu_3 \rightarrow \nu_1} \psi_{\text{h.c.}}(E_\nu^{\prime}, E_\nu (1+z))\\
& +~n_{\bar{\nu}_3}(E_\nu^{\prime}, z) \Gamma_{\bar{\nu}_3 \rightarrow \nu_1} \psi_{\text{h.f.}}(E_\nu^{\prime},E_\nu (1+z)) \\
& +~n_{\nu_2}(E_\nu^{\prime}, z) \Gamma_{\nu_2 \rightarrow \nu_1} \psi_{\text{h.c.}}(E_\nu^{\prime}, E_\nu (1+z))\\
& +~n_{\bar{\nu}_2}(E_\nu^{\prime}, z) \Gamma_{\bar{\nu}_2 \rightarrow \nu_1} \psi_{\text{h.f.}}(E_\nu^{\prime},E_\nu (1+z)) \Big] \Bigr\}.
\end{aligned}
\end{equation}
where h.c. and h.f. spectra are given in Eqs.\eqref{eq:psiSH} respectively.

{\it IO:}
the masses of $\nu_2$ and $\nu_1$ are quasi-degenerate, and the mass of $\nu_3$ is considered to be much smaller, i.e. $m_2 \simeq m_1 \gg m_3 \simeq 0$. 
The neutrino spectra are given, accordingly, from Eqs.\eqref{eq:psiQD}-\eqref{eq:psiSH}.Therefore, in this case, the DSNB flux for the mass eigenstates on Earth (z = 0) is given by the following expressions:
\begin{equation}\label{eq:decayIO_n2}
n_{\nu_2} (E_{\nu}) = \int_0^{\infty} {dz \over{H(z)} } R_{SN}(z) Y_{\nu_2} \bigl(E_{\nu} (1 + z) \bigl) e^{- \Gamma_{\nu_2}\chi(z)} \ , 
\end{equation}
\begin{equation}\label{eq:decayIO_n1}
\begin{aligned}
n_{\nu_1}\left(E_\nu\right) &=\int_0^{\infty} \frac{d z}{H(z)}\Big[R_{SN}(z) Y_{\nu_1}\bigl(E_\nu(1+z)\bigl) \\
&+ n_{\nu_2}(E_\nu (1+z), z)\Gamma_{\nu_2 \rightarrow \nu_1}\Big]\times e^{-\Gamma_{\nu_1} \chi(z)},
\end{aligned}
\end{equation}
\begin{equation}\label{eq:decayIO_n3}
\begin{aligned}
n_{\nu_3}\left(E_\nu\right) &=\int_0^{\infty} \frac{d z}{H(z)} \Bigl \{R_{SN}(z) Y_{\nu_3}\bigl(E_\nu(1+z)\bigl) \\
&+\int_{E_\nu (1+z)}^{\infty} dE_\nu^{\prime} \Big[ n_{\nu_2}(E_\nu^{\prime}, z) \Gamma_{\nu_2 \rightarrow \nu_3} \psi_{\text{h.c.}}(E_\nu^{\prime}, E_\nu (1+z))\\
& +~n_{\bar{\nu}_2}(E_\nu^{\prime}, z) \Gamma_{\bar{\nu}_2 \rightarrow \nu_3} \psi_{\text{h.f.}}(E_\nu^{\prime},E_\nu (1+z)) \\
& +~n_{\nu_1}(E_\nu^{\prime}, z) \Gamma_{\nu_1 \rightarrow \nu_3} \psi_{\text{h.c.}}(E_\nu^{\prime}, E_\nu (1+z))\\
& +~n_{\bar{\nu}_1}(E_\nu^{\prime}, z) \Gamma_{\bar{\nu}_1 \rightarrow \nu_3} \psi_{\text{h.f.}}(E_\nu^{\prime},E_\nu (1+z)) \Big] \Bigr\}.
\end{aligned}
\end{equation}

From these expressions, one can obtain the equations in the effective $2 \nu$ formalism by setting $\Gamma_{\nu_3 \rightarrow \nu_2} = 0 $ for NO QD (and also $\Gamma_{\bar{\nu}_3 \rightarrow \nu_2} = 0 $ for NO SH)  and
$ \Gamma_{\nu_2 \rightarrow \nu_1} = 0$ for IO.


\begin{thebibliography}{99}

\bibitem{Super-Kamiokande:1998kpq}
Y.~Fukuda \textit{et al.} [Super-Kamiokande],
Phys. Rev. Lett. \textbf{81} (1998), 1562-1567
[arXiv:hep-ex/9807003 [hep-ex]].

\bibitem{SNO:2001kpb}
Q.~R.~Ahmad \textit{et al.} [SNO],
Phys. Rev. Lett. \textbf{87} (2001), 071301
[arXiv:nucl-ex/0106015 [nucl-ex]].

\bibitem{KamLAND:2002uet}
K.~Eguchi \textit{et al.} [KamLAND],
Phys. Rev. Lett. \textbf{90} (2003), 021802
[arXiv:hep-ex/0212021 [hep-ex]].

\bibitem{Kamiokande-II:1987idp}
K.~Hirata \textit{et al.} [Kamiokande-II],
Phys. Rev. Lett. \textbf{58} (1987), 1490-1493.

\bibitem{Bionta:1987qt}
R.~M.~Bionta, G.~Blewitt, C.~B.~Bratton, D.~Casper, A.~Ciocio, R.~Claus, \textit{et al.}
Phys. Rev. Lett. \textbf{58} (1987), 1494.

\bibitem{Alekseev:1988gp}
E.~N.~Alekseev, L.~N.~Alekseeva, I.~V.~Krivosheina and V.~I.~Volchenko,
Phys. Lett. B \textbf{205} (1988), 209-214.

\bibitem{Ando:2004hc}
S.~Ando and K.~Sato,
New J. Phys. \textbf{6} (2004), 170
[arXiv:astro-ph/0410061 [astro-ph]].

\bibitem{Beacom:2010kk}
J.~F.~Beacom,
Ann. Rev. Nucl. Part. Sci. \textbf{60} (2010), 439-462
[arXiv:1004.3311 [astro-ph.HE]].

\bibitem{Lunardini:2010ab}
C.~Lunardini,
Astropart. Phys. \textbf{79} (2016), 49-77
[arXiv:1007.3252 [astro-ph.CO]].

\bibitem{Super-Kamiokande:2002hei}
M.~Malek \textit{et al.} [Super-Kamiokande],
Phys. Rev. Lett. \textbf{90} (2003), 061101
[arXiv:hep-ex/0209028 [hep-ex]].

\bibitem{Super-Kamiokande:2013ufi}
H.~Zhang \textit{et al.} [Super-Kamiokande],
Astropart. Phys. \textbf{60} (2015), 41-46
[arXiv:1311.3738 [hep-ex]].

\bibitem{Super-Kamiokande:2021jaq}
K.~Abe \textit{et al.} [Super-Kamiokande],
Phys. Rev. D \textbf{104} (2021) no.12, 122002
[arXiv:2109.11174 [astro-ph.HE]].

\bibitem{KamLAND:2011bnd}
A.~Gando \textit{et al.} [KamLAND],
Astrophys. J. \textbf{745} (2012), 193
[arXiv:1105.3516 [astro-ph.HE]].

\bibitem{Borexino:2019wln}
M.~Agostini \textit{et al.} [Borexino],
Astropart. Phys. \textbf{125} (2021), 102509
[arXiv:1909.02422 [hep-ex]].

\bibitem{SNO:2006dke}
B.~Aharmim \textit{et al.} [SNO],
Astrophys. J. \textbf{653} (2006), 1545-1551
[arXiv:hep-ex/0607010 [hep-ex]].

\bibitem{Lunardini:2008xd}
C.~Lunardini and O.~L.~G.~Peres,
JCAP \textbf{08} (2008), 033
[arXiv:0805.4225 [astro-ph]].

\bibitem{Suliga:2021hek}
A.~M.~Suliga, J.~F.~Beacom and I.~Tamborra,
Phys. Rev. D \textbf{105} (2022) no.4, 043008
[arXiv:2112.09168 [astro-ph.HE]].

\bibitem{Ando:2002ky}
S.~Ando, K.~Sato and T.~Totani,
Astropart. Phys. \textbf{18} (2003), 307-318
[arXiv:astro-ph/0202450 [astro-ph]].

\bibitem{Galais:2009wi}
S.~Galais, J.~Kneller, C.~Volpe and J.~Gava,
Phys. Rev. D \textbf{81} (2010), 053002
[arXiv:0906.5294 [hep-ph]].

\bibitem{Chakraborty:2010fft}
S.~Chakraborty, S.~Choubey and K.~Kar,
Phys. Lett. B \textbf{702} (2011), 209-215
[arXiv:1006.3756 [hep-ph]].

\bibitem{Priya:2017bmm}
A.~Priya and C.~Lunardini,
JCAP \textbf{11} (2017), 031
[arXiv:1705.02122 [astro-ph.HE]].

\bibitem{Moller:2018kpn}
K.~Moller, A.~M.~Suliga, I.~Tamborra and P.~B.~Denton,
JCAP \textbf{05} (2018), 066
[arXiv:1804.03157 [astro-ph.HE]].

\bibitem{Horiuchi:2017qja}
S.~Horiuchi, K.~Sumiyoshi, K.~Nakamura, T.~Fischer, A.~Summa, T.~Takiwaki, \textit{et al.},
Mon. Not. Roy. Astron. Soc. \textbf{475} (2018) no.1, 1363-1374
[arXiv:1709.06567 [astro-ph.HE]].

\bibitem{Kresse:2020nto}
D.~Kresse, T.~Ertl and H.~T.~Janka,
Astrophys. J. \textbf{909} (2021) no.2, 169
[arXiv:2010.04728 [astro-ph.HE]].

\bibitem{Tabrizi:2020vmo}
Z.~Tabrizi and S.~Horiuchi,
JCAP \textbf{05} (2021), 011
[arXiv:2011.10933 [hep-ph]].

\bibitem{Lunardini:2009ya}
C.~Lunardini,
Phys. Rev. Lett. \textbf{102} (2009), 231101
[arXiv:0901.0568 [astro-ph.SR]].

\bibitem{Nakazato:2015rya}
K.~Nakazato, E.~Mochida, Y.~Niino and H.~Suzuki,
Astrophys. J. \textbf{804} (2015) no.1, 75
[arXiv:1503.01236 [astro-ph.HE]].

\bibitem{Horiuchi:2020jnc}
S.~Horiuchi, T.~Kinugawa, T.~Takiwaki, K.~Takahashi and K.~Kotake,
Phys. Rev. D \textbf{103} (2021) no.4, 043003
[arXiv:2012.08524 [astro-ph.HE]].

\bibitem{Beacom:2003nk}
J.~F.~Beacom and M.~R.~Vagins,
Phys. Rev. Lett. \textbf{93} (2004), 171101
[arXiv:hep-ph/0309300 [hep-ph]].

\bibitem{JUNO:2022lpc}
A.~Abusleme \textit{et al.} [JUNO],
[arXiv:2205.08830 [hep-ex]].

\bibitem{Hyper-Kamiokande:2018ofw}
K.~Abe \textit{et al.} [Hyper-Kamiokande],
[arXiv:1805.04163 [physics.ins-det]].

\bibitem{DUNE:2016hlj}
R.~Acciarri \textit{et al.} [DUNE],
[arXiv:1601.05471 [physics.ins-det]].

\bibitem{Mathews:2014qba}
G.~J.~Mathews, J.~Hidaka, T.~Kajino and J.~Suzuki,
Astrophys. J. \textbf{790} (2014), 115
[arXiv:1405.0458 [astro-ph.CO]].

\bibitem{Schilbach:2018bsg}
T.~S.~H.~Schilbach, O.~L.~Caballero and G.~C.~McLaughlin,
Phys. Rev. D \textbf{100} (2019) no.4, 043008
[arXiv:1808.03627 [astro-ph.HE]].

\bibitem{Mathews:2019klh}
G.~J.~Mathews, L.~Boccioli, J.~Hidaka and T.~Kajino,
Mod. Phys. Lett. A \textbf{35} (2020) no.25, 2030011
[arXiv:1907.10088 [astro-ph.HE]].


\bibitem{Duan:2010bg}
H.~Duan, G.~M.~Fuller and Y.~Z.~Qian,
Ann. Rev. Nucl. Part. Sci. \textbf{60} (2010), 569-594
[arXiv:1001.2799 [hep-ph]].

\bibitem{Mirizzi:2015eza}
A.~Mirizzi, I.~Tamborra, H.~T.~Janka, N.~Saviano, K.~Scholberg, R.~Bollig, \textit{et al.},
Riv. Nuovo Cim. \textbf{39} (2016) no.1-2, 1-112
[arXiv:1508.00785 [astro-ph.HE]].

\bibitem{Horiuchi:2018ofe}
S.~Horiuchi and J.~P.~Kneller,
J. Phys. G \textbf{45} (2018) no.4, 043002
[arXiv:1709.01515 [astro-ph.HE]].

\bibitem{Volpe:2015rla}
C.~Volpe,
Int. J. Mod. Phys. E \textbf{24} (2015) no.09, 1541009
[arXiv:1506.06222 [astro-ph.SR]].

\bibitem{Wolfenstein:1977ue}
L.~Wolfenstein,
Phys. Rev. D \textbf{17} (1978), 2369-2374.

\bibitem{Mikheev:1986wj}
S.~P.~Mikheev and A.~Y.~Smirnov,
Nuovo Cim. C \textbf{9} (1986), 17-26.

\bibitem{Nakazato:2013maa}
K.~Nakazato,
Phys. Rev. D \textbf{88} (2013) no.8, 083012
[arXiv:1306.4526 [astro-ph.HE]].

\bibitem{Gonzalez-Garcia:2008mgl}
M.~C.~Gonzalez-Garcia and M.~Maltoni,
Phys. Lett. B \textbf{663} (2008), 405-409
[arXiv:0802.3699 [hep-ph]].

\bibitem{Berryman:2014qha}
J.~M.~Berryman, A.~de Gouvea and D.~Hernandez,
Phys. Rev. D \textbf{92} (2015) no.7, 073003
[arXiv:1411.0308 [hep-ph]].

\bibitem{SNO:2018pvg}
B.~Aharmim \textit{et al.} [SNO],
Phys. Rev. D \textbf{99} (2019) no.3, 032013
[arXiv:1812.01088 [hep-ex]].

\bibitem{deGouvea:2019goq}
A.~de Gouv\^ea, I.~Martinez-Soler and M.~Sen,
Phys. Rev. D \textbf{101} (2020) no.4, 043013
[arXiv:1910.01127 [hep-ph]].

\bibitem{Kachelriess:2000qc}
M.~Kachelriess, R.~Tomas and J.~W.~F.~Valle,
Phys. Rev. D \textbf{62} (2000), 023004
[arXiv:hep-ph/0001039 [hep-ph]].

\bibitem{Farzan:2002wx}
Y.~Farzan,
Phys. Rev. D \textbf{67} (2003), 073015
[arXiv:hep-ph/0211375 [hep-ph]].

\bibitem{Escudero:2019gfk}
M.~Escudero and M.~Fairbairn,
Phys. Rev. D \textbf{100} (2019) no.10, 103531
[arXiv:1907.05425 [hep-ph]].

\bibitem{Berryman:2014yoa}
J.~M.~Berryman, A.~de Gouv\^ea, D.~Hern\'andez and R.~L.~N.~Oliveira,
Phys. Lett. B \textbf{742} (2015), 74-79
[arXiv:1407.6631 [hep-ph]].

\bibitem{Chattopadhyay:2021eba}
D.~S.~Chattopadhyay, K.~Chakraborty, A.~Dighe, S.~Goswami and S.~M.~Lakshmi,
Phys. Rev. Lett. \textbf{129} (2022) no.1, 011802
[arXiv:2111.13128 [hep-ph]].

\bibitem{Ando:2003ie}
S.~Ando,
Phys. Lett. B \textbf{570} (2003), 11
[arXiv:hep-ph/0307169 [hep-ph]].

\bibitem{Fogli:2004gy}
G.~L.~Fogli, E.~Lisi, A.~Mirizzi and D.~Montanino,
Phys. Rev. D \textbf{70} (2004), 013001
[arXiv:hep-ph/0401227 [hep-ph]].

\bibitem{DeGouvea:2020ang}
A.~De Gouv\^ea, I.~Martinez-Soler, Y.~F.~Perez-Gonzalez and M.~Sen,
Phys. Rev. D \textbf{102} (2020), 123012
[arXiv:2007.13748 [hep-ph]].



\bibitem{Keil:2002in}
M.~T.~Keil, G.~G.~Raffelt and H.~T.~Janka,
Astrophys. J. \textbf{590} (2003), 971-991
[arXiv:astro-ph/0208035 [astro-ph]].



\bibitem{Dighe:1999bi}
A.~S.~Dighe and A.~Y.~Smirnov,
Phys. Rev. D \textbf{62} (2000), 033007
[arXiv:hep-ph/9907423 [hep-ph]].


\bibitem{Duan:2009cd}
H.~Duan and J.~P.~Kneller,
J. Phys. G \textbf{36} (2009), 113201
[arXiv:0904.0974 [astro-ph.HE]].

\bibitem{Barranco:2017lug}
J.~Barranco, A.~Bernal and D.~Delepine,
J. Phys. G \textbf{45} (2018) no.5, 055201
[arXiv:1706.03834 [astro-ph.CO]].

\bibitem{DiValentino:2021izs}
E.~Di Valentino, O.~Mena, S.~Pan, L.~Visinelli, W.~Yang, A.~Melchiorri, \textit{et al.},
Class. Quant. Grav. \textbf{38} (2021) no.15, 153001
[arXiv:2103.01183 [astro-ph.CO]].

\bibitem{Salpeter:1955it}
E.~E.~Salpeter,
Astrophys. J. \textbf{121} (1955), 161-167.

\bibitem{Ziegler:2022ivq}
J.~J.~Ziegler, T.~D.~P.~Edwards, A.~M.~Suliga, I.~Tamborra, S.~Horiuchi, S.~Ando and K.~Freese,
[arXiv:2205.07845 [astro-ph.GA]].

\bibitem{Yuksel:2008cu}
H.~Yuksel, M.~D.~Kistler, J.~F.~Beacom and A.~M.~Hopkins,
Astrophys. J. Lett. \textbf{683} (2008), L5-L8
[arXiv:0804.4008 [astro-ph]].


\bibitem{Madau:2014bja}
P.~Madau and M.~Dickinson,
Ann. Rev. Astron. Astrophys. \textbf{52} (2014), 415-486
[arXiv:1403.0007 [astro-ph.CO]].

\bibitem{Baldry:2003xi}
I.~K.~Baldry and K.~Glazebrook,
Astrophys. J. \textbf{593} (2003), 258-271
[arXiv:astro-ph/0304423 [astro-ph]].

\bibitem{Horiuchi:2008jz}
S.~Horiuchi, J.~F.~Beacom and E.~Dwek,
Phys. Rev. D \textbf{79} (2009), 083013
[arXiv:0812.3157 [astro-ph]].

\bibitem{Horiuchi:2011zz}
S.~Horiuchi, J.~F.~Beacom, C.~S.~Kochanek, J.~L.~Prieto, K.~Z.~Stanek and T.~A.~Thompson,
Astrophys. J. \textbf{738} (2011), 154-169
[arXiv:1102.1977 [astro-ph.CO]].

\bibitem{Lunardini:2005jf}
C.~Lunardini,
Astropart. Phys. \textbf{26} (2006), 190-201
[arXiv:astro-ph/0509233 [astro-ph]].

\bibitem{Sumiyoshi:2007pp}
K.~Sumiyoshi, S.~Yamada and H.~Suzuki,
Astrophys. J. \textbf{667} (2007), 382-394
[arXiv:0706.3762 [astro-ph]].

\bibitem{Hudepohl:2013zsj}
L. Hudepohl,
Munich, Tech. U. (2013).

\bibitem{Kim:1990km}
C.~W.~Kim and W.~P.~Lam,
Mod. Phys. Lett. A \textbf{5} (1990), 297-299.

\bibitem{ParticleDataGroup:2020ssz}
P.~A.~Zyla \textit{et al.} [Particle Data Group],
PTEP \textbf{2020} (2020) no.8, 083C01


\bibitem{Beacom:2002cb}
J.~F.~Beacom and N.~F.~Bell,
Phys. Rev. D \textbf{65} (2002), 113009
[arXiv:hep-ph/0204111 [hep-ph]].

\bibitem{Capozzi:2021fjo}
F.~Capozzi, E.~Di Valentino, E.~Lisi, A.~Marrone, A.~Melchiorri and A.~Palazzo,
Phys. Rev. D \textbf{104} (2021) no.8, 083031
[arXiv:2107.00532 [hep-ph]].

\bibitem{BQ}
B. Quilain, Private communication.

\bibitem{JUNO:2015zny}
F.~An \textit{et al.} [JUNO],
J. Phys. G \textbf{43} (2016) no.3, 030401
[arXiv:1507.05613 [physics.ins-det]].

\bibitem{Strumia:2003zx}
A.~Strumia and F.~Vissani,
Phys. Lett. B \textbf{564} (2003), 42-54
[arXiv:astro-ph/0302055 [astro-ph]].

\bibitem{GMP}
The cross section is available on SNOwGLobES:
https://github.com/SNOwGLoBES/snowglobes/tree
/master/xscns.



\bibitem{KH}
H. Kunxian,
PhD Thesis, Kyoto University (2015).

\bibitem{Barenboim:2020vrr}
G.~Barenboim, J.~Z.~Chen, S.~Hannestad, I.~M.~Oldengott, T.~Tram and Y.~Y.~Y.~Wong,
JCAP \textbf{03} (2021), 087
[arXiv:2011.01502 [astro-ph.CO]].


\bibitem{Chen:2022idm}
J.~Z.~Chen, I.~M.~Oldengott, G.~Pierobon and Y.~Y.~Y.~Wong,
Eur. Phys. J. C \textbf{82} (2022) no.7, 640
[arXiv:2203.09075 [hep-ph]].


\end{thebibliography}
\end{document}